\documentclass[aip,reprint]{revtex4-1}

\usepackage[utf8]{inputenc}

\usepackage{xr}
\usepackage{dcolumn}
\usepackage{siunitx}
\usepackage{mathtools}
\usepackage{upgreek}
\usepackage{physics}
\usepackage{bm}
\usepackage{amssymb}
\usepackage[version=4]{mhchem}

\usepackage[colorlinks=true,linkcolor=blue,citecolor=blue,urlcolor=blue]{hyperref}

\newcommand{\ie}{i.e.\ }

\newcommand{\etal}{et al.\ }

\renewcommand{\rm}{\mathrm}

\newcommand{\vbr}{\vb{r}}
\newcommand{\vbo}{\boldsymbol{\upomega}}
\newcommand{\vbp}{\vb{p}}

\newcommand{\RI}{\vb{R}_I}
\newcommand{\nel}{n_\mathrm{e}}
\newcommand{\nc}{n_\mathrm{c}}
\newcommand{\sol}{\mathrm{sol}}
\newcommand{\vdW}{\mathrm{vdW}}
\newcommand{\solv}{\mathrm{solv}}
\newcommand{\ion}{\mathrm{ion}}
\newcommand{\diel}{\mathrm{diel}}
\newcommand{\cav}{\mathrm{cav}}
\newcommand{\rot}{\mathrm{rot}}
\newcommand{\pol}{\mathrm{pol}}
\newcommand{\vac}{\mathrm{s}}
\newcommand{\eps}{\epsilon_0}
\newcommand{\Eloc}{\mathcal{E}}
\newcommand{\vbEloc}{\vb*{\mathcal{E}}}
\newcommand{\bulk}{\mathrm{b}}

\newcommand{\intr}{\int \dd[3]{\vbr}}
\newcommand{\intO}{\int \frac{\dd{\vbo}}{4\pi}}

\newcommand{\oparg}[1]{\{#1\}}
\newcommand{\bigv}{\vphantom{\qty\big()}}
\newcommand{\Bigv}{\vphantom{\qty\Big()}}

\DeclareMathOperator\erfc{erfc}

\DeclareSIUnit{\e}{e}
\DeclareSIUnit{\molar}{M}
\DeclareSIUnit{\angstrom}{\text {\AA}}
\DeclareSIUnit{\bar}{bar}

\newcommand{\SM}{\hyperref[SM]{Supporting Information}}

\draft % marks overfull lines with a black rule on the right

\makeatletter

\newcommand*{\addFileDependency}[1]{% argument=file name and extension
\typeout{(#1)}% latexmk will find this if $recorder=0
\@addtofilelist{#1}
\IfFileExists{#1}{}{\typeout{No file #1.}}
}\makeatother

\newcommand*{\myexternaldocument}[2][]{%
\externaldocument[#1]{#2}%
\addFileDependency{#2.tex}%
\addFileDependency{#2.aux}%
}
\myexternaldocument[SI-]{SI}

\begin{document}

%\preprint{}

\title{A nonlocal and nonlinear implicit electrolyte model for plane wave density functional theory} %Title of paper



\author{S M Rezwanul Islam}
\affiliation{Department of Chemical Engineering, Louisiana State University, Baton Rouge, Louisiana 70803, USA}
\author{Foroogh Khezeli}
\affiliation{Department of Chemical Engineering, Louisiana State University, Baton Rouge, Louisiana 70803, USA}
\author{Stefan Ringe}
\affiliation{Department of Chemistry, Korea University, Seoul 02841, Republic of Korea}
\author{Craig Plaisance}
%\thanks{}
\affiliation{Department of Chemical Engineering, Louisiana State University, Baton Rouge, Louisiana 70803, USA}


\date{\today}

\begin{abstract}
We have developed and implemented an implicit electrolyte model in the Vienna Ab initio Simulation Package (VASP) that includes nonlinear dielectric and ionic responses as well as a nonlocal definition of the cavities defining the spatial regions where these responses can occur. The implementation into the existing VASPsol code is numerically efficient and exhibits robust convergence, requiring computational effort only slightly higher than the original linear polarizable continuum model (LPCM). The nonlinear+nonlocal model is able to reproduce the characteristic `double hump' shape observed experimentally for the differential capacitance of an electrified metal interface while preventing the `leakage' of the electrolyte into regions of space too small to contain a single water molecule or solvated ion. The model also gives a reasonable prediction of molecular solvation free energies as well as the self-ionization free energy of water and the absolute electron chemical potential of the standard hydrogen electrode. All of this, combined with the additional ability to run constant potential density functional theory calculations, should enable the routine computation of activation barriers for electrocatalytic processes.
\end{abstract}


\maketitle %\maketitle must follow title, authors, abstract and \pacs


\section{Introduction} \label{sec:intro}

Understanding chemical processes occurring at the electrochemical interface is crucial to the design of improved electrocatalysts for processes such as the oxygen evolution reaction, \ce{CO2} reduction \cite{RingeCO2,RingeDL,ChenEF,ShinCO2,Marcandalli,MonteiroCO2},  
or the hydrogen evolution reaction (HER) \cite{Ringecation}.
The physics of the solid-liquid interface also plays a major role in nanoparticle synthesis, chemical reactions at electrode surfaces, and other areas that are important to energy applications \cite{VASPsol1,VASPsol2,20}. 
In order to better understand the processes occurring at such interfaces, it is invaluable to have accurate and computationally inexpensive methods for modeling them. This is particularly true for the routine calculation of activation barriers for electrocatalytic processes, which are often ignored in computational studies that instead only consider the thermodynamics of catalytic intermediates \cite{CHE}. Development of such models is an ongoing challenge in which significant progress has been made, but further improvement is still required \cite{3}.

A fundamental understanding of the electrochemical interface via experimental studies is challenging due to the heterogeneous nature and the complexities of the solid-liquid systems involved \cite{VASPsol1,4}. Computational studies can provide these insights directly at the atomic level \cite{VASPsol1,5,ShinEDL,Deissenbeckdielectric,LePt111}; 
however, a complete atomic-level understanding of electrochemical interfaces by quantum chemical simulation is hampered by the computationally demanding task of sampling a large number of degrees of freedom associated with the electrolyte \cite{VASPsol1,20}.
Computational models for the electrode-electrolyte interface are required to accurately describe the change in interfacial charge distribution with potential and predict the mechanisms and kinetics of electrochemical reactions. Calculating the charge on a given surface at a given potential requires approximating a thermodynamic average over all solvent and electrolyte configurations. This explicit treatment is computationally expensive due to the large number of configurations and large system sizes required to appropriately capture the equilibrium properties of the system \cite{VASPsol1,3}. While it is possible to develop coarse-grained models based on molecular mechanics, these models have the disadvantage that they cannot straightforwardly be parameterized for a different system making them less promising for use in material screening.
%CP what do you mean by "Coarse-grained models based on molecular mechanics"? Do you have a ref for this type of approach?

Implicit solvation models on the other hand resemble a highly coarse-grained scheme with exceptional computational efficiency, but often acceptable accuracy. In particular, implicit solvation methods have already been shown to be able to explain several phenomena at electrified solid-liquid interfaces, such as cation effects \cite{RingeCO2} or surface reconstruction \cite{Huangpotential}, and have been also used for computational screening of electrocatalysts \cite{RingeEC}. Therefore, it is essential to develop efficient and accurate implicit computational frameworks for the study of solid-liquid interfaces.

Implicit electrolyte approaches model the surface quantum mechanically and represent the electrolyte implicitly using a continuum electrostatic screening scheme \cite{4}. These types of models were originally developed for the solvation of small molecules in non-periodic systems \cite{Cramer1,Cramer2,Cramer3}.
Implicit solvation models for plane wave density functional theory (DFT) codes were first developed by Fattebert and Gygi \cite{6}. 
%%SR at this point an explanation of JDFT is needed. I would rather then move JDFT to a later point with more explanation what it really is.
%CP I just took out the part about JDFT since we are not using this framework to obtain the new model. I've never been a fan of writing mini-reviews in the intro to my papers
In these models, the spatial extent of the continuum electrolyte is derived from the surface electron density resulting in a numerically stable implementation for periodic systems within a plane wave basis \cite{3}. Later these models were extended by Marzari \etal \cite{8} to include cavitation and dispersion interactions, resulting in the self-consistent continuum solvation (SCCS) model. Recently, Mathew \etal implemented an almost identical model developed by Gunceler \etal \cite{Guncelar1}, the linear polarizable vontinuum model (LPCM), into the Vienna Ab initio Simulation Package (VASP). The resulting implementation, called VASPsol \cite{VASPsol1,VASPsol2}, was the first implicit solvation method made available in a widely used plane wave DFT code. This model also includes ionic screening in the electrolyte by the linear Poisson-Boltzmann model, allowing for the study of charged electrochemical interfaces under an applied external potential.

Despite the utility of the LPCM model implemented in VASPsol, there are still several shortcomings that limit the accuracy of the results for modeling the charged electrochemical interface. First, as with most implicit solvation models available in commonly used plane wave DFT codes, the LPCM model describes the volume of space occupied by the electrolyte in terms of a solvent cavity function \cite{9}. The value of the cavity function at a particular point in space is typically determined only by the solute (\ie surface) electron density at the same point in space, giving a purely local definition of the cavity. This ignores the finite size of water molecules and solvated ions, in some cases allowing the cavity to penetrate into regions of space that would otherwise be too small to accommodate an entire water molecule or solvated ion. This phenomenon, of `solvent leakage' is especially noticeable when simulating explicit bulk water itself, where implicit water enters into the interstitial spaces between the explicit water molecules \cite{solvent-aware}. By instead using a nonlocal definition of the cavity, as in the solvent-aware interface \cite{solvent-aware} implemented in the Environ package for Quantum Espresso or in the SaLSA model \cite{SaLSA} implemented in the JDFTx code \cite{JDFTx}, this solvent leakage is prevented.

%%SR it sounds like these are limitations of the SCCS model only. I think it is better to speak about the limitations of the current VASP model because Environ/QE do have non-local models and all these things implemented. Otherwise, it could make them in a bad mood if they become the reviewer.
%%CP agreed, but i thought it was pretty clearly written that these limitations are only associated with the SCCS model (the model that VASPsol is based on)
The second limitation of the LPCM model is that both the dielectric and ionic responses in the electrolyte are linear. In the presence of strong electric fields associated with highly charged electrodes, both of these responses become nonlinear and lead to a characteristic `double hump' shape of the differential capacitance curve \cite{Sundararaman1}. The nonlinear effects arise from saturation of the dielectric and ionic responses in the presence of high electric fields, as well as the exponential enhancement of the ionic response at moderate field strengths due to Boltzmann statistics. A linear model such as the LPCM is not capable of reproducing these effects and thus does not reproduce the decrease in capacitance observed experimentally for high electrode polarization \cite{Sundararaman1}.
%%SR citation for this?
%%CP added
Models including nonlinear dielectric \cite{Guncelar1,Guncelar2} and ionic \cite{Guncelar1,Guncelar2,Ringe} responses have been developed, but are not available in widely used periodic DFT codes such as VASP. Furthermore, the nonlinear models have not to our knowledge been combined with nonlocal models of the cavity.

In this paper, we develop an accurate and computationally efficient implicit electrolyte model that can account for both the nonlinear and nonlocal effects described above. We have implemented this electrolyte model within the framework of the original VASPsol code, and thus name it VASPsol++. 
%%SR above it was called VASPsol2
%%CP that was just the bibtex label, no?
It is found to be both numerically efficient and robust, with computational cost only slightly exceeding that of the original VASPsol method despite the significantly higher complexity of the model. 
%%SR I think this sounds a bit negative. The model complexity is also much higher, so the computational costs are of course higher. I think this should be stressed here.
%%CP I rewrote a bit
We show that the resulting model is able to overcome the issues with other implicit electrolyte models that are described above, giving good results for describing a variety of solvated systems including molecules, ion, and metal electrode interfaces.

This paper is organized as follows. We begin by describing the theoretical framework in terms of the cavity definitions, the free energy functional, and the ground state solution in Section \ref{sec:theory}. Then in Section \ref{sec:implementation}, we describe details of the numerical implementation such as the efficient and robust techniques used to solve the resulting nonlinear Poisson-Boltzmann equation and integration with the self-consistent field cycle in VASP. Finally, we present the results of several case studies on systems ranging from electrified metal surfaces to solvated molecules and ions in Section \ref{sec:results}.

\section{Theoretical framework} \label{sec:theory}

In this section, the theoretical framework underlying the solvation model will be described. We start out by discussing the nonlocal definitions of the dielectric and ionic cavities that we have developed to remedy the issue of `solvent leakage'. We then describe the nonlinear free energy functional that characterizes the solute/solvent system. Application of the variational principle leads to the governing equations for the ground state polarization density, ionic concentrations, electrostatic potential, and electron density, along with a simplified expression for the ground state free energy of the solute/solvent system.

\subsection{Dielectric and ionic cavity definitions} \label{sec:cavity}

As with most other implicit solvation methods, dielectric and ionic responses are limited to regions of space that are physically accessible to water molecules and solvated ions, respectively. The water molecules and ions are typically treated as point-like dipoles and charges such that the dielectric and ionic regions are characterized by smooth cavity functions. This cavity function effectively scales the bulk dielectric and ionic responses by modulating the densities of the molecules and ions of the implicit electrolyte \cite{5,19}, decaying to zero in the region of space occupied by the solute and approaching unity in the bulk electrolyte. Ideally, different cavities are desired for the dielectric and ionic responses; however, due to a combination of numerical challenges and difficulties in parameterization, most solvation models only employ a single cavity for both types of responses \cite{19}.

Most implicit solvation methods define the cavity based on either a radial cutoff or an electron density cutoff. Cavities based on radial cutoffs are highly flexible and can deliver high accuracy, but are difficult to parameterize since a different cutoff radius must be defined for each type of atom. On the other hand, cavities based on an electron density cutoff can in principle be constructed from only a single parameter (the cutoff density) and are able to adapt self-consistently to the electron density \cite{9,19,20}.

A shortcoming of both types of cavity definitions is that neither takes into account the finite size of water molecules and solvated ions. Values of the cutoff density that realistically represent the approach of water molecules and ions to a planar surface or a nanoparticle also allow these species to enter concave regions that are too small to contain a single water molecule or solvated ion. This leads to the phenomenon of `solvent leakage' whereby the electrolyte enters regions of space that it physically cannot enter when accounting for the finite sizes of the electrolyte species. An extreme example of this phenomenon can be seen in a simulation of explicit bulk water. Physically, implicit water should not be present at all as the entire simulation volume is already occupied by explicit bulk water; however, implicit solvation models with a local cavity definition are found to place a significant amount of implicit water inside of the explicit water \cite{solvent-aware}. Approaches have been introduced to address the issue of solvent leakage, such as the SaLSA method of Sundararaman \etal \cite{SaLSA} or the solvent-aware interface of Andreussi \etal \cite{solvent-aware} which both make the solvent cavity a nonlocal function of the solute electron density. While both of these approaches are effective at eliminating solvent leakage, we propose an alternative cavity definition that closely represents the underlying physics of the solute-electrolyte system.

\begin{figure}
    \includegraphics[width=.45\textwidth]{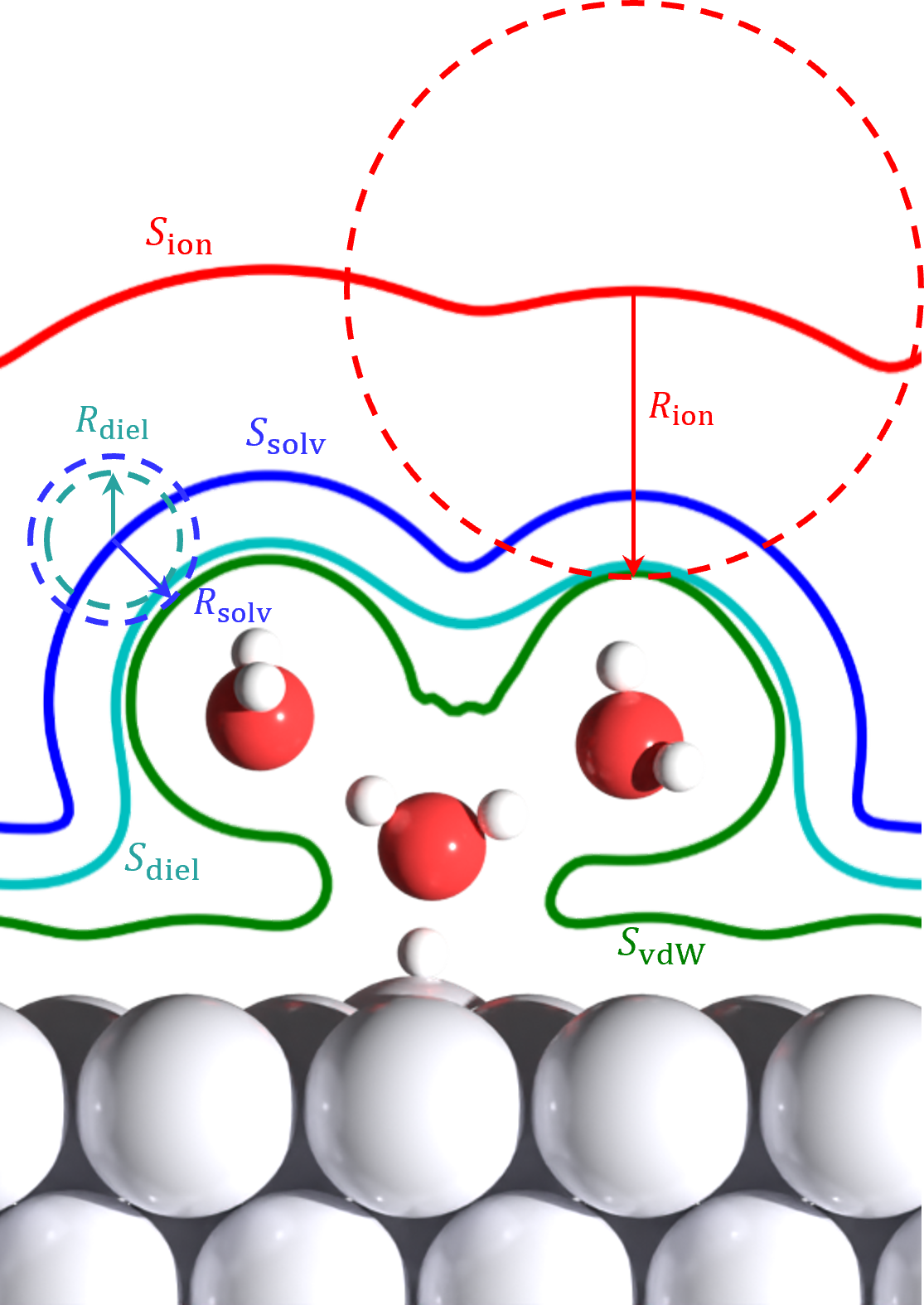}
    \caption{Schematic demonstrating construction of the different cavity functions for the transition state of the Volmer reaction occurring on a Pt(111) surface. The surface of the van der Waals cavity, $S_\vdW$, corresponds to the isosurface where the solute electron density is equal to the cutoff electron density $\nc$. The solvent and ionic cavities, $S_\solv$ and $S_\ion$, are displaced outwards from the van der Waals cavity by distances equal to the solvent and ionic radii, $R_\solv$ and $R_\ion$, respectively. The dielectric cavity $S_\diel$ is displaced inwards from the solvent cavity by a distance equal to the dielectric radius $R_\diel$ of the solvent.}
    \label{fig:cavity_hand}
\end{figure}

To illustrate our proposed cavity definition, consider the system depicted in Figure \ref{fig:cavity_hand}. The solute can be characterized by a \emph{van der Waals cavity} denoted $S_\vdW$ that corresponds to the region of space in which the van der Waals volume of a solvent molecule cannot penetrate. The van der Waals volume of a solvent molecule is characterized by the region of space extending a distance $R_\solv$ (the solvent radius) from the geometric center of the molecule. Extending the van der Waals cavity of the solute outwards by a distance $R_\solv$, the \emph{solvent  cavity} $S_\solv$ is obtained. This cavity corresponds to the region of space occupied by the \emph{centers} of the solvent molecules and is often referred to as the \emph{solvent accessible region}. 
%%SR For clarity, I would suggest to draw a solute and also a solvent molecule in Fig. 1 including the different radii as well
%%CP Agreed, we'll work on this but it might take some time. Might include in the revisions if it takes too long
%%SR further, the cavity in Fig. 1 seems to be not smooth but have kinks, I would rather take an actual cavity from the simulations and draw the 2D cut of the cavity to have smooth lines. You can then also indicate the parameter values in the caption of that figure
%%CP Do you have any suggestions on visualizers that can do this? What did you typically use? Rezwan made a smoother version, but I agree that an actual isocontour would be better.
Using this definition, the solvent centers will fill all regions of space in which the van der Waals volume of the solvent molecule does not overlap with the van der Waals cavity of the solute. Finally, we must account for the fact that a solvent molecule like water is not a point dipole but rather has a finite charge distribution. This means that the dielectric response will actually extend closer to the solute than does the solvent center cavity $S_\solv$. The corresponding \emph{dielectric cavity} is obtained by extending the solvent center cavity back towards the solute a distance $R_\diel$ (the dielectric radius).

In addition to the solvent and dielectric cavities, we also define an \emph{ionic cavity} $S_\ion$ that represents the region of space accessible to the centers of solvated ions in the electrolyte. This cavity is obtained from the van der Waals cavity in a similar way as the solvent cavity, but using the ionic radius $R_\ion$ instead of the solvent radius $R_\solv$.

Construction of the cavities described above is not straightforward in the plane wave representation that is required for incorporation into periodic DFT codes. First, the cavities must be smooth enough to represent on the same plane wave grid as the solute electron density. 
%%SR check the grammar of the previous sentence (and 2x represent). Maybe: First, the cavities must be smooth enough to be represented with the same plane wave grid as the solute electron density.
%%CP fixed
Second, functional derivatives of the cavities with respect to the solute electron density must be smooth enough that the effective cavity contribution to the Kohn-Sham potential can be included in the self-consistent field iterations without leading to numerical instabilities.

In general, each cavity $S(\vbr)$ is constructed from an effective electron density $n(\vbr)$. The first step is to define the dimensionless logarithm $x$ of the effective electron density,
\begin{equation}\label{eq:x}
    x \equiv \ln(\frac{n}{\nc})
    \quad .
\end{equation}
The same cutoff electron density $\nc$ is used for all cavities. The cavity $S$ is constructed from $x$ using a shape function based on the complementary error function, as is done in the original VASPsol implementation \cite{VASPsol1,VASPsol2},
\begin{equation} \label{eq:S}
    S \equiv \frac12 \erfc\qty(\frac{x}{\sigma\sqrt{2}})
    \quad ,
\end{equation}
where $\sigma$ is a parameter that controls the sharpness of the cavity function.
%%SR citation missing. please cite my the ionic parameter work
%%CP I think you use a different definition in your paper, right? We'll still cite it somewhere else (probably in the results and intro)
To more compactly represent a cavity function $S(\vbr)$ in terms of an effective electron density $n(\vbr)$ using eqs \eqref{eq:x} and \eqref{eq:S} we write it as,
\begin{equation} \label{eq:S_n}
    S(\vbr) = S\oparg{n}(\vbr)
    \quad .
\end{equation}

Having defined a general cavity function $S(\vbr)$ in terms of an effective electron density $n(\vbr)$, we now discuss how the specific cavity functions $S_\vdW(\vbr)$, $S_\solv(\vbr)$, $S_\diel(\vbr)$, and $S_\ion(\vbr)$ are obtained from the solute electron density $\nel(\vbr)$. The procedure begins with the relatively straightforward construction of the van der Waals cavity $S_\vdW(\vbr)$ from the solute electron density,
\begin{equation} \label{eq:S_vdw}
    S_\vdW(\vbr) = S\oparg{\nel}(\vbr)
    \quad .
\end{equation}
The transition of the van der Waals cavity from zero inside the solute to unity in the bulk electrolyte is centered at the isosurface of the solute electron density corresponding to the cutoff density $\nc$. We note that this is the only cavity used in the original VASPsol implementation where it is used there for both the dielectric and ionic cavities.

\begin{figure}
    \includegraphics[width=.45\textwidth]{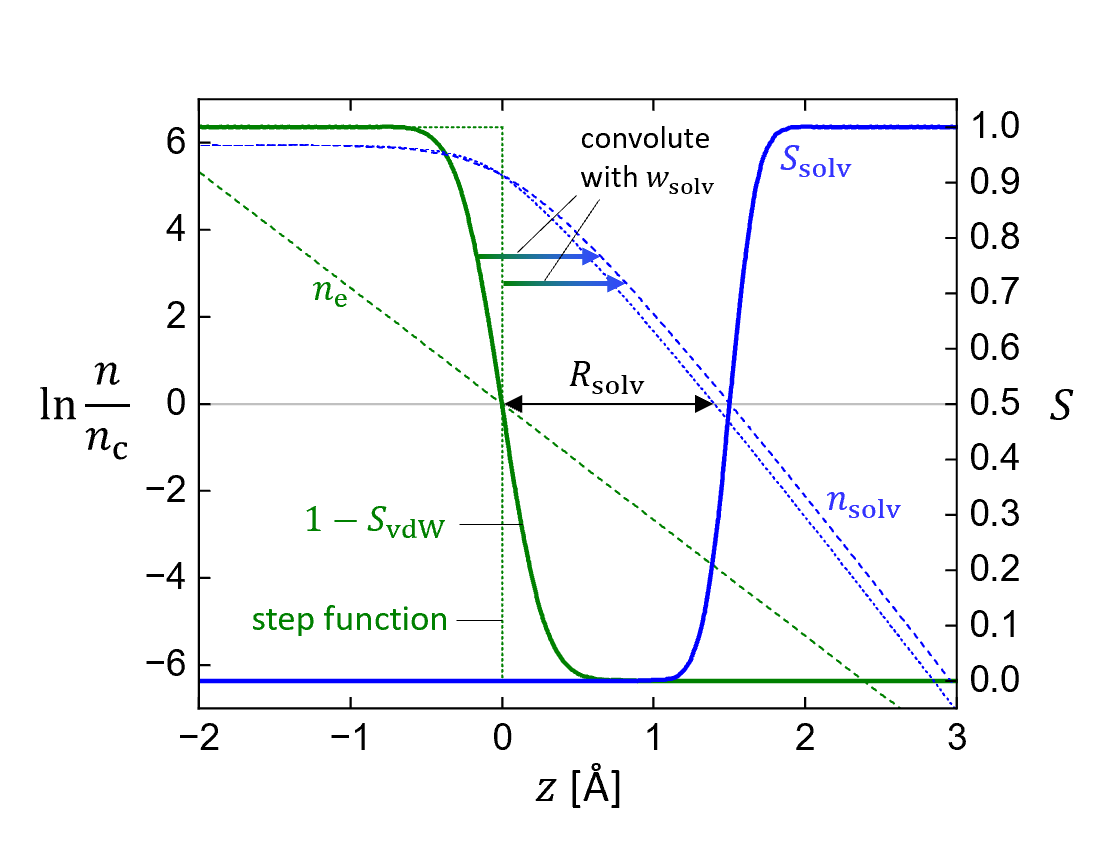}
    \caption{Schematic demonstrating construction of $S_\vdW$ and $S_\solv$ from effective electron densities. The van der Waals cavity $S_\vdW$ is constructed from the solute electron density $\nel$ according to eq \eqref{eq:S_vdw}, while the solvent cavity $S_\solv$ is constructed from the effective electron density $n_\solv$ according to eq \eqref{eq:S_solv}. The latter is obtained from the van der Waals cavity using eq \eqref{eq:n_solv} by convolution with a density kernel $w_\solv$. The resulting effective electron density decays to a value of $\nc$ at a distance of approximately $R_\solv$ from the surface of the van der Waals cavity. The distance is exactly equal to $R_\solv$ if the van der Waals cavity is represented by a planar step function, as shown by the dotted lines.}
    \label{fig:cavity_construction}
\end{figure}

The solvent cavity should ideally be displaced a distance $R_\solv$ from the van der Waals cavity. The most computationally convenient way to construct such a cavity is by defining it in terms of an effective electron density $n_\solv(\vbr)$ using eq \eqref{eq:S},
\begin{equation} \label{eq:S_solv}
    S_\solv(\vbr) = S\oparg{n_\solv}(\vbr)
    \quad .
\end{equation}
The effective electron density has no physical meaning; it is only used computationally to define the cavity functions by eq \eqref{eq:S_n}. This is depicted graphically in Figure \ref{fig:cavity_construction}. The effective electron density should ideally have an isosurface corresponding to the cutoff density $\nc$ that is displaced outwards from the edge of the van der Waals cavity by a distance equal to the solvent radius $R_\solv$. Such an effective electron density can be constructed by convoluting the complement of the van der Waals cavity ($1-S_\vdW$) with a density kernel $w_\solv$,
\begin{equation} \label{eq:n_solv}
    n_\solv = \nc w_\solv \ast \qty(1 - S_\vdW)
    \quad .
\end{equation}
%%SR In the SI, from eq. 3 to 4, how do you calculate this integral? Because S depends on n which you do not have at that point? I think it will be better to write this step-by-step.
%%CP Yes, it did not make it clear enough. S is a step function as written in the SI text. I made this more clear in the SI now
This and other density kernels are taken to be exponential functions decaying over the length $b = a/\sigma$,
\begin{equation} \label{eq:w}
    w_i(\vbr) = \qty(\frac{2b}{2b+R_i})\qty(\frac{1}{4\pi b^3})\exp(-\frac{r-R_i}{b})
    \quad ,
\end{equation}
where $r = \abs{\vbr}$. The index $i$ indicates which cavity the density kernel is being used to define, with $i = (\solv, \ion, \diel, \cav)$. A prefactor is chosen so that the effective electron density attains a value of $\nc$ a distance $R_i$ away from the $S=0.5$ isosurface of a planar cavity function, as depicted in Figure \ref{fig:cavity_construction} and detailed in the \SM. 
Thus, if the interface is planar then the solvent cavity will be displaced exactly $R_\solv$ from the van der Waals cavity. For concave interfaces, the solvation cavity will be slightly further away while for convex interfaces it will be slightly closer. Using a smaller value for the decay length of the density kernel reduces this deviation, thus we use the smallest value $a$ that can be numerically accommodated by the FFT grid.

The ionic cavity is constructed in an analogous way to the solvent cavity but using the ionic radius $R_\ion$ to construct an effective electron density $n_\ion$,
\begin{equation} \label{eq:S_ion}
    S_\ion(\vbr) = S\oparg{n_\ion}(\vbr)
    \quad ,
\end{equation}
\begin{equation} \label{eq:n_ion}
    n_\ion = \nc w_\ion \ast \qty(1 - S_\vdW)
    \quad .
\end{equation}
Thus, it will ideally be displaced a distance $R_\ion$ from the van der Waals cavity. The dielectric cavity should ideally be displaced inwards a distance $R_\diel$ from the solvent cavity. It is also constructed from an effective electron density $n_\diel(\vbr)$ according to,
\begin{equation}  \label{eq:S_diel}
    S_\diel(\vbr) = 1 - S\oparg{n_\diel}(\vbr)
    \quad ,
\end{equation}
\begin{equation}  \label{eq:n_diel}
    n_\diel = \nc w_\diel \ast S_\solv
    \quad .
\end{equation}

Finally, we define a fifth cavity $S_\cav(\vbr)$ that is used to compute the surface area appearing in the expression for the cavity formation energy discussed in the next section. This cavity is constructed from the solvent cavity in an analogous way to the dielectric cavity, but using a displacement $R_\cav$ rather than $R_\diel$,
\begin{equation} \label{eq:S_cav}
    S_\cav(\vbr) = 1 - S\oparg{n_\cav}(\vbr)
    \quad ,
\end{equation}
\begin{equation} \label{eq:n_cav}
    n_\cav = \nc w_\cav \ast S_\solv
    \quad .
\end{equation}

To simplify the development in the next section, we introduce the notation $S_\diel\oparg{\nel}(\vbr)$, $S_\ion\oparg{\nel}(\vbr)$, and $S_\cav\oparg{\nel}(\vbr)$ to indicate the cavities obtained from the solute electron density by chaining together eqs \eqref{eq:S_vdw} -- \eqref{eq:n_solv} with \eqref{eq:S_diel} -- \eqref{eq:n_diel} (for $S_\diel$), \eqref{eq:S_ion} -- \eqref{eq:n_ion} (for $S_\ion$), and \eqref{eq:S_cav} -- \eqref{eq:n_cav} (for $S_\cav$).

\subsection{Free energy functional} \label{sec:Atot}

Having specified the cavity definitions for the regions of space containing the dielectric and ionic responses, we can now fully describe the model in terms of a free energy functional of the combined solute/electrolyte system,
%%SR I guess this one is the same as from JDFT? Then it should be cited or mentioned that it has been adapted from it.
%%SR In general this section seems like you derived everything by yourself, so proper citations are needed to the relevant literature if you have not.
%%CP This is my own formulation of the free energy function, but it is quite similar to the one in the Gunceler paper (10.1088/0965-0393/21/7/074005). The major difference is that they do not use phi as a degree of freedom, which leads to issues with net charge in periodic systems. This formulation (like yours and the VASPsol formulation) do not have this problem. I also include the interactions between the bound and ionic charges and the solute in Adiel and Aion, respectively, as this makes the minimization notationally simpler. So it's somewhat of a combination between the Gunceler, VASPsol, and your formulations. I've added text to point this out.
\begin{equation}
\label{eq:Atot}
\begin{split}
    A_\rm{tot} &= A_\rm{TXC}[\nel] 
    + \intr \qty[e\nel(\vbr) + \rho_Z\oparg{\RI}(\vbr)] \phi(\vbr) \\
    &- \frac{\eps}{8\pi} \intr \abs{\grad\phi(\vbr)}^2 
    - \mu_\rm{e} \qty[\intr \nel(\vbr) - N_\rm{e}] \\
    &+ A_\diel[\nel,\phi,\rho_\rot,\vbp_\pol] 
    + A_\ion[\nel,\phi,\theta_+,\theta_-,\theta_\vac] \\
    &+ A_\cav[\nel]
    \quad.
\end{split}
\end{equation}
%%SR should the eps0/8pi term not be with a negative sign in front? Or is this due to the different sign convention because in my case it always had the negative sign in front
%%CP Yes, you are right. I fixed it
%%SR also in my case it was not eps0/8pi but epsr(r)*eps0/8pi and the sp(r) is in the integral. I thought that maybe you included this in Adiel, but I could not really see it there. If it really is the same, some comment on that would be great, as this is the form most implicit solvation models use. Like, for which limit your Adiel converges into this form.
%%CP There is no epsr(r) in this model b/c the dielectric response is nonlinear and is defined from a statistical mechanics model. In the bulk linear limit, epsr arises as discussed in eqs 39-45. The specific form you are referring to would arise from the "eps0/8pi" term and Adiel calculated with the dielectric ground state in the linear bulk limit. This latter term contributes the "(epsr(r)-1)*eps0/8pi" term
This functional form is inspired by the nonlinear polarizable continuum model proposed by Gunceler \etal \cite{Guncelar1}, although we decompose it differently into the individual terms. Additionally, we treat the electrostatic potential $\phi(\vbr)$ as an explicit degree of freedom. This avoids issues arising from charged surfaces in periodic systems that are encountered in the formulation of Gunceler \etal, and is inspired from the free energy functional used in the original VASPsol implementation \cite{VASPsol1,VASPsol2} and the modified Poisson-Boltzmann model developed by Ringe \etal \cite{Ringe}.

The free energy functional depends on eight degrees of freedom -- the nuclear coordinates of the solute atoms $\RI$, the solute electron density $\nel(\vbr)$, the electrostatic potential $\phi(\vbr)$, the rotational distribution $\rho_\rot(\vbr,\vbo)$ and internal polarization $\vbp_\pol(\vbr)$ of the solvent molecules, and the electrolyte `site' occupancies of cations $\theta_+(\vbr)$, anions $\theta_-(\vbr)$, or solvent $\theta_\vac(\vbr)$. It is composed of six contributions plus a constraint on the total number of electrons $N_\rm{e}$ in the system enforced by specifying the electron chemical potential $\mu_\rm{e}$.

The first term $A_\rm{TXC}$ accounts for the kinetic, exchange, and correlation energy of the electrons in the solute, while the second accounts for the interaction of the electrostatic potential with the solute ($\rho_Z(\vbr)$ is the nuclear charge density of the solute using an `electron is positive' sign convention). The third term is the self-energy of the electric field, while the fourth term implements the constraint on the number of electrons in the solute. Together, these first four terms comprise the Kohn-Sham free energy functional in the absence of the electrolyte.

The free energy of the electrolyte is contained in the remaining three terms of the free energy functional. The first two account for the free energy to alter the rotational distribution and polarization of the solvent molecules ($A_\diel$) and the spatial distribution of the ions ($A_\ion$) from the bulk distributions in response to the electric field. The last term, $A_\cav$, quantifies the free energy required to form the solute cavity from the bulk electrolyte.

\subsubsection{Dielectric free energy} \label{sec:Adiel}
The dielectric free energy functional is inspired from the functional proposed by Gunceler \etal \cite{Guncelar1}, although we employ different notation and groupings of terms. It is a functional of both the solute electron density $\nel(\vbr)$ (via $S_\diel(\vbr)$) and the polarization degrees of freedom of the electrolyte ($\rho_\rot(\vbr,\vbo)$ and $\vbp_\pol(\vbr)$, where $\vbo$ is an angular direction). It is convenient to define the molecular dielectric free energy $\bar{A}_\diel(\vbr)$ of a solvent molecule at position $\vbr$ so that the dependence on $\nel(\vbr)$ can be separated out,
\begin{equation}
    A_\diel = n_\rm{mol} \intr S_\diel\oparg{\nel}(\vbr) \bar{A}_\diel\oparg{\phi,\rho_\rot,\vbp_\pol}(\vbr)
    \quad ,
\end{equation}
with $n_\rm{mol}$ being the bulk solvent molecular density. The molecular dielectric free energy is written in terms of a rotational contribution $\bar{A}_\rot$ that accounts for the rotational entropy of the molecules, an internal polarization contribution $\bar{A}_\pol$ that accounts for the energy to distort the electrons and nuclei in a single solvent molecule, a self-interaction + correlation term $\bar{A}_\rm{sic}$, and the interaction with the electric field,
\begin{equation}
\label{eq:Adiel}
\begin{split}
    \bar{A}_\diel(\vbr) &= \bar{A}_\rot\oparg{\rho_\rot}(\vbr) + \bar{A}_\pol\oparg{\vbp_\pol}(\vbr) \\
    &+ \bar{A}_\rm{sic}\oparg{\rho_\rot,\vbp_\pol}(\vbr) + \vbp(\vbr) \vdot \qty(w_\rm{b}\ast\grad\phi)(\vbr)
    \quad .
\end{split}
\end{equation}
In this expression, we have defined the total polarization $\vbp$ and the rotational polarization $\vbp_\rot$ of a solvent molecule,
\begin{equation}
    \vbp \equiv \vbp_\rot + \vbp_\pol
    \quad ,
\end{equation}
\begin{equation} \label{eq:prot}
    \vbp_\rot(\vbr) \equiv p_\rm{mol} \intO \hat{\vb{u}}(\vbo) \rho_\rot(\vbr,\vbo)
    \quad ,
\end{equation}
where $\hat{\vb{u}}(\vbo)$ is a unit vector in the direction $\vbo$.

The rotational free energy of a solvent molecule at point $\vbr$ is written as a configurational free energy integral over all orientations $\vbo$ with the rotational distribution function $\rho_\rot(\vbr,\vbo)$,
\begin{equation}
\begin{split}
    \bar{A}_\rot(\vbr) &= \frac{1}{\beta} \intO \qty[\rho_\rot(\vbr,\vbo)(\ln{\rho_\rot(\vbr,\vbo)} - 1) + 1] \\
    &- \lambda_\rot(\vbr) \intO \qty(\rho_\rot(\vbr,\vbo) - 1)
    \quad .
\end{split}
\end{equation}
The normalization condition on the rotational distribution function is enforced at each point in space using the Lagrange multiplier $\lambda_\rot(\vbr)$. The polarization free energy of a solvent molecule is written as a quadratic form in terms of the internal polarization $\vbp_\pol$,
\begin{equation}
    \bar{A}_\pol(\vbr)  = \frac{2\pi}{\eps \alpha_\pol} p^2_\pol(\vbr)
    \quad ,
\end{equation}
where $\alpha_\pol$ is the molecular polarizability. This polarizability is obtained from the condition that the bulk dielectric constant matches the experimental optical dielectric constant in the high-field limit as discussed in Section \ref{sec:Adiel_min}. The self-interaction + correlation term $\bar{A}_\rm{sic}$ accounts for the fact that a molecule does not interact with its own dipole and that correlations exist between the rotational distributions of nearby solvent molecules. It is given by,
\begin{equation}
    \bar{A}_\rm{sic}(\vbr)  = -\frac{2\pi}{\eps \alpha_\rm{sic}} p^2(\vbr)
    \quad ,
\end{equation}
where the empirical parameter $\alpha_\rm{sic}$ is obtained from the condition that the bulk solvent exhibits the experimental dielectric constant in the low-field limit as discussed in Section \ref{sec:Adiel_min}.

The interaction with the electric field is actually written in terms of a convolution of the electric field with a normalized Gaussian function $w_\rm{b}$ having width $a$ (the same $a$ used in eq \eqref{eq:w}). This serves to smooth out the polarization so that it can be represented on the FFT grid. Without this convolution, the polarization and bound charge exhibit high frequency oscillations arising from truncation of the FFT. This convolution is unique to our implementation and is not found in the implementation of Gunceler \etal \cite{Guncelar1}. In the limit $a\to0$, our dielectric free energy functional becomes mathematically equivalent to theirs.

\subsubsection{Ionic free energy} \label{sec:Aion}

The ionic free energy functional uses the size-modified Poisson-Boltzmann form proposed by Borukhov \etal \cite{MPB} and used in the model of Ringe \etal \cite{Ringe}. Like the dielectric free energy, the ionic free energy is also a functional of both the solute electron density (via $S_\ion(\vbr)$) and the occupational distributions of electrolyte species ($\theta_+(\vbr)$, $\theta_-(\vbr)$, $\theta_\vac(\vbr)$). Again, it is convenient to separate out the dependence on $\nel(\vbr)$ by defining a `molecular' ionic free energy $\bar{A}_\ion(\vbr)$ analogous to $\bar{A}_\diel(\vbr)$ so that the total ionic free energy can be written as,
\begin{equation}
    A_\ion = n_\rm{max} \intr S_\ion\oparg{\nel}(\vbr) \bar{A}_\ion\oparg{\theta_+,\theta_-,\theta_\vac}(\vbr)
    \quad .
\end{equation}
The expression for $\bar{A}_\ion(\vbr)$ is based on a translationally invariant lattice gas model of the electrolyte \cite{MPB} where each `site' can be occupied by either a cation, an anion, or solvent. The volume of each site corresponds to the hydrated volume of an ion $V_\ion = \frac{4\pi}{3}R_\ion^3$, which also defines the maximum ion concentration $n_\rm{max} = V_\ion^{-1}$. The occupations of an electrolyte `site' at position $\vbr$ are denoted $\theta_+(\vbr)$ for cations, $\theta_-(\vbr)$ for anions, and $\theta_\vac(\vbr)$ for solvent. The resulting `site' free energy consists of three terms characterizing the entropy and chemical potential $\mu_i$ of each occupancy state, a term for the interaction between the ions and the electrostatic potential, and a constraint on the sum of the occupancies,
\begin{equation}
\label{eq:Aion}
\begin{split}
    \bar{A}_\ion(\vbr) &= \sum_{i=\qty{+,-,\vac}} \theta_i(\vbr)\qty[\frac{1}{\beta} \ln{\theta_i(\vbr)} - \mu_i] \\
    &- \lambda_\ion(\vbr)\qty[\sum_{i=\qty{+,-,\vac}} \theta_i(\vbr)  - 1] + \bar{\rho}_\ion(\vbr) \phi(\vbr)
    \quad .
\end{split}
\end{equation}
The constraint that the occupancies of all three states must sum to unity at each point in space is enforced using the Lagrange multiplier $\lambda_\ion(\vbr)$. In this expression, we make use of the average charge of the `lattice site' at position $\vbr$ defined as,
\begin{equation} \label{eq:rhobar_ion}
    \bar{\rho}_\ion(\vbr) \equiv ze \qty(\theta_-(\vbr) - \theta_+(\vbr))
    \quad ,
\end{equation}
where $z$ is the formal charge of the cations and anions in a symmetric $z$:$z$ electrolyte.

\subsubsection{Cavity formation free energy} \label{sec:Acav}

The cavity formation free energy accounts for the free energy required to form the solute cavity from the bulk electrolyte as well as for the dispersion interactions between the solute and the surrounding electrolyte. We use the empirical form suggested by Marzari \etal \cite{8} that is proportional to the surface area of the cavity according to,
\begin{equation}
    A_\cav = \tau \intr \abs{\grad S_\cav\oparg{\nel}(\vbr)}
    \quad .
\end{equation}
The surface area is computed by integrating over the gradient of the cavity function $S_\cav(\vbr)$ and is scaled by the effective surface tension $\tau$.

\subsection{Minimization of the free energy functional} \label{sec:minimization}

As described in the last section, the combined solute/electrolyte system is described by eight degrees of freedom. The first two ($\nel(\vbr)$ and $\RI$) characterize the solute. The dielectric response is characterized by $\rho_\rot(\vbr)$ and $\vbp_\pol(\vbr)$, while the ionic response is characterized by $\theta_+(\vbr)$, $\theta_-(\vbr)$, and $\theta_\vac(\vbr)$. All parts of the system are coupled by the electrostatic potential $\phi(\vbr)$.

The ground state of the system corresponds to a stationary point of the total free energy functional given by eq \eqref{eq:Atot}. We first show that analytical solutions exists for the dielectric and ionic response degrees of freedom. Using these analytical solutions, we then obtain a nonlinear Poisson-Boltzmann equation as a stationary point with respect to variations in the electrostatic potential. Finally, we obtain corrections to the Kohn-Sham potential and the Hellmann-Feynman forces from variations of the free energy functional with respect to the solute electron density and nuclear coordinates, respectively.

\subsubsection{Minimization of the dielectric free energy with respect to the rotational distribution function and internal polarization} \label{sec:Adiel_min}

The ground state rotational distribution $\rho_\rot(\vbr,\vbo)$ and internal polarization $\vbp_\pol(\vbr)$ of the solvent molecules correspond to a minimum of the dielectric free energy of a solvent molecule, $\bar{A}_\diel(\vbr)$, at each point in space. Minimizing with respect to $\vbp_\pol(\vbr)$ leads to the governing equation for the internal polarization,
\begin{equation}
    \pdv{\bar{A}_\diel(\vbr)}{\vbp_\pol} = \frac{4\pi}{\eps\alpha_\pol} \vbp_\pol(\vbr) - \vbEloc(\vbr) = 0
    \quad ,
\end{equation}
where the local electric field $\vbEloc(\vbr)$ felt by a solvent molecule at point $\vbr$ is defined as,
\begin{equation} \label{eq:Eloc}
    \vbEloc(\vbr) \equiv  -\qty(w_b\ast\grad\phi)(\vbr) + \frac{4\pi}{\eps \alpha_\rm{sic}} \vbp(\vbr)
    \quad .
\end{equation}
The difference between the macroscopic electric field $-\grad\phi(\vbr)$ and the local electric field is due to the self-interaction + correlation term $\bar{A}_\rm{sic}$. The resulting ground state internal polarization is,
\begin{equation} \label{eq:ppol_opt}
    \vbp_\pol(\vbr) = \frac{\eps}{4\pi} \alpha_\pol \vbEloc(\vbr)
    \quad ,
\end{equation}
representing a linear response to the local field.

\begin{figure}
    \includegraphics[width=.45\textwidth]{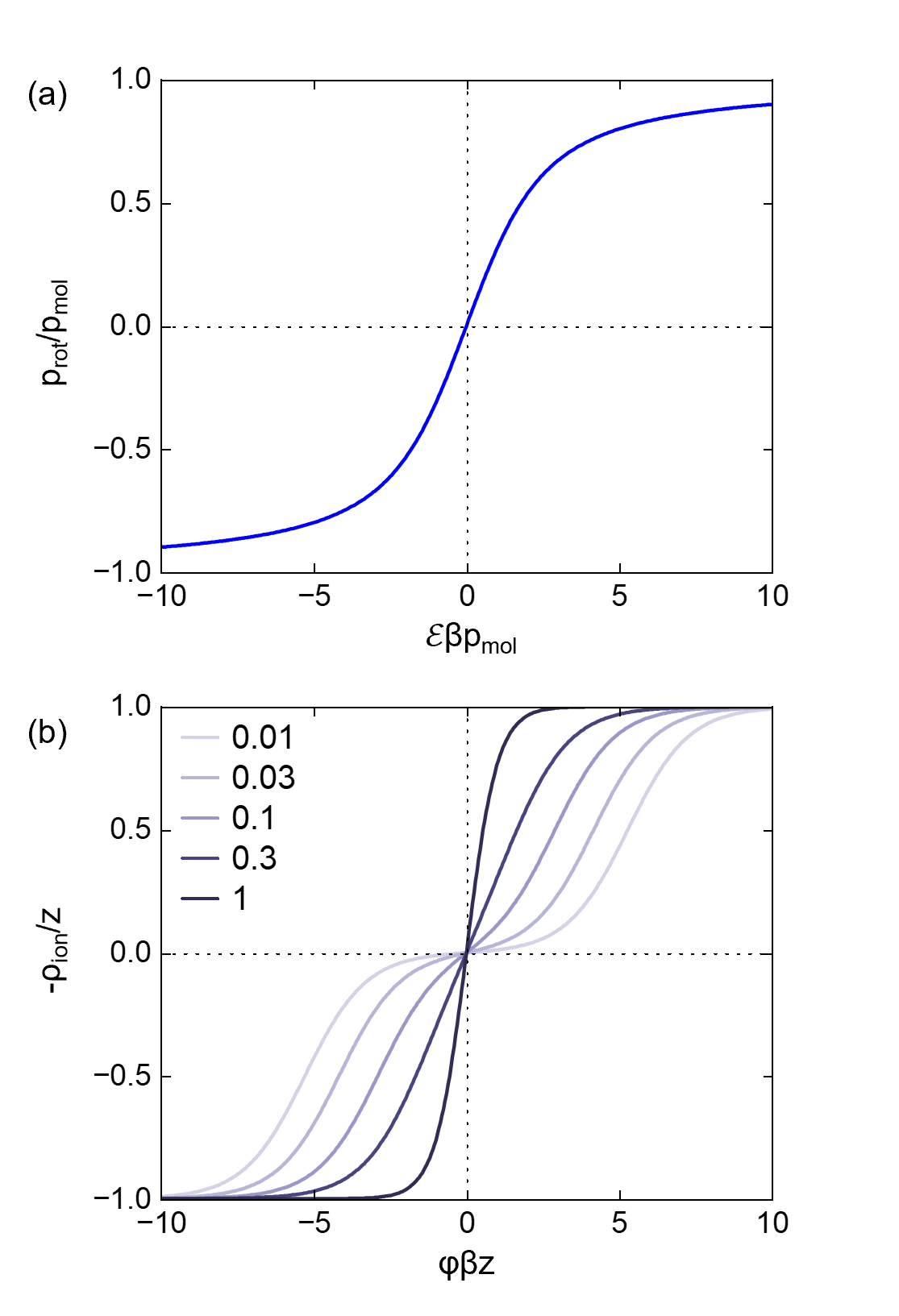}
    \caption{(a) Dielectric response function with respect to dimensionless local field strength. (b) Ionic response function with respect to dimensionless electrostatic potential, plotted for the labeled values of $\theta_\ion^\bulk$.}
    \label{fig:grot_gion}
\end{figure}

Similarly, minimizing with respect to $\rho_\rot(\vbr,\vbo)$ yields a governing equation for the rotational distribution function,
\begin{equation}
\begin{split}
    4\pi \fdv{\bar{A}_\diel(\vbr)}{\rho_\rot(\vbr,\vbo)} &= \frac{1}{\beta} \ln{\rho_\rot(\vbr,\vbo)} \\
    &- p_\rm{mol} \hat{\vb{u}}(\vbo) \vdot \vbEloc(\vbr) -  \lambda_\rot(\vbr)  = 0
    \quad .
\end{split}
\end{equation}
The Lagrange multiplier $\lambda_\rot(\vbr)$ at each point in space is determined by the condition that the rotational distribution function be normalized to unity, resulting in,
\begin{equation} \label{eq:lambda_rot}
    \lambda_\rot(\vbr) = -\frac{1}{\beta} \ln(\frac{\sinh(\bigv\beta p_\rm{mol}\Eloc(\vbr))}{\beta p_\rm{mol}\Eloc(\vbr)})
    \quad .   
\end{equation}
This leads to the ground state expression for $\rho_\rot(\vbr,\vbo)$,
\begin{equation}
    \rho_\rot(\vbr,\vbo) = \frac{\beta p_\rm{mol}\Eloc(\vbr) \exp(\bigv\beta p_\rm{mol} \hat{\vb{u}}(\vbo) \vdot \vbEloc(\vbr))}{\sinh(\bigv\beta p_\rm{mol}\Eloc(\vbr))}
    \quad .
\end{equation}
The rotational polarization $\vbp_\rot(\vbr)$ is given by the average dipole of a solvent molecule at position $\vbr$ according to eq \eqref{eq:prot},
\begin{equation} \label{eq:prot_opt}
    \vbp_\rot(\vbr) = \frac{\eps}{4\pi} \alpha_\rot^0 g_\rot\big(\beta p_\rm{mol}\Eloc(\vbr)\big) \vbEloc(\vbr)
    \quad ,
\end{equation}
where $\alpha_\rot^0$ is the low-field rotational polarizability,
\begin{equation}
    \alpha_\rot^0 = \frac{1}{3}\qty(\frac{4\pi}{\eps}) \beta p_\rm{mol}^2
    \quad ,
\end{equation}
and $g_\rot(x)$ is the rotational dielectric saturation function,
\begin{equation}
    g_\rot(x) = \frac{3}{x^2} (x \coth{x} - 1)
    \quad .
\end{equation}
This latter function accounts for saturation of the rotational response at high electric fields, as can be seen in Figure \ref{fig:grot_gion}(a),
%%SR better to introduce a) and b) for b etter referencing here
%%TODO@SMRI add (a) and (b) labels to figure
%%SMRI done
%%CP done
and is responsible for the nonlinear dielectric response. This function is absent in the dielectric response of a linear electrolyte model such as in refs \citenum{VASPsol1,VASPsol2}.

Having obtained the ground state rotational distribution function and internal polarization, we can substitute these into the expression for $\bar{A}_\diel(\vbr)$ in eq \eqref{eq:Adiel} to get the ground state dielectric free energy $\lambda_\diel(\vbr)$ of a solvent molecule at position $\vbr$,
\begin{equation}
\begin{split}
    \lambda_\diel(\vbr) &\equiv \min_{\substack{\rho_\rot(\vbr,\vbo) \\ \vbp_\pol(\vbr)}} \oparg{\bar{A}_\diel(\vbr)} \\
    &= \lambda_\rot(\vbr) + \lambda_\pol(\vbr) + \lambda_\rm{sic}(\vbr)
    \quad .
\end{split}
\end{equation}
This expression is written in terms of the ground state molecular free energies for rotation $\lambda_\rot(\vbr)$, internal polarization $\lambda_\pol(\vbr)$, and self-interaction + correlation $\lambda_\rm{sic}(\vbr)$. The rotational contribution is equivalent to the Lagrange multiplier in the rotational free energy functional given by eq \eqref{eq:lambda_rot}, having the nonlinear form,
\begin{equation}
    \lambda_\rot(\vbr) = -\frac{1}{\beta} \ln(\frac{\sinh(\bigv\beta p_\rm{mol}\Eloc(\vbr))}{\beta p_\rm{mol}\Eloc(\vbr)})
    \quad .   
\end{equation}
The latter two contributions have quadratic forms given by,
\begin{equation}
    \lambda_\pol = -\frac12 \frac{4\pi}{\eps\alpha_\pol} p_\pol^2
    \quad ,
\end{equation}
and,
\begin{equation}
    \lambda_\rm{sic} = \frac12 \frac{4\pi}{\eps\alpha_\rm{sic}} p^2
    \quad .
\end{equation}

The parameters $\alpha_\pol$ and $\alpha_\rm{sic}$ are determined by the condition that the dielectric response must match the bulk dielectric constants in the linear response regime. In the low-field limit, the rotational and internal polarization densities $\vb{P} = n_\rm{mol}\vbp$ based on eqs \eqref{eq:prot_opt} and \eqref{eq:ppol_opt} become,
\begin{align}
    \vb{P}_\rot(\vbr) &\rightarrow n_\rm{mol}\frac{\eps}{4\pi} \alpha_\rot^0 \vbEloc(\vbr)
    \quad , \\
    \vb{P}_\pol(\vbr) &\rightarrow n_\rm{mol}\frac{\eps}{4\pi} \alpha_\pol \vbEloc(\vbr)
    \quad ,
\end{align}
with the local field defined in eq \eqref{eq:Eloc} becoming,
\begin{equation}
    \vbEloc(\vbr) \rightarrow -\qty[1 - \frac{ \alpha_\rot^0+\alpha_\pol}{ \alpha_\rm{sic}}]^{-1} \grad\phi(\vbr)
    \quad .
\end{equation}
These polarization densities can also be written in terms of the bulk static and optical dielectric constants of the solvent $\epsilon_\bulk$ and $\epsilon_\infty$,
\begin{align}
    \vb{P}_\rot(\vbr) &\rightarrow - \frac{\eps}{4\pi} \qty(\epsilon_\bulk - \epsilon_\infty) \grad\phi(\vbr)
    \quad , \\
    \vb{P}_\pol(\vbr) &\rightarrow - \frac{\eps}{4\pi} \qty(\epsilon_\infty - 1) \grad\phi(\vbr)
    \quad .
\end{align}
Matching the two sets of expressions for $\vb{P}_\rot(\vbr)$ and $\vb{P}_\pol(\vbr)$ yields expressions for $\alpha_\pol$ and $\alpha_\rm{sic}$ in terms of the bulk dielectric constants,
\begin{equation}
    \alpha_\pol = \alpha_\rot^0 \qty(\frac{\epsilon_\infty - 1}{\epsilon_\bulk - \epsilon_\infty})
    \quad ,
\end{equation}
\begin{equation}
    \frac{1}{\alpha_\rm{sic}} = \qty(\frac{\epsilon_\bulk - \epsilon_\infty}{\alpha_\rot^0} - n_\rm{mol}) \qty(\frac{1}{\epsilon_\bulk -1})
    \quad .
\end{equation}

\subsubsection{Minimization of the ionic free energy with respect to ion concentrations} \label{sec:Aion_min}

Likewise, the ground state electrolyte species occupations $\theta_i(\vbr)$ ($i = \qty{+,-,\vac}$) correspond to a minimum of the ionic `site' free energy $\bar{A}_\ion(\vbr)$ at each point in space. Minimizing with respect to $\theta_\pm(\vbr)$ leads to governing equations for the ionic species occupations,
\begin{equation}
    \pdv{\bar{A}_\ion(\vbr)}{\theta_\pm(\vbr)} = \frac{1}{\beta} \ln{\theta_\pm(\vbr)} \pm z \phi(\vbr) - \mu_\pm - \lambda_\ion(\vbr) = 0
    \quad .
\end{equation}
Similarly, minimizing with respect to $\theta_\vac$ yields a governing equation for $\theta_\vac$,
\begin{equation}
    \pdv{\bar{A}_\ion(\vbr)}{\theta_\vac(\vbr)} = \frac{1}{\beta} \ln{\theta_\vac(\vbr)} - \mu_\vac - \lambda_\ion(\vbr) = 0
    \quad .
\end{equation}
The chemical potentials are determined so that $\theta_\pm \to \theta_\pm^\bulk$ and $\theta_\vac \to 1-\theta_\ion^\bulk$ when $\phi \to 0$, where $\theta_\pm^\bulk$ and $\theta_\ion^\bulk \equiv 2\theta_\pm^\bulk$ are the the volume fractions of cations/anions and total ions in the bulk electrolyte.
The Lagrange multiplier at each point in space is obtained by the condition that the species occupations sum to unity giving,
\begin{equation} \label{eq:lambda_ion}
    \lambda_\ion(\vbr) = -\frac{1}{\beta} \ln(\Bigv 1 - \theta_\ion^\bulk +\theta_\ion^\bulk \cosh(\bigv\beta z \phi(\vbr)))
    \quad ,
\end{equation}
The resulting ground state occupancies are then,
\begin{equation}
    \theta_\pm(\vbr) = \frac{\theta_\pm^\bulk \exp(\bigv\mp \beta z \phi(\vbr))}{1 - \theta_\ion^\bulk + \theta_\ion^\bulk \cosh(\bigv\beta z \phi(\vbr))}
    \quad ,
\end{equation}
and,
\begin{equation}
    \theta_\vac(\vbr) = \frac{1 - \theta_\ion^\bulk}{1 - \theta_\ion^\bulk + \theta_\ion^\bulk \cosh(\bigv\beta z \phi(\vbr))}
    \quad .
\end{equation}
The average charge of an electrolyte `lattice site' at position $\vbr$ (defined by eq \eqref{eq:rhobar_ion}) is then given by,
\begin{equation}
    \bar{\rho}_\ion(\vbr) = - \frac{\eps}{4\pi} \alpha_\ion^\bulk g_\ion\big(\beta z \phi(\vbr)\big) \phi(\vbr)
    \quad ,
\end{equation}
where $\alpha_\ion^\bulk$ is the ionic response constant in the bulk electrolyte,
\begin{equation}
    \alpha_\ion^\bulk = \qty(\frac{4\pi}{\eps}) \theta_\ion^\bulk \beta z^2e^2
    \quad ,
\end{equation}
and $g_\ion(x)$ is the nonlinear enhancement function for the ionic response,
\begin{equation}
    g_\ion(x) = \frac{1}{x}\qty(\frac{\sinh{x}}{1 - \theta_\ion^\bulk + \theta_\ion^\bulk \cosh{x}})
    \quad .
\end{equation}
The enhancement function accounts for the exponential nature of the Poisson-Boltzmann response of the ionic distributions at moderate values of the electrostatic potential. In addition, it describes ionic saturation that occurs at high values of the electrostatic potential. Both of these behaviors are illustrated in Figure \ref{fig:grot_gion}(b) for different values of $\theta_\ion^\bulk$.

Substituting the electrolyte species distributions into the expression for $\bar{A}_\ion(\vbr)$ in eq \eqref{eq:Aion} gives the ground state ionic free energy $\lambda_\ion(\vbr)$ of an electrolyte `site' at position $\vbr$,
\begin{equation}
\begin{split}
    \lambda_\ion(\vbr) &\equiv \min_{\theta_i(\vbr)} \oparg{\bar{A}_\ion(\vbr)} \\
    &=-\frac{1}{\beta} \ln(\Bigv 1 - \theta_\ion^\bulk +\theta_\ion^\bulk \cosh(\bigv\beta z \phi(\vbr)))
    \quad .
\end{split}
\end{equation}
It can be seen that this is equivalent to the Lagrange multiplier in the ionic free energy functional given by eq \eqref{eq:lambda_ion}.

\subsubsection{Maximization with respect to the electrostatic potential to obtain the nonlinear Poisson-Boltzmann equation} \label{sec:phi_min}

Maximizing the total free energy with respect to $\phi(\vbr)$ yields a nonlinear Poisson-Boltzmann equation that can be solved to obtain the electrostatic potential,
%%SR Maximizing? It should be minimizing right?
%%CP No, the free energy functional is a maximum wrt to phi, minimum wrt everything else. I checked this explicitly in the implementation. Makes sense too b/c the "eps0/8pi" term has a negative sign in eq 14. In vacuum, you could get an arbitrarily negative free energy
\begin{equation} \label{eq:NLPB}
    -\frac{\eps}{4\pi} \laplacian \phi(\vbr) = \rho_\sol(\vbr) + \rho_\solv\oparg{\phi}(\vbr)
    \quad ,
\end{equation}
The solute charge density $\rho_\sol(\vbr) \equiv e\,\nel(\vbr) + \rho_Z(\vbr)$ is the sum of the electron and nuclear charge densities, while the electrolyte charge density $\rho_\solv(\vbr) \equiv \rho_\rm{b}(\vbr) + \rho_\ion(\vbr)$ is the sum of the bound charge density $\rho_\rm{b}(\vbr)$ arising from the dielectric response,
\begin{equation} \label{eq:rhob}
    \rho_\rm{b}(\vbr) \equiv -n_\rm{mol} \qty(w_\rm{b}\ast\div\qty(S_\diel\,\vbp))(\vbr)
    \quad ,
\end{equation}
and the ionic charge density $\rho_\ion(\vbr)$ arising from the ionic response,
\begin{equation} \label{eq:rhoion}
    \rho_\ion(\vbr) \equiv n_\rm{max} S_\ion(\vbr) \bar{\rho}_\ion(\vbr)
    \quad .
\end{equation}
Since the electrolyte charge density is a nonlinear function of the electrostatic potential, eq \eqref{eq:NLPB} must be solved numerically as discussed in Section \ref{sec:NLPB_solver}.

\subsubsection{Minimization with respect to the solute electron density to obtain the electrolyte correction to the Kohn-Sham potential} \label{sec:nel_min}

Substituting the ground state dielectric and ionic free energies into eq \eqref{eq:Atot} gives the total free energy functional when the electrolyte is in its ground state,
\begin{equation}
\label{eq:Atot_opt}
\begin{split}
    A_\rm{tot} &= A_\rm{TXC} + \intr \phi(\vbr) \rho_\sol(\vbr) \\
    &+ \frac{\eps}{8\pi} \intr \phi(\vbr) \laplacian \phi(\vbr) + A_\cav \\
    &+ n_\rm{mol} \intr S_\diel(\vbr) \lambda_\diel(\vbr) \\
    &+ n_\rm{max} \intr S_\ion(\vbr) \lambda_\ion(\vbr)
    \quad .
\end{split}
\end{equation}
Taking the functional derivative with respect to $\nel(\vbr)$ then gives a governing equation for the solute electron density,
\begin{equation} \label{eq:KS}
   \fdv{A_\rm{TXC}}{\nel(\vbr)} + e\phi_\sol(\vbr) + v_\rm{corr}(\vbr) = \mu_\rm{e}
   \quad ,
\end{equation}
where $\phi_\sol(\vbr)$ is the electrostatic potential of the solute in vacuum obtained from solving eq \eqref{eq:NLPB} with $\rho_\solv(\vbr)=0$. This expression contains the electrolyte correction to the Kohn-Sham potential,
\begin{equation} \label{eq:vcorr}
    v_\rm{corr}(\vbr) = e\phi_\solv(\vbr) + v_\solv(\vbr)
    \quad ,
\end{equation}
which consists of the electrostatic potential $\phi_\solv(\vbr)$ due to the electrolyte charge density $\rho_\solv(\vbr)$ in addition to a term $v_\solv(\vbr)$ arising from the dependence of the cavity functions $S_\diel(\vbr)$, $S_\ion(\vbr)$, and $S_\cav(\vbr)$ on the solute electron density. This latter correction is composed of terms due to variations of the dielectric free energy ($v_\diel(\vbr)$), the ionic free energy ($v_\ion(\vbr)$), and the cavity formation free energy ($v_\cav(\vbr)$) with respect to the solute electron density,
\begin{equation}
\label{Vsolv}
\begin{aligned}
    v_\solv(\vbr) &&&= &&\fdv{A_\cav}{\nel(\vbr)} &&+ &&\fdv{A_\diel}{\nel(\vbr)} &&+ &&\fdv{A_\ion}{\nel(\vbr)} \\
    &&&= &&v_\cav(\vbr) &&+ &&v_\diel(\vbr) &&+ &&v_\ion(\vbr)
    \quad .
\end{aligned}
\end{equation}

Derivation of these three potential correction terms is a strenuous exercise in variational calculus due to the complex dependence of the cavity shape on multiple nested functions and convolutions of the solute electron density. The full details are given in the \SM.
All three potential corrections are expressed in terms of a general cavity potential functional defined as,
\begin{equation} \label{eq:vS}
    v_S\oparg{\mathcal{A},n} \equiv \fdv{n} \int \dd[3]{\vbr'} S\oparg{n}(\vbr') \, \mathcal{A}(\vbr')
    \equiv \frac{1}{n} S'\mathcal{A}
    \quad ,
\end{equation}
where $\mathcal{A}(\vbr)$ is some free energy density and $S'$ is the cavity function derivative,
\begin{equation} \label{eq:Sprime}
    S' \equiv \dv{S}{x} \equiv -\frac{1}{\sqrt{2\pi}\sigma}\exp(-\frac{x^2}{2\sigma^2})
    \quad .
\end{equation}
To simplify the notation, the dependence of $S$, $S'$, and $x$ on the effective electron density $n$ is not explicitly written out.

The potential correction $v_\diel(\vbr)$ is then computed according to,
\begin{subequations}
\begin{align}
    v_\diel''(\vbr) &= v_S\oparg{-n_\rm{mol}\lambda_\diel,n_\diel}(\vbr) \\
    v_\diel'(\vbr) &= v_S\oparg{\nc \qty(w_\diel \ast v_\diel''),n_\solv}(\vbr) \\
    v_\diel(\vbr) &= v_S\oparg{-\nc \qty(w_\solv \ast v_\diel'),n_\vdW}(\vbr)
    \quad ,
\end{align}
\end{subequations}
while $v_\ion(\vbr)$ is computed as,
\begin{subequations}
\begin{align}
    v_\ion'(\vbr) &= v_S\oparg{n_\rm{max}\lambda_\ion,n_\ion}(\vbr) \\
    v_\ion(\vbr) &= v_S\oparg{-\nc \qty(w_\ion \ast v_\ion'),n_\vdW}(\vbr)
    \quad ,
\end{align}
\end{subequations}
and $v_\cav(\vbr)$ is computed as,
\begin{subequations}
\begin{align}
    v_\cav''(\vbr) &= v_S\oparg{-n_\rm{mol}\lambda_\cav,n_\cav}(\vbr) \\
    v_\cav'(\vbr) &= v_S\oparg{\nc \qty(w_\cav \ast v_\cav''),n_\solv}(\vbr) \\
    v_\cav(\vbr) &= v_S\oparg{-\nc \qty(w_\solv \ast v_\cav'),n_\vdW}(\vbr)
    \quad .
\end{align}
\end{subequations}
The latter is computed in terms of a quantity $\lambda_\cav(\vbr)$ that plays an analogous role to $\lambda_\diel(\vbr)$ and $\lambda_\ion(\vbr)$ in the expressions for $v_\diel(\vbr)$ and $v_\ion(\vbr)$,
\begin{equation}
    \lambda_\cav(\vbr) = -\frac{\tau}{n_\rm{mol}} \div \frac{\grad S_\cav}{\abs{\grad S_\cav}}(\vbr)
    \quad .
\end{equation}

\subsubsection{Minimization with respect to the solute ionic positions to obtain the atomic forces} \label{sec:forces}

Finally, we take the partial derivative of the free energy with respect to the ionic positions $\vb{R}_I$ of the solute to obtain the atomic forces. In doing so, we have to consider that the solute electron density $\nel(\vbr)$ is actually composed of the valence electron density $n_\rm{v}(\vbr)$ and the electron density $n_\rm{c}(\vbr)$ of the cores,
\begin{equation}
    \nel(\vbr) = n_\rm{v}(\vbr) + n_\rm{c}(\vbr)
    \quad .
\end{equation}
The valence electron density is a variational quantity and thus has no explicit dependence on the ionic positions. However, the core electron density of each ion moves with the position of that ion so that it has an explicit dependence on the ionic positions in pseudopotential methods. This means that variations in the ionic positions will lead to variation in the vdW cavity function $S_\vdW$ since it explicitly depends on the solute electron density. The resulting expression for the force on atom $I$ is,
\begin{equation} \label{eq:F}
\begin{split}
    \vb{F}_I = -\pdv{A_\rm{tot}}{\vb{R}_I} = &-\intr \phi(\vbr) \pdv{\rho_Z(\vbr)}{\vb{R}_I} \\
    &-\intr v_\solv(\vbr) \pdv{\rho_\rm{c}(\vbr)}{\vb{R}_I}
    \quad ,
\end{split}
\end{equation}
where $\rho_Z(\vbr)$ is the nuclear charge density of the solute. The second term on the right hand side arises from the explicit dependence of the core electron density on the ionic positions. This term actually turns out to be negligible from a practical standpoint as long as the vdW cavity does not overlap with the core electron density. This is due to the fact that $v_\solv(\vbr)$ vanishes when $S_\vdW(\vbr)$ vanishes, as can be seen from examining eqs \eqref{eq:vS} and \eqref{eq:Sprime}. Since the core electrons do not come close to the vdW surface in practice, we neglect this term in the computation of atomic forces.

\section{Implementation} \label{sec:implementation}

We have implemented the implicit electrolyte model described above into the Vienna Ab initio Simulation Package (VASP), a widely used parallel plane-wave DFT code supporting both ultrasoft pseudopotentials \cite{12,13} and the projector-augmented wave method \cite{14}. The numerical efficiency and parallel scalability of VASP make it one of the most popular DFT codes for studying surface processes in an electrochemical environment. Nonetheless, VASPsol \cite{VASPsol1,VASPsol2} is the only implicit solvation method that interfaces with VASP and it provides less functionality than those implemented in other DFT codes such as JDFTx \cite{JDFTx} and Quantum Espresso Environ. We therefore significantly extend the capabilities of performing implicit electrolyte calculations within VASP by implementation of the nonlinear+nonlocal model. Since ours is an extension of the original VASPsol implementation, we call it VASPsol++.

Efficient implementation of the nonlinear+nonlocal electrolyte model is made possible by two key strategies. First, the many gradient calculations and convolutions are carried out in reciprocal space making use of fast Fourier transforms to convert to and from the real space representation. Second, the solution of the nonlinear Poisson-Boltzmann equation -- which represents almost all of the computational expense of the method -- is carried out efficiently and robustly by an approach based on Newton's method with an additional line search.

\subsection{Newton's method for solving the nonlinear Poisson-Boltzmann equation} \label{sec:NLPB_solver}

%%SR You should cite my original JCTC paper reporting the Newton method in this section, as the original idea comes from that paper (although of course the derivations change according to your model).
%%CP Yes, it's so obvious for me I forgot to actually cite it :D
The nonlinear Poisson-Boltzmann equation is a second order partial differential equation that determines the electrostatic potential at every point in space. In reciprocal space it becomes a set of coupled nonlinear algebraic equations, one for each point on the reciprocal lattice. The numerical solution of a nonlinear system of equations such as this is challenging and requires sophisticated methods in order to be both efficient and robust. The solution process is significantly aided by the fact that the Jacobian of the electrolyte charge density response $\rho_\solv$ can be expressed analytically, allowing for the use of Newton's method rather than the less efficient quasi-Newton or nonlinear conjugate gradient methods. This is inspired by the implementation of Ringe \etal \cite{Ringe}, although we have extended it to work with a nonlinear dielectric response and have added a line search step for numerical stability.

\begin{figure}
    \includegraphics[width=.45\textwidth]{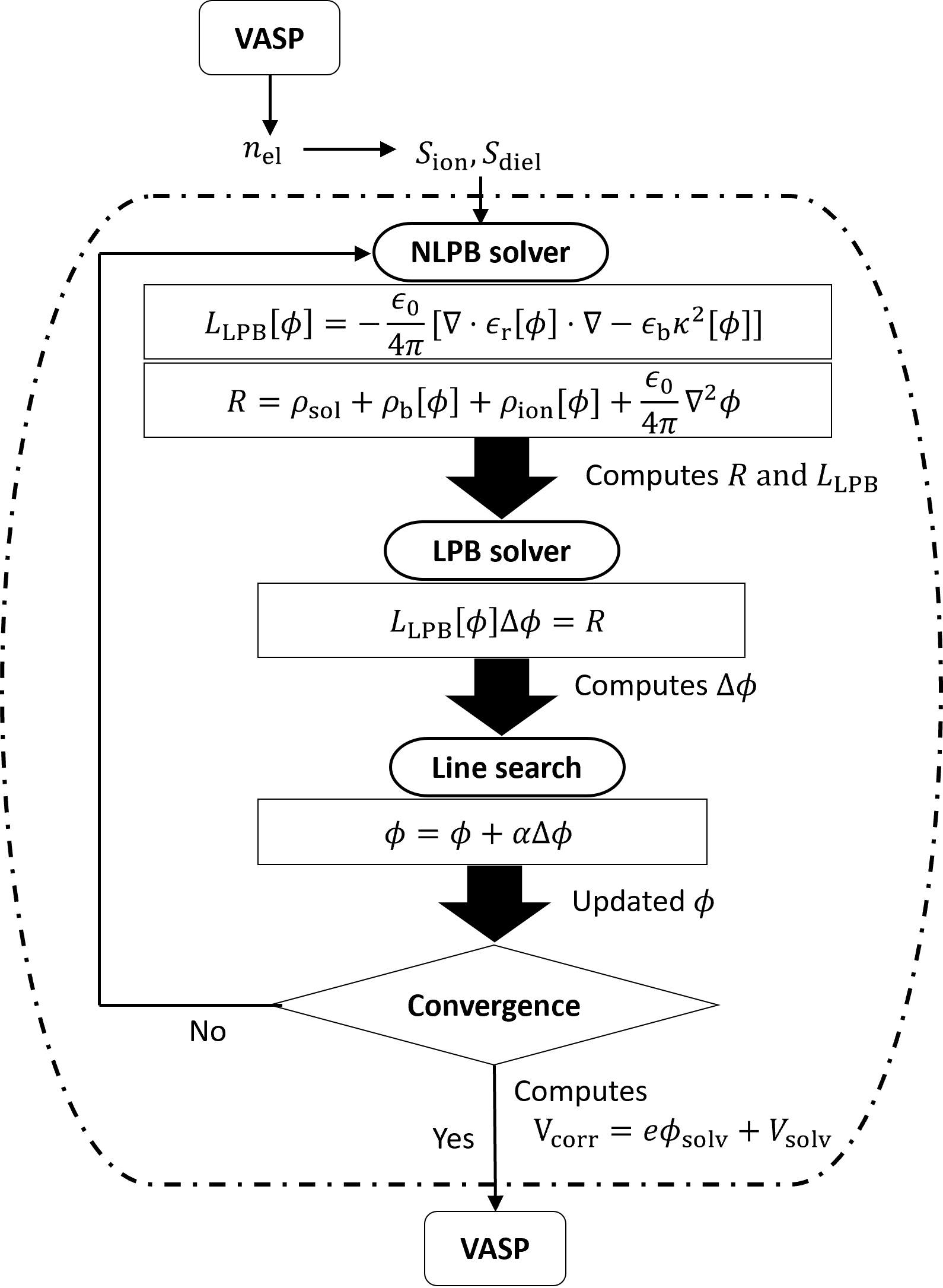}
    \caption{Algorithm used for solving the NLPB equation in VASP.}
    \label{fig:algorithm}
\end{figure}

For the purpose of numerical solution, the nonlinear Poisson-Boltzmann (NLPB) equation in eq \eqref{eq:NLPB} is written in the form,
\begin{equation}
    R\oparg{\phi} \equiv \rho_\sol - (\hat{L}_\rm{P}\phi -\rho_\solv\oparg{\phi}) = 0
    \quad ,
\end{equation}
where $R$ is the NLPB residual and the Poisson operator is defined as,
\begin{equation}
    \hat{L}_\rm{P} \equiv -\frac{\eps}{4\pi} \laplacian
    \quad .
\end{equation}
To simplify notation, we have omitted explicit dependence of all quantities on the position $\vbr$.
The electrolyte charge density $\rho_\solv$ consists of the bound charge density $\rho_\rm{b}$ and the ionic charge density $\rho_\ion$. Both of these quantities map to the electrostatic potential according to eqs \eqref{eq:rhob} and \eqref{eq:rhoion} and can be linearly expanded about a particular solution of the potential $\phi$ according to,
\begin{equation}
\begin{split}
    &\rho_\rm{b}\oparg{\phi + \Delta\phi} \approx \rho_\rm{b}\oparg{\phi} \\
    &\quad+ \frac{\eps}{4\pi} \div \qty\big[w_\rm{b} \ast \vb*{\chi}\oparg{\phi} \vdot \qty\big(w_\rm{b} \ast \grad\qty(\Delta\phi))]
    \quad ,
\end{split}
\end{equation}
and,
\begin{equation}
    \rho_\ion\oparg{\phi + \Delta\phi} \approx \rho_\ion\oparg{\phi} - \frac{\eps}{4\pi} \epsilon_\bulk \kappa^2\oparg{\phi} \Delta\phi
    \quad .
\end{equation}

The linearized dielectric response at each point in space is characterized by a local susceptibility tensor $\vb*{\chi}$,
\begin{equation}
    \vb*{\chi}\oparg{\vbEloc} = \chi_\perp \vb{I} + \qty(\chi_\parallel - \chi_\perp) \qty(\hat{\vbEloc} \otimes \hat{\vbEloc})
    \quad ,
\end{equation}
where $\hat{\vbEloc}(\vbr)$ is a unit vector pointing in the same direction as $\vbEloc(\vbr)$. The anisotropic form follows from the nonlinearity of the full dielectric response so that the susceptibility parallel to the local field ($\chi_\parallel$) is lower than the susceptibility perpendicular to it ($\chi_\perp$),
\begin{align}
    \chi_\perp &\equiv n_\rm{mol} S_\diel \, \qty(\alpha_\perp^{-1} - \alpha_\rm{sic}^{-1})^{-1}
    \quad , \\
    \chi_\parallel &\equiv n_\rm{mol} S_\diel \, \qty(\alpha_\parallel^{-1} - \alpha_\rm{sic}^{-1})^{-1}
    \quad .
\end{align}
The two components of the susceptibility tensor are in turn defined in terms of polarizabilities in the perpendicular and parallel directions,
\begin{align}
    \alpha_\perp &\equiv \alpha_\pol + \alpha_\rot^0 g_\rot(\beta p_\rm{mol} \Eloc)
    \quad , \\
    \alpha_\parallel &\equiv \alpha_\perp + \alpha_\rot^0 g'_\rot(\beta p_\rm{mol} \Eloc) \beta p_\rm{mol} \Eloc
    \quad ,
\end{align}
where $g'_\rot(x) = \dv{x} g_\rot(x)$. Likewise, the linearized ionic response at each point in space is characterized by a local effective inverse Debye length $\kappa$,
\begin{equation}
    \epsilon_\bulk\kappa^2\oparg{\phi} = n_\rm{max} S_\ion \alpha_\ion
    \quad .
\end{equation}
This is defined in terms of the intrinsic ionic response $\alpha_\ion$,
\begin{equation}
    \alpha_\ion \equiv \alpha_\ion^\bulk \qty[g_\ion(\beta z\phi) + g'_\ion(\beta z\phi)\beta z\phi]
    \quad ,
\end{equation}
where $g'_\ion(x) = \dv{x} g_\ion(x)$. Finally, we can define a linearized Poisson-Boltzmann operator according to,
\begin{equation}
    \hat{L}_\rm{LPB}\oparg{\phi} \equiv - \frac{\eps}{4\pi} [\div \vb*{\epsilon}_\rm{r}\oparg{\phi} \vdot \grad - \epsilon_\bulk \kappa^2\oparg{\phi}]
    \quad ,
\end{equation}
with the local dielectric operator defined as,
\begin{equation}
    \vb*{\epsilon}_\rm{r}\oparg{\phi} \equiv 1 + w_\rm{b} \ast \vb*{\chi}\oparg{\phi} \vdot w_\rm{b} \ast
    \quad .
\end{equation}

With these definitions, we can construct an algorithm based on Newton's method to solve the NLPB equation. At each iteration $i$ we first compute the residual $R_i$ from the current solution of the potential, $\phi_i$,
\begin{equation}
    R_i = \rho_\sol + \rho_\rm{b}\oparg{\phi_i} +\rho_\ion\oparg{\phi_i} - \hat{L}_\rm{P}\phi_i
    \quad .
\end{equation}
We then solve the linearized Poisson-Boltzmann (LPB) equation,
\begin{equation}
    \hat{L}_\rm{LPB}\oparg{\phi_i} \Delta\phi_i = R_i
    \quad ,
\end{equation}
to obtain a search direction $\Delta\phi_i$. This is done by a modified conjugate gradient method employing the inverse Poisson operator as a preconditioner. The modification involves how the $G=0$ component of $\Delta\phi_i$ is determined and results in faster convergence. Specifically, the $G=0$ component of $\Delta\phi_i$ is determined in the initial step so that the total ionic charge in the electrolyte balances the total charge on the solute (the $G=0$ component of the LPB residual). Each subsequent step direction of the conjugate gradient method is then modified so that cell neutrality is maintained within the linearized response model. Further details are given in the \SM.

Convergence of the LPB solver is obtained when the RMS of the LPB residual $r \equiv R_i - \hat{L}_\rm{LPB} \Delta\phi_i$ is a factor of 10 smaller than the RMS of the NLPB residual $R_i$. This ensures that the Newton step direction $\Delta\phi_i$ is sufficiently accurate relative to the NLPB residual while minimizing the number of iterations spent in the LPB solver.

Finally, the potential is updated according to,
\begin{equation}
    \phi_{i+1} = \phi_i + \alpha \Delta\phi_i
    \quad ,
\end{equation}
where the step length $\alpha$ is determined by performing a backtracking line search. Further details are given in the \SM. The line search makes the solution method absolutely convergent by ensuring that the total free energy given by eq \eqref{eq:Atot_opt} is continuously increasing. Since the free energy is a concave functional of the electrostatic potential, this ensures that each step is guaranteed to bring the current guess closer to the solution. In the absence of a line search, cases were found where the solution method diverged due the the pathological form of the nonlinear dielectric and ionic response functions.

\subsection{Integration into the self-consistent field method} \label{sec:SCF_integration}

Solution of the NLPB equation to obtain the electrostatic potential is part of the self-consistent field (SCF) method employed by VASP to determine the solute electron density. Specifically, the SCF condition takes the form,
\begin{equation}
    R_\rm{SCF} \equiv n_\rm{out}\oparg{\nel} - \nel = 0
    \quad ,
\end{equation}
where the $n_\rm{out}\oparg{\nel}$ is the solute electron density corresponding to solution of the Kohn-Sham equation in which the Kohn-Sham potential is determined from $\nel$ according to eqs \eqref{eq:NLPB}, \eqref{eq:KS}, and \eqref{eq:vcorr}.

At each SCF step, convergence of the NLPB solver is obtained when the NLPB residual is a factor of 10 less than the RMS of the SCF residual $R_\rm{SCF}$. It should be noted that all of the residuals ($R_\rm{SCF}$, $R$, and $r$) have the same units of charge density so should be directly comparable.

The total free energy expression in eq \eqref{eq:Atot_opt} can be written in a more convenient form,
\begin{equation} \label{eq:Atot_scf}
    A_\rm{tot} = A_\sol + A_\solv
    \quad ,
\end{equation}
by defining the free energy of the solute in the absence of solvation,
\begin{equation}
    A_\sol = A_\rm{TXC} + \frac12 \intr \phi_\sol(\vbr) \rho_\sol(\vbr)
    \quad .
\end{equation}
The solute electrostatic potential $\phi_\sol(\vbr)$ is related to the solute charge density $\rho_\sol(\vbr)$ by the Poisson equation,
\begin{equation}
    -\frac{\eps}{4\pi} \phi_\sol = \rho_\sol - q_\sol
    \quad ,
\end{equation}
where $q_\sol$ is the solute charge using an `electron is positive' convention. The solvation component of the free energy is then given by,
\begin{equation} \label{eq:Asolv}
\begin{split}
    A_\solv &= q_\sol\expval{\phi_\solv} + A_\cav \\
    &+ \frac{\eps}{8\pi} \intr \phi_\solv(\vbr) \laplacian \phi_\solv(\vbr) \\
    &+ n_\rm{mol} \intr S_\diel(\vbr) \lambda_\diel(\vbr) \\
    &+ n_\rm{max} \intr S_\ion(\vbr) \lambda_\ion(\vbr)
    \quad ,
\end{split}
\end{equation}
where $\phi_\solv(\vbr) = \phi(\vbr) - \phi_\sol(\vbr)$ is the electrostatic potential arising from the dielectric and ionic response of the electrolyte. The average value of this potential is denoted $\expval{\phi_\solv}$.

The free energy calculated in the main VASP program is,
\begin{equation} \label{ground state out of VASPsol}
    A_\rm{VASP} = A_\sol + \intr v_\rm{corr}(\vbr) n_\rm{val}(\vbr)
    \quad ,
\end{equation}
where the last term is due to the correction $v_\rm{corr}$ to the Kohn-Sham potential that arises from the induced charge density in the electrolyte and the dependence of the cavity functions on the solute electron density. It is seen that the VASP free energy must be corrected by,
\begin{equation}
    A_\rm{corr} = A_\solv - \intr v_\rm{corr}(\vbr) n_\rm{val}(\vbr)
    \quad ,
\end{equation}
to obtain the free energy given by eq \eqref{eq:Atot_scf}.

After convergence of the SCF cycle, the atomic forces are computed by eq \eqref{eq:F}. The only difference between this expression and the expression in the main VASP program is that the total potential $\phi(\vbr)$ is used in place of the solute potential $\phi_\sol(\vbr)$. Thus, the only required modification is to add the electrolyte potential $\phi_\solv(\vbr)$ to the solute potential when computing the forces.

\subsection{Constant potential calculations} \label{sec:constant_pot}

When the electrolyte model incorporates ionic screening, it becomes possible to carry out calculations at constant electrochemical potential instead of at constant solute charge \cite{VASPsol2,GCDFT}. The form of the ionic free energy implicitly sets the electrostatic potential of the bulk electrolyte to zero, as shown in ref \citenum{VASPsol2}. Therefore, the Fermi level of the solute calculated in VASP is referenced to the bulk electrolyte level rather than to the average potential in the unit cell (as is the case for calculations without ionic screening in the electrolyte). Methodologically, this occurs because solution of the NLPB equation sets the $G=0$ component of the electrostatic potential such that the total charge in the unit cell (solute+ionic) is zero.

We have therefore implemented a modification to VASP that allows for performing calculations at constant electron chemical potential. Since convergence issues have been reported for these types of calculations when the solute charge is updated during the SCF cycle \cite{SundararamanGCDFT}, we choose to instead update the charge between SCF cycles (\ie during the geometric update) as is done in ref \citenum{Kastlunger}.
During each geometric update step $i$, the number of electrons on the solute is updated according to,
\begin{equation}
    N_{\rm{e},i+1} = N_{\rm{e},i} - C \qty(\varepsilon_{\rm{F},i} - \mu_\rm{e})
    \quad ,
\end{equation}
where $\varepsilon_\rm{F}$ is the Fermi level of the surface (solute) and $\mu_\rm{e}$ is the specified electrochemical potential with respect to bulk electrolyte. The constant $C$ is the approximate capacitance of the surface that is initially set to a value of \SI{1}{\per\eV} and then updated at each step $i>1$ according to,
\begin{equation}
    C = \frac{N_{\rm{e},i}-N_{\rm{e},i-1}}{\varepsilon_{\rm{F},i}-\varepsilon_{\rm{F},i-1}}
    \quad .
\end{equation}
The update of $C$ is only done when $\varepsilon_{\rm{F},i}-\varepsilon_{\rm{F},i-1} > 0.1$ eV; otherwise the value from the previous iteration is kept.

When performing constant potential calculations, the free energy given by eqs \eqref{eq:Atot} and \eqref{eq:Atot_opt} is not variational with respect to the solute charge. Instead, one must define the Landau (or grand) potential by accounting for a reservoir of electrons at chemical potential $\mu_\rm{e}$ \cite{SundararamanGCDFT},
\begin{equation}
    \Omega_\rm{tot} = A_\rm{tot} - q_\sol\mu_\rm{e}
    \quad ,
\end{equation}
where the solute charge $q_\sol$ is a variational parameter. The Landau potential can be used directly to compute the free energy changes of processes occurring at constant potential \cite{SundararamanGCDFT}.

\subsection{Parameterization of the electrolyte model} \label{sec:parameterization}

Several parameters must be specified in order to fully define the electrolyte model. The default values of these parameters are listed in Table \ref{tab:parameters}. Some of these are taken from experimental values for water such as the bulk static and optical dielectric constants ($\epsilon_\bulk$ and $\epsilon_\infty$), the dipole moment of water $p_\rm{mol}$, and the bulk molecular density of water $n_\rm{mol}$. The parameter $\sigma$ that determines the width of the cavity function in eq \eqref{eq:S} is taken as \num{0.6} from the original VASPsol implementation. The parameter $a$ that appears in eq \eqref{eq:w} is set to the smallest values that still eliminates most of the FFT truncation error. Since the typical FFT grid spacing used in VASP is around \SI{0.1}{\angstrom}, we use a value slightly larger than this of \SI{0.125}{\angstrom}.

\begin{table}
\caption{Default values of the parameters used in the electrolyte model.}
\label{tab:parameters}
\begin{ruledtabular}
\begin{tabular}{lllllll}
    $\nc$\footnotemark[1] & $\sigma$ & $a$\footnotemark[3] & $R_\solv$\footnotemark[3] & $R_\diel$\footnotemark[3] & $R_\ion$\footnotemark[3] \\
    \num{0.015} & \num{0.6} & \num{0.125} & \num{1.40} & \num{1.00} & \num{4.00} \\
    $\tau$\footnotemark[2] & $n_\rm{mol}$\footnotemark[1] & $p_\rm{mol}$\footnotemark[4] & $\epsilon_\bulk$ & $\epsilon_\infty$ \\
    \num{0.879} & \num{0.0335} & \num{0.50} & \num{78.4} & \num{1.78} \\
\end{tabular}
\end{ruledtabular}
\footnotetext[1]{\unit{\angstrom^{-3}}}
\footnotetext[2]{\unit{\milli\eV\per\angstrom^2}}
\footnotetext[3]{\unit{\angstrom}}
\footnotetext[4]{\unit{\e\angstrom}}
\end{table}

The electron density cutoff $\nc$ specifies the value of the solute electron density corresponding to the van der Waals cavity surface. To determine this value, the surface area of the van der Waals cavity $S_\cav$ for a set of small molecules was computed, as detailed in the \SM. These surface areas were then compared against van der Waals surface areas computed by an overlapping spheres model using reported atomic van der Waals radii. A value of $\nc = \SI{0.015}{\angstrom^{-3}}$ was found to provide the closest match.

The effective surface tension $\tau$ characterizes the free energy to form a cavity in the electrolyte plus the dispersion interactions between the electrolyte and the solute. This parameter was fit by comparing calculated solvation free energies of alkane molecules in water against the experimentally measured values. Since alkanes are nonpolar, their interaction with the solvent water is expected to be dominated by the cavity formation free energy. The optimal value of $\tau$ is strongly dependent on the values of the solvent and dielectric radii $R_\solv$ and $R_\diel$. A value of 1.4 \AA\ was used for the solvent radius since this is close to the effective hard sphere radius of water (1.385 \AA) \cite{CANDLE}. Using this value of $R_\solv$, the optimal values of $\tau$ for different values of $R_\diel$ are reported in Table \ref{tab:tau} where it can be seen that $\tau$ becomes more positive as $R_\diel$ increases. This behavior will be discussed in the results section.

\begin{table}
\caption{Values of the effective surface tension $\tau$ optimized for different values of $R_\diel$.}
\label{tab:tau}
\begin{ruledtabular}
\begin{tabular}{lrrrrrr}
    $R_\diel$\footnote{\unit{\angstrom}} & \num{0.7} & \num{0.9} & \num{1.0} & \num{1.1} & \num{1.3} & \num{1.5} \\
    $\tau$\footnote{\unit{\milli\eV\per\angstrom^2}} & \num{0.536} & \num{0.698} & \num{0.879} & \num{1.20} & \num{2.78} & \num{8.11} \\
\end{tabular}
\end{ruledtabular}
\end{table}

While fitting $\tau$, we also examined the effect of using different values of $R_\cav$ on the ability of the model to reproduce the experimental alkane solvation free energies. It was found that using a value of $R_\cav$ = 0 gave a better fit than any positive value, which corresponds to the cavity $S_\cav$ used for computing the cavity formation free energy being equivalent to the solvent cavity $S_\solv$.

The remaining two parameters, $R_\diel$ and $R_\ion$, are more ambiguous to parameterize than the previously discussed parameters. Their effect on computed properties such as the point of zero charge and capacitance of a metal surface and solvation free energies of polar molecules will be discussed in the next section.

\section{Results} \label{sec:results}

To demonstrate the capabilities of the nonlocal and nonlinear electrolyte model, we have applied it to examine several systems of interest to electrochemistry. The first application is for modeling the electrochemical interface at the Au(111) surface to illustrate the shapes of the different cavities, the dielectric and ionic responses, and the variation of electrostatic potential across the interface. We then demonstrate the differences between the linear and nonlinear electrolyte models, showing that both the dielectric and ionic responses must be nonlinear in order to qualitatively reproduce the characteristic `double hump' differential capacitance curve observed experimentally \cite{Sundararaman1}. Lastly, we explore the effect of the dielectric and ionic radii used in the model, showing that the properties of the interface are far more sensitive to the former than the latter. It is however found that the interfacial capacitance is underpredicted and the work function is overpredicted for all reasonable values of these parameters. We speculate why this is the case and how the model could be improved to give better predictions.

The second application is to a water bilayer on a Pt(111) surface in the presence of an implicit electrolyte. It is found that the local cavity definition in the original VASPsol implementation allows water to unphysically penetrate into the water bilayer since it does not take into account the finite size of a single water molecule. The nonlocal cavity definition eliminates this `solvent leakage' as long as a large enough value is used for the solvent radius.

Finally, we compute solvation free energies for a large set of organic molecules, the self-solvation and self-ionization free energies of water, and the absolute chemical potential of the standard hydrogen electrode. Comparable results are obtained between the nonlinear+nonlocal model and the original linear+local VASPsol model. We will discuss how different values of the dielectric radius are required to reproduce the experimental values of these different quantities and how this may be related to limitations of the model.

\subsection{Van der Waals cavities of neutral atoms} \label{sec:vdW_cavities_atoms}

First, we examine several neutral atoms to ascertain how well the computed vdW cavities correspond to reported vdW radii. The radial distance to the surface of the vdW cavity, defined as the isosurface where the cavity function has a value of 0.5, is reported for atomic H, C, O, N, Pt, and Au in Table \ref{tab:vdW_radii}. It can be seen that all of the values are within \SI{0.1}{\angstrom} of the reported vdW radii except for atomic H, which is over predicted by \SI{0.23}{\angstrom}. This could possibly be due to curvature effects since the small radius (thus higher curvature) of H would lead to less overlap of electron density with a probe atom. This is somewhat corroborated by the under prediction observed for Pt and Au, which have larger radii (thus lower curvature) so would be expected to have greater overlap of electron density with a probe atom. One must keep in mind however that vdW radii for transition metal atoms are less well defined than those for main group elements.

\begin{table}
\caption{Van der Waals radii (\unit{\angstrom}) computed for atoms from the vdW cavity.}
\label{tab:vdW_radii}
\begin{ruledtabular}
\begin{tabular}{ldddddd}
    & \mbox{H} & \mbox{C} & \mbox{N} & \mbox{O} & \mbox{Pt} & \mbox{Au} \\ \hline
    computed & 1.33 & 1.72 & 1.64 & 1.56 & 2.03 & 2.02 \\
    literature & 1.10\footnotemark[1] & 1.70\footnotemark[1] & 1.55\footnotemark[1] & 1.52\footnotemark[1] & 2.13\footnotemark[2] & 2.14\footnotemark[2] \\
\end{tabular}
\end{ruledtabular}
\footnotetext[1]{Ref \citenum{vdW1}}
\footnotetext[2]{Ref \citenum{vdW2}}
\end{table}

\subsection{Modeling the electrochemical interface at an Au(111) electrode} \label{sec:Au111}

To demonstrate the performance of the implicit electrolyte model for describing the electrochemical interface, we apply it to an Au(111) electrode in an aqueous 1:1 electrolyte. This electrode was chosen since it is the least likely to have chemical interactions with the electrolyte due to the inertness of the Au(111) surface. The system was modeled as a 6-layer slab with each layer containing a \numproduct{2 x 2} surface unit cell. Further details are given in the \SM.

\subsubsection{Shapes of the vdW, solvent, ionic, and dielectric cavities} \label{sec:Au111_cavities}

\begin{figure}
    \includegraphics[width=.45\textwidth]{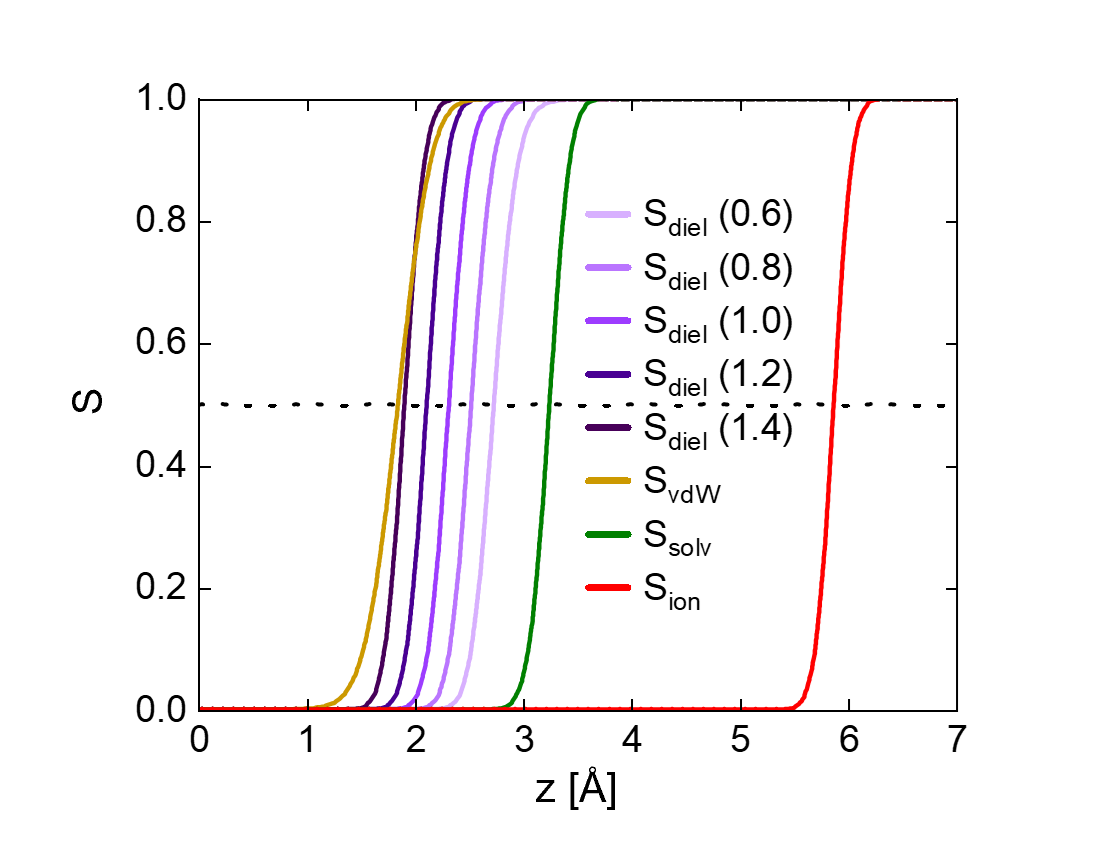}
    \caption{Van der Waals cavity $S_\vdW$ computed for $\nc = \SI{0.015}{\angstrom^{-3}}$, solvent cavity $S_\solv$ computed for $R_\solv = \SI{1.4}{\angstrom}$, ionic cavity $S_\ion$ computed for $R_\ion = \SI{4}{\angstrom}$, and dielectric cavities $S_\diel$ computed for the labeled values of $R_\diel$. The horizontal axis corresponds to the distance normal to the plane of the surface.}
    \label{fig:cavities}
\end{figure}

Figure \ref{fig:cavities} shows the shape of the different cavity functions (averaged in the plane of the surface) for the uncharged electrode along the direction normal to the surface. The transition of the vdW cavity (the isosurface where $S_\vdW=0.5$) is centered \SI{1.85}{\angstrom} from the plane of the surface (defined as the plane passing through the centers of the top layer of Au atoms), which is closer than the reported vdW radius of Au \qtyrange{2.10}{2.14}{\angstrom}. This would be expected since the vdW cavity will extend closer to the plane of the surface in the regions between metal atoms. In fact, the transition of the vdW cavity directly above a surface Au atom is centered \SI{2.04}{\angstrom} from the surface plane, in much better agreement with the vdW radius.
%%SR I do not quite get how the vdW radius in your case is calculated. Is this not a fittable parameter in your model?
%%CP It is the distance of the isosurface corresponding to Svdw=0.5 from the center of the atom. I tried to explain better in the text

The centers of the solvent and ionic cavities are displaced by distances of \qtylist{1.40;4.02}{\angstrom}, respectively, outwards from the center of the vdW cavity, while the center of the dielectric cavity is displaced a distance of \SI{0.93}{\angstrom} inwards from the center of the solvent cavity. These cavity positions correspond well to the specified solvent, ionic, and dielectric radii of \qtylist{1.4; 4.0; 1.0}{\angstrom}, indicating that the method for constructing these cavities works properly.

\subsubsection{Surface, bound, and ionic charge distributions and electrostatic potential} \label{sec:Au111_charges}

\begin{figure}
    \includegraphics[width=.45\textwidth]{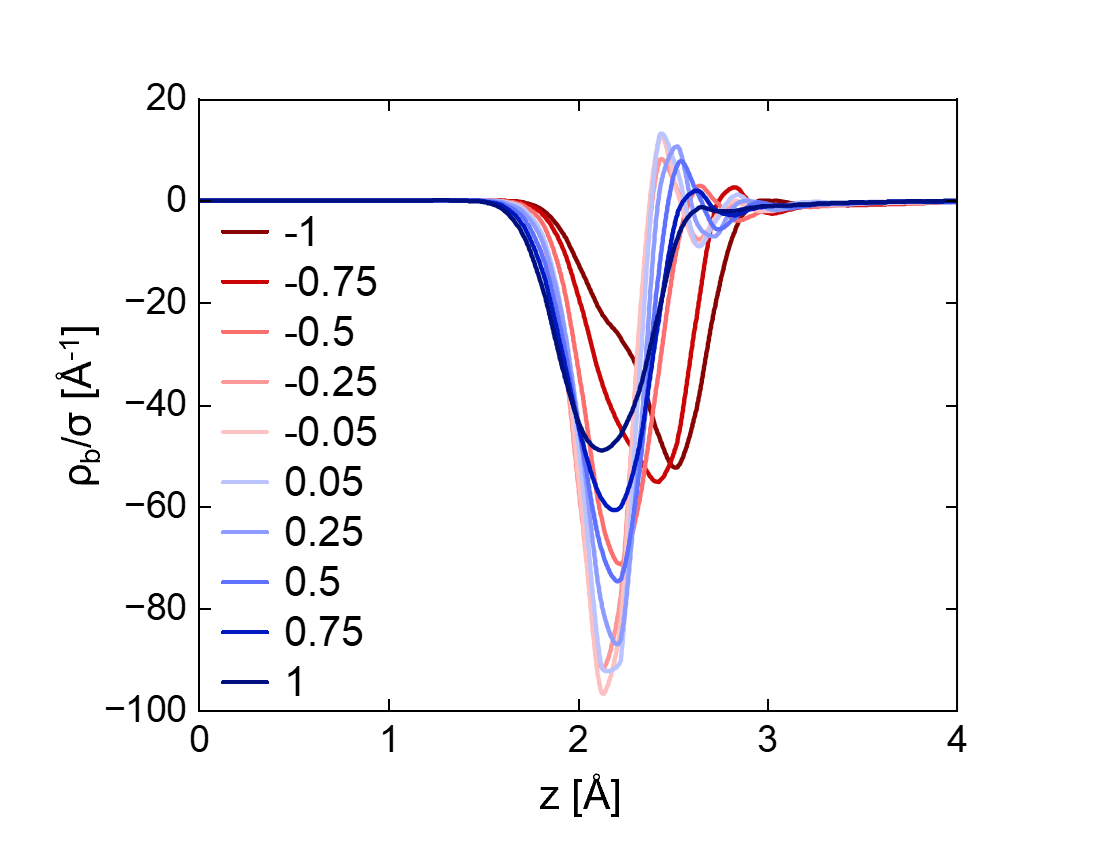}
    \caption{Bound charge distributions for the Au(111) surface in an aqueous \SI{1}{\molar} 1:1 electrolyte calculated at different surface charge densities. The bound charge has been normalized by the surface charge density and each curve is labeled by the charge (in units of \unit{\e}) in a \numproduct{2 x 2} unit cell of a six-layer slab. The horizontal axis corresponds to the distance normal to the plane of the surface.}
    \label{fig:bound_charge}
\end{figure}

Figure \ref{fig:bound_charge} depicts the surface and bound charge density profiles in a \SI{1}{\molar} 1:1 electrolyte for Au(111) electrodes with varying charges on the surface. The corresponding profiles for the neutral surface have been subtracted and the resulting profiles have been normalized by the magnitude of the surface charge to facilitate comparison of the shapes at different surface charges. One can see that the profile of the surface charge, defined as the distribution of the electron density that accumulates on the surface as it is charged, is similar for all three surface charge states. The surface charge peaks about \SI{1.15}{\angstrom} from the plane passing through the centers of the surface atoms, which is consistent with other calculations showing the surface charge extending approximately \SI{1}{\angstrom} past the centers of the surface atoms \cite{Sundararaman3}.

\begin{figure}
    \includegraphics[width=.45\textwidth]{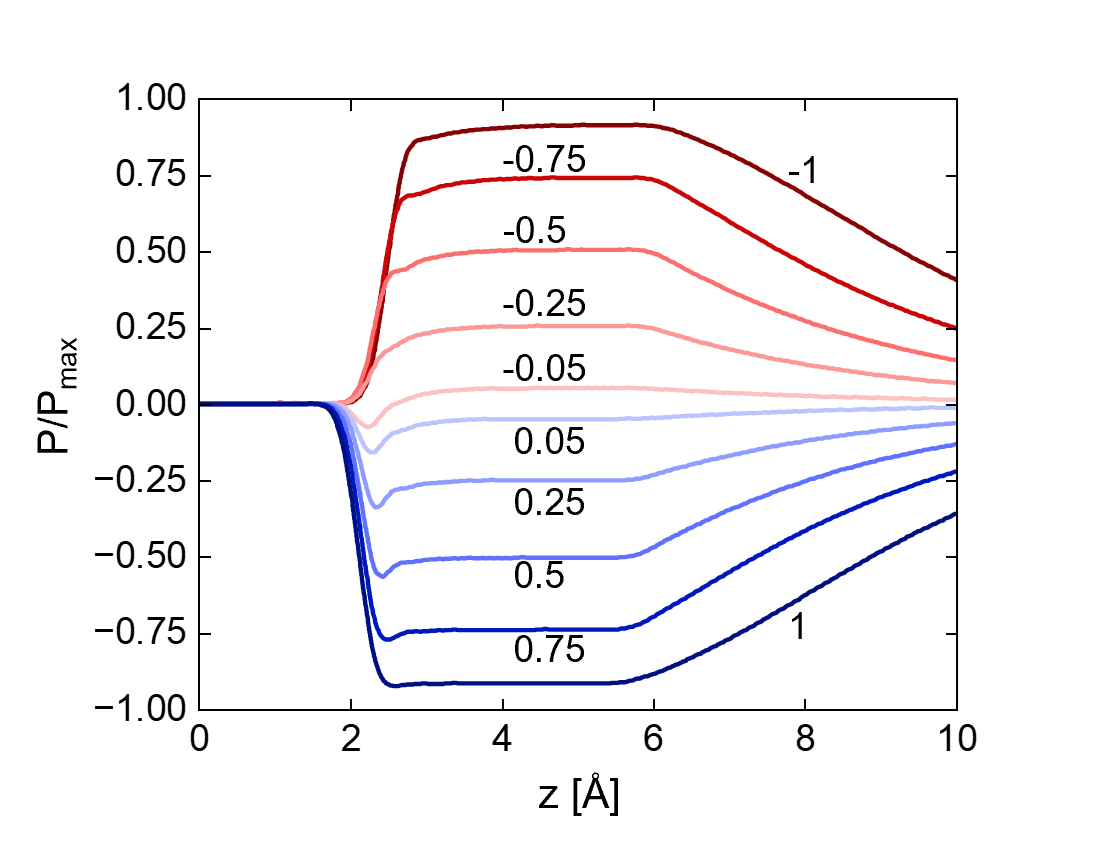}
    \caption{Polarization density $P$ relative to $P_\rm{max} = n_\rm{mol} p_\rm{mol}$ for the Au(111) surface in an aqueous \SI{1}{\molar} 1:1 electrolyte calculated at different surface charge densities. Each curve is labeled by the charge (in units of \unit{\e}) in a \numproduct{2 x 2} unit cell of a six-layer slab. The horizontal axis corresponds to the distance normal to the plane of the surface.}
    \label{fig:polarization}
\end{figure}

The bound charge density begins around \SI{1.5}{\angstrom} from the plane of the surface and peaks at a distance of \qtyrange{2.2}{2.5}{\angstrom}. The peak is located close to \SI{2.2}{\angstrom} from the surface plane for anodic polarization but moves away towards \SI{2.5}{\angstrom} for increasingly cathodic polarization. Close to the potential of zero charge (PZC), the bound charge peak has a width at half maximum of approximately \SI{0.3}{\angstrom}, mainly due to the Gaussian smoothing function $w_\rm{b}$ in eq \eqref{eq:rhob} using $a = \SI{0.125}{\angstrom}$. Further from the PZC, the bound charge peak becomes wider and shorter due to dielectric saturation in the high electric field. This can be seen in the polarization density plotted in Figure \ref{fig:polarization}, which approaches the limiting value of $P_\rm{max} = n_\rm{mol} p_\rm{mol}$ for highly charged surfaces.

\begin{figure}
    \includegraphics[width=.45\textwidth]{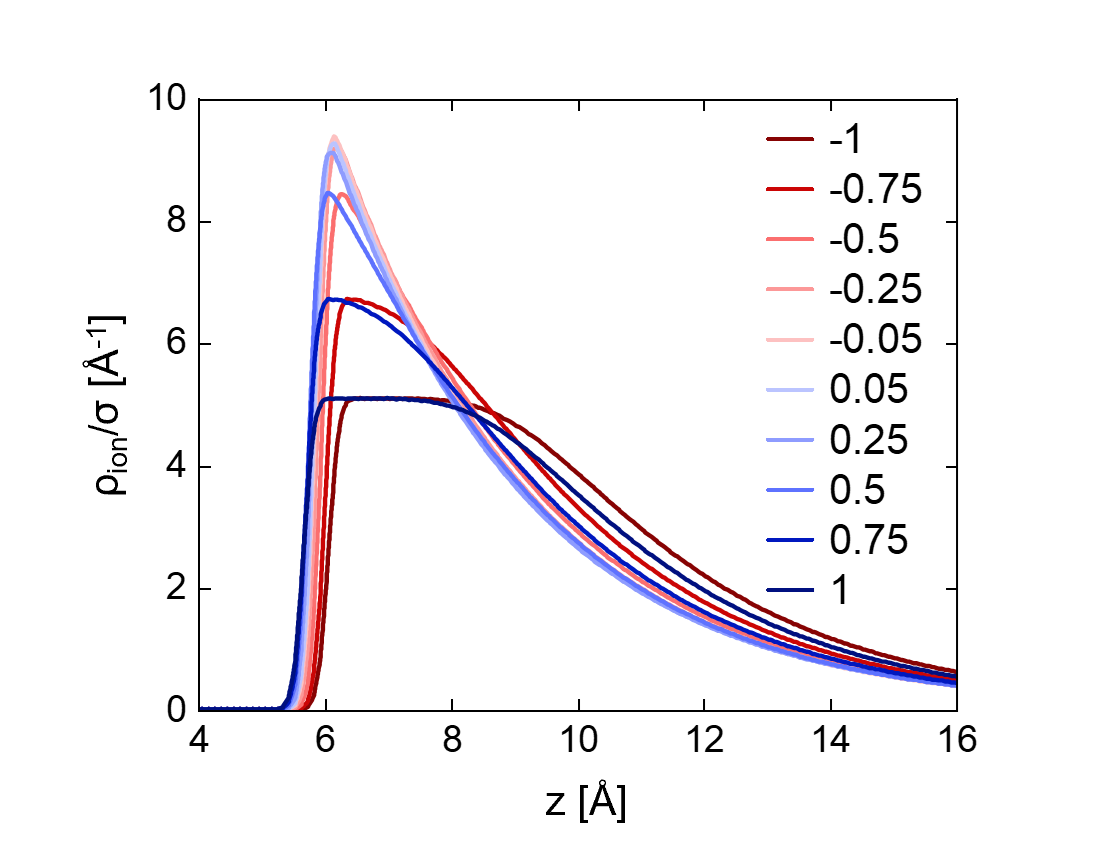}
    \caption{Ionic charge distributions for the Au(111) surface in an aqueous \SI{1}{\molar} 1:1 electrolyte calculated at different surface charge densities. The ionic charge has been normalized by the surface charge density and each curve is labeled by the charge (in units of \unit{\e}) in a \numproduct{2 x 2} unit cell of a six-layer slab. The horizontal axis corresponds to the distance normal to the plane of the surface.}
    \label{fig:ionic_charge}
\end{figure}

The ionic charge density, shown in Figure \ref{fig:ionic_charge}, peaks approximately \SI{6}{\angstrom} from the surface plane and exhibits exponential decay characteristic of the diffuse layer, with a Debye length of approximately \SI{3}{\angstrom}. Far from the PZC, the ionic charge density exhibits a layer where the ion concentration approaches the saturation limit $n_\rm{max}$. Additionally, close to the PZC the ionic charge density is almost fully screened by the bound charge, while far from the PZC an appreciable amount remains unscreened close to the edge of the ionic cavity due to dielectric saturation. This can be seen in Figure \ref{SI-fig:SI_ionic_charge} in the \SM. Figure \ref{SI-fig:log_ionic_cavity} in the \SM shows that the nonlinear region extends about \SI{6}{\angstrom} into the ionic cavity at the highest surface charges examined. Noticeably, there is no region exhibiting the super-exponential decay that arises at lower bulk electrolyte concentrations. This is due to the fact that a bulk electrolyte concentration of \SI{1}{\molar} is already \SI{44}{\percent} of the saturation limit ($\theta_\rm{b}^\ion = \num{0.44}$) for $R_\ion = \SI{4}{\angstrom}$. At this value of $\theta_\rm{b}^\ion$, the ionic response function does not exhibit exponential enhancement near the PZC, as can be seen in Figure \ref{fig:grot_gion}.

\begin{figure}
    \includegraphics[width=.45\textwidth]{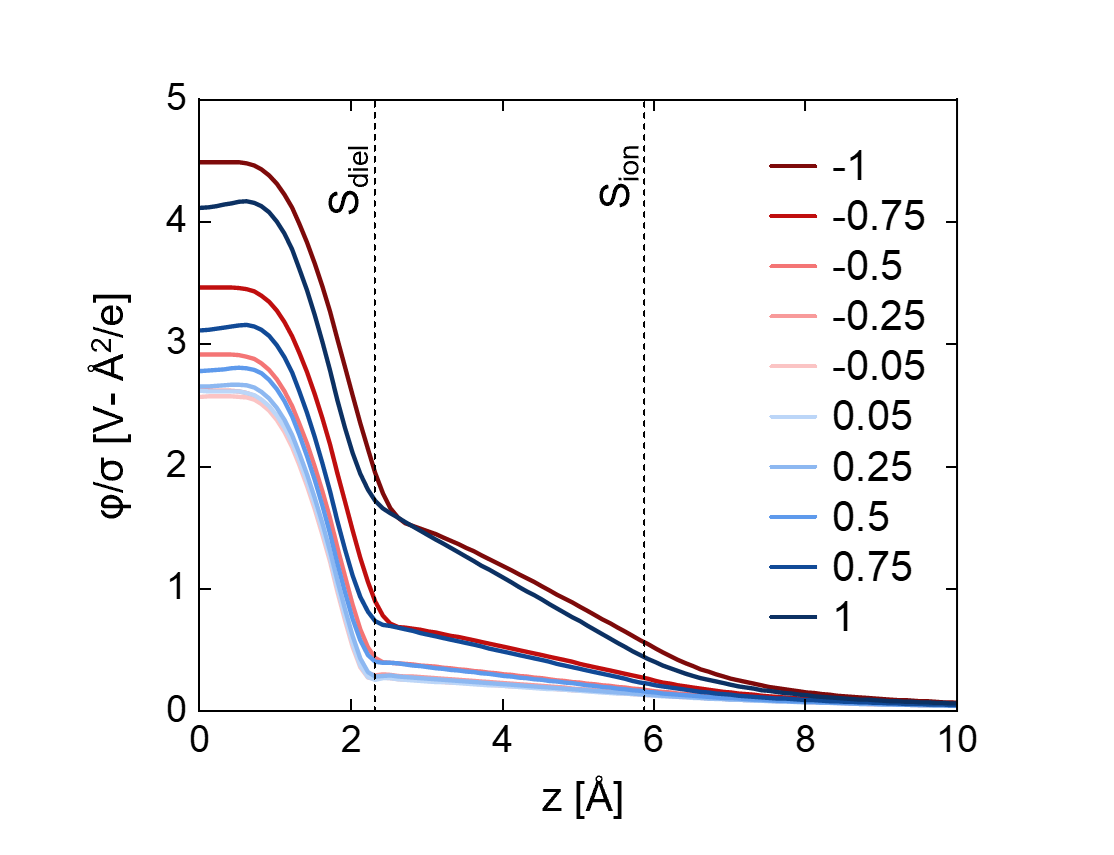}
    \caption{Electrostatic potential with respect to the bulk electrolyte plotted for the Au(111) surface in an aqueous \SI{1}{\molar} 1:1 electrolyte calculated at different surface charge densities. The electrostatic potential has been normalized by the surface charge density and each curve is labeled by the charge (in units of \unit{\e}) in a \numproduct{2 x 2} unit cell of a six-layer slab. The horizontal axis corresponds to the distance normal to the plane of the surface.}
%%SR In this figure, how about drawing also lines for the solvent radii to show how they coincide with the kinks in the potential?
%%TODO@SMRI
%%SMRI done
%%CP Agreed, I think I was originally planning to do that
%%TODO@SMRI change y-axis label to phi/sigma (V-Ang^2/e)
%%SMRI done
    \label{fig:potential}
\end{figure}

The electrostatic potential with respect to the bulk electrolyte (and normalized by the surface charge) is plotted in Figure \ref{fig:potential}, where it can be seen to exhibit four characteristic regions. The potential is nearly constant in the region inside the surface, although it exhibits Friedel oscillations deeper inside the slab. Around \SI{1}{\angstrom} outwards from the surface plane, the potential begins to drop off rapidly until the edge of the dielectric cavity. This region corresponds to the \textit{vacuum gap} \cite{Sundararaman1,Sundararaman2} and makes the largest contribution to the overall potential drop since dielectric screening is not present. The third region corresponds to the \textit{solvent gap} \cite{Sundararaman1}, and extends from the edge of the dielectric cavity to the edge of the ionic cavity. The potential drops off more gradually in this region since dielectric screening is present. However, for high polarization the potential drop in this region becomes significant and comparable to the drop in the vacuum gap. This is due to dielectric saturation that occurs at high electric field strengths in the nonlinear dielectric model. The last region occurs inside the ionic cavity. At low polarization the potential drops exponentially in this region, corresponding to the diffuse layer, while at high polarization there is a region near the edge of the ionic cavity where the potential drops super-exponentially due to dielectric and ionic saturation. This can be seen more clearly in Figure \ref{SI-fig:log_ionic_cavity} in the \SM. 
%For moderate polarization at lower bulk electrolyte concentrations, we have found instead a region of sub-exponential potential drop due to the presence of an enhancement region in the ionic response curve.

% capacitance at 1 \AA separation is 8.85 uF/cm2

\subsubsection{Effect of linear and nonlinear dielectric and ionic responses on the differential capacitance curve} \label{sec:Au111_L_vs_NL}

\begin{figure}
    \includegraphics[width=.45\textwidth]{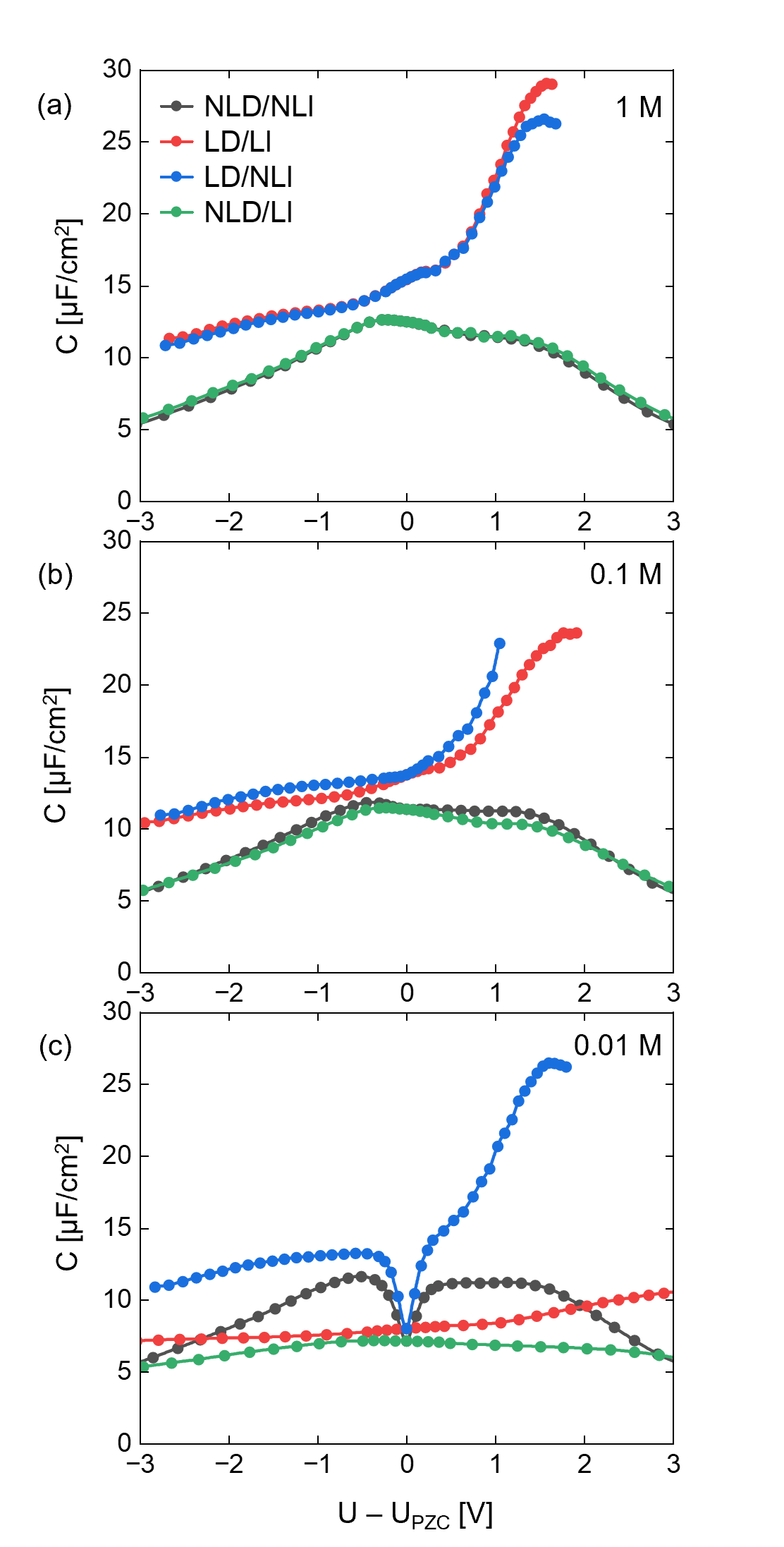}
    \caption{Differential capacitance computed for the Au(111) surface in an aqueous 1:1 electrolyte at concentrations of (top) \SI{1}{\molar}, (middle) \SI{0.1}{\molar}, and (bottom) \SI{0.01}{\molar}. The calculations were performed using different combinations of linear (LD) / nonlinear (NLD) dielectric response and linear (LI) / nonlinear (NLI) ionic response.}
    \label{fig:linear_vs_nonlinear}
\end{figure}

We now examine the effect of nonlinear versus linear dielectric and ionic responses on the differential capacitance curve. These are shown in Figure \ref{fig:linear_vs_nonlinear} for electrolyte concentrations of \qtylist{0.01;0.1;1}{\molar}. At the highest concentration, the curves having linear and nonlinear ionic responses coincide over most of the potential range except at highly anodic potentials when a linear dielectric model is used. This can be explained by the fact that the nonlinear ionic response does not exhibit an enhancement region in a \SI{1}{\molar} electrolyte with $R_\ion = \SI{4}{\angstrom}$ since the bulk concentration is already \SI{44}{\percent} of the saturation limit.
%Ionic saturation at high polarization in the nonlinear model also does not have an appreciable effect because it is overwhelmed by the dielectric saturation that is also occurring.
At the lowest electrolyte concentration, however, the differential capacitance deviates significantly between the linear and nonlinear ionic response. Since the bulk electrolyte concentration is only \SI{0.44}{\percent} of the saturation limit, the nonlinear ionic response curves exhibit a pronounced enhancement region that causes the differential capacitance to increase as the potential deviates from the PZC.

Even in the \SI{1}{\molar} electrolyte, there are significant differences in the capacitance between models with a linear and nonlinear dielectric response. The difference is most pronounced at high polarization, particularly on the anodic side, and is due to a rapid increase in the differential capacitance at anodic polarization for models with a linear dielectric response. The rapid increase arises from movement of the dielectric cavity closer to the plane of the surface as polarization varies from anodic to cathodic. This in turn occurs because the vdW cavity moves closer to the surface as the electron density tail begins to decay faster at increasingly anodic potentials. The result is an increase in differential capacitance as the effective width of the vacuum gap shrinks. Due to the inverse dependence of capacitance on this width, the effect is more pronounced for narrower gaps at anodic polarization. This is enhanced even further by additional dielectric polarization due to penetration of the dielectric cavity further into the electron density of the surface, as discussed in the next paragraph. The behavior is not observed in models with a nonlinear dielectric response because the response close to the surface is already beginning to saturate even at the PZC, as discussed in the next paragraph.

%%SR So why does the non-local dielectric model somehow not show the asymmetry of the capacitance? Because eventually both local and non-local dielectric models are based on the same dielectric function definition. Can this be described better in the text?
%%CP Do you mean the nonlinear dielectric model? I added a sentence to explain this

Surprisingly, there is also a difference in capacitance between the two models even at the PZC where one would expect nonlinear effects to be absent. This occurs due to dielectric polarization that arises not from charge on the surface but from the aforementioned penetration of the dielectric cavity into the electron density of the surface. The result is dielectric polarization at the PZC, which is dampened in the nonlinear dielectric model due to dielectric saturation. This is likely a spurious effect, as it is not possible for bound charge density to overlap with the surface electron density due to Pauli repulsion. In reality, Pauli repulsion would compress the tail of the surface electron density; however, this interaction is not present in any implicit electrolyte model that we are aware of.

\subsubsection{Effect of dielectric and ionic radii on the properties of the interface} \label{sec:Au111_radii}

\begin{figure}
    \includegraphics[width=.45\textwidth]{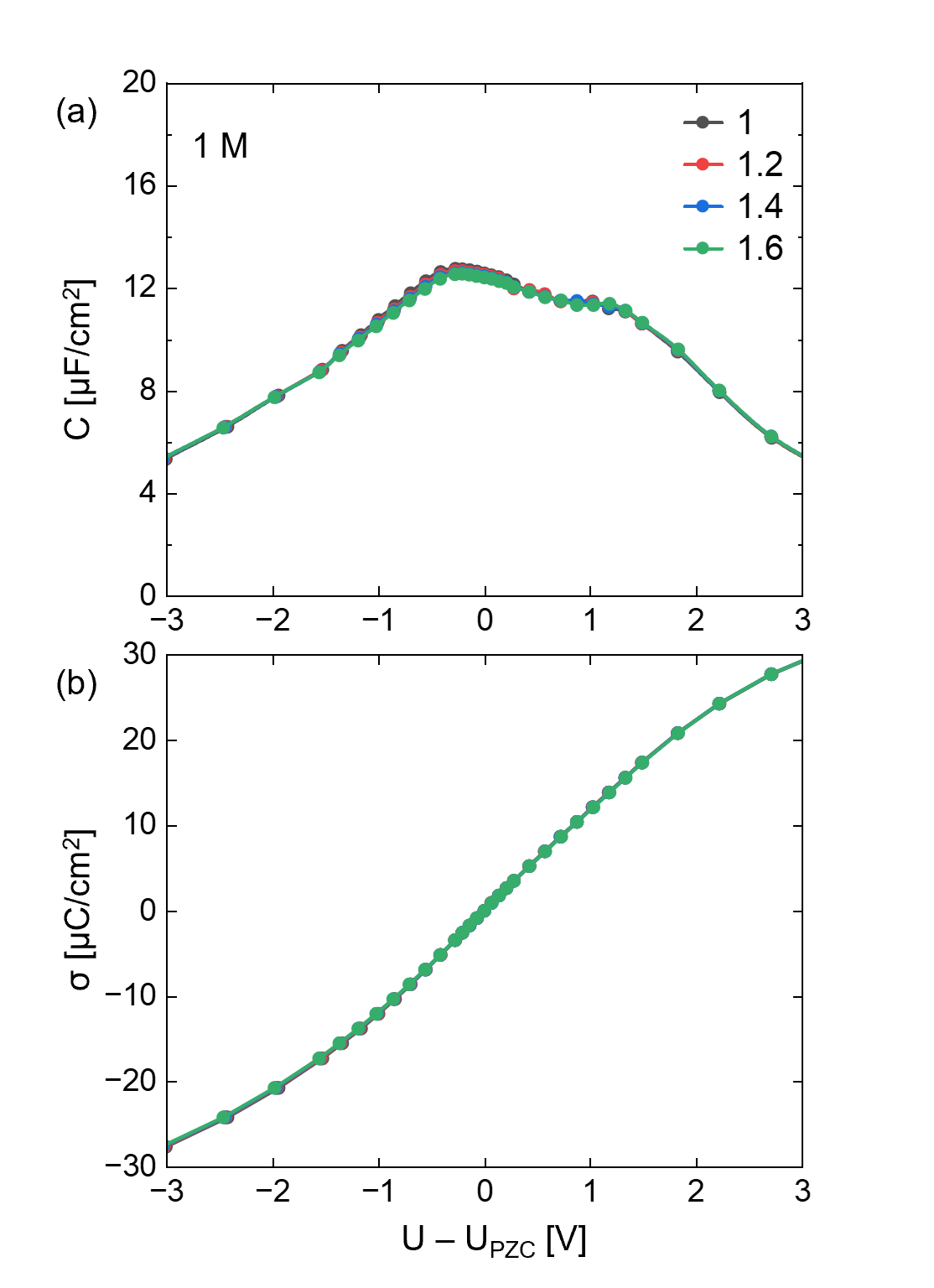}
    \caption{Differential capacitance and charging curves computed for the Au(111) surface in an aqueous \SI{1}{\molar} 1:1 electrolyte with indicated values of the solvent radius $R_\solv$. The difference $R_\solv - R_\diel$ is fixed to \SI{0.4}{\angstrom}.}
    \label{fig:capacitance_rsolv}
\end{figure}

\begin{figure}
    \includegraphics[width=.45\textwidth]{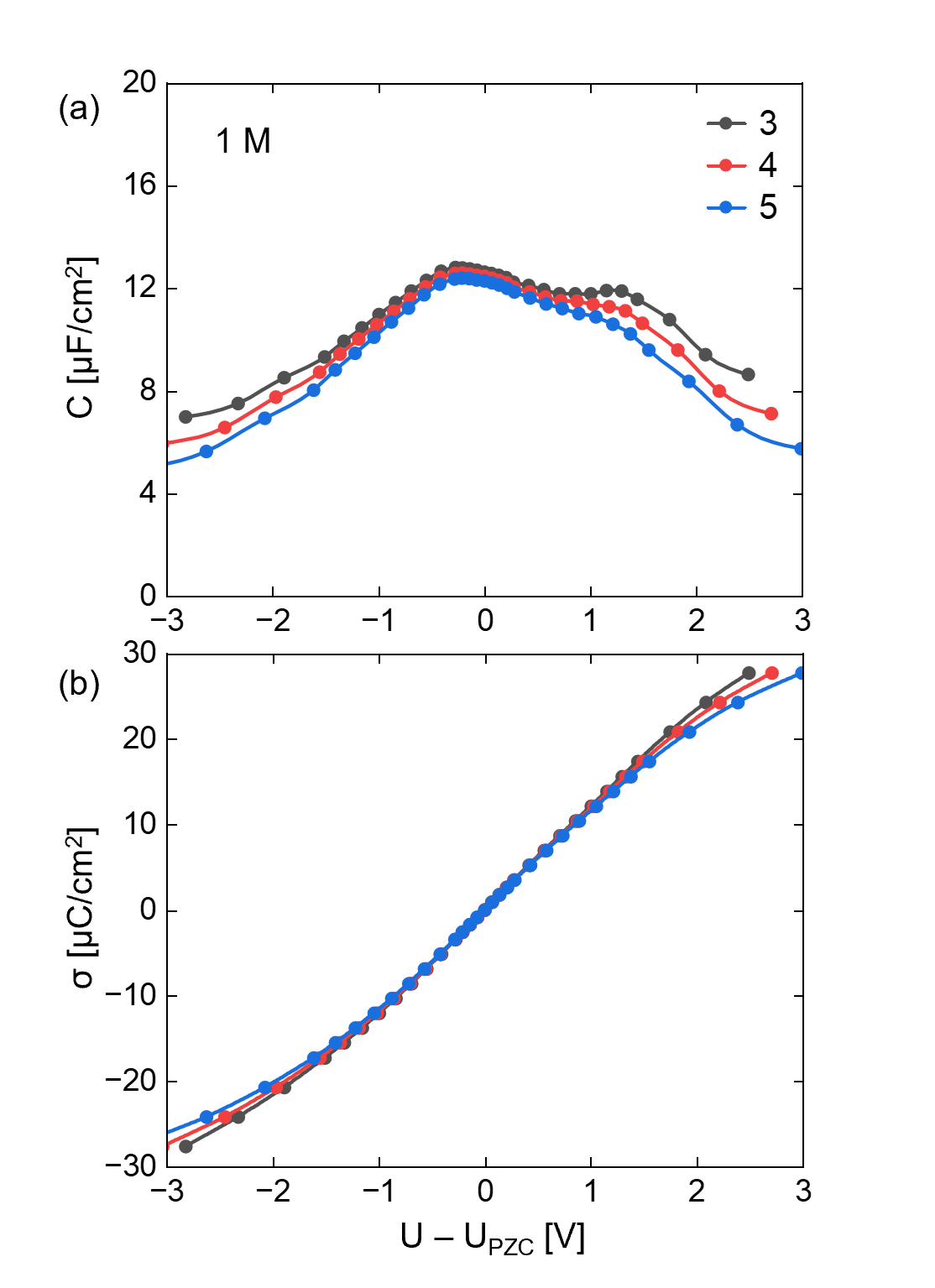}
    \caption{Differential capacitance and charging curves computed for the Au(111) surface in an aqueous \SI{1}{\molar} 1:1 electrolyte with indicated values of the ionic radius $R_\ion$.}
    \label{fig:capacitance_rion}
\end{figure}

\begin{figure}
    \includegraphics[width=.45\textwidth]{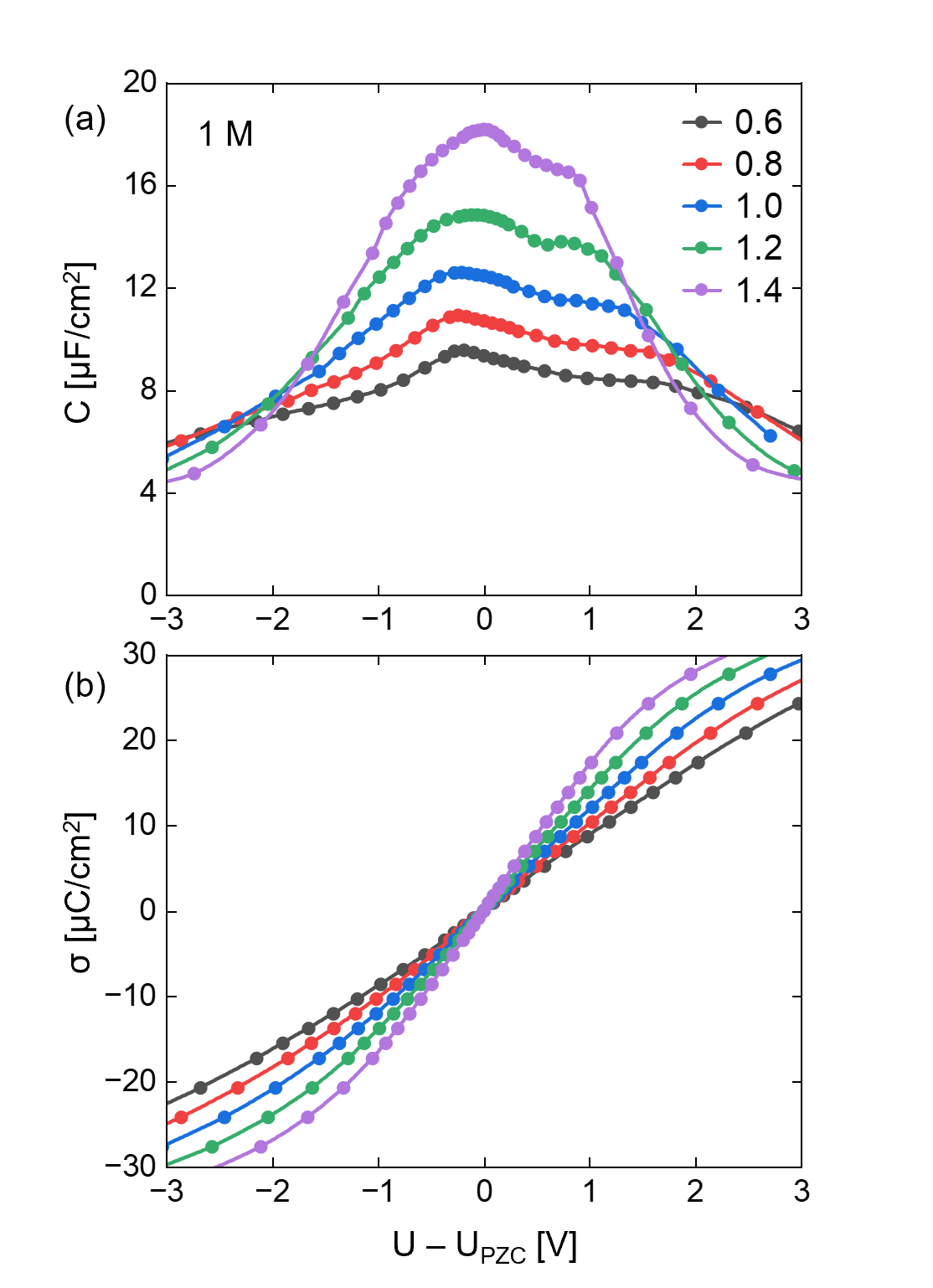}
    \caption{Differential capacitance and charging curves computed for the Au(111) surface in an aqueous \SI{1}{\molar} 1:1 electrolyte with indicated values of the dielectric radius $R_\diel$.}
    \label{fig:capacitance_rdiel}
\end{figure}

Figures \ref{fig:capacitance_rsolv} - \ref{fig:capacitance_rdiel} depict the surface charge density and differential capacitance versus potential for the fully nonlinear model computed with different values of the solvent radius $R_\solv$, the ionic radius $R_\ion$, and the dielectric radius $R_\diel$. When $R_\solv$ is varied, the difference $R_\solv-R_\diel$ is fixed at a value of \SI{0.4}{\angstrom} so that the dielectric cavity is maintained at a constant distance from the edge of the vdW cavity. As can be seen in Figure \ref{fig:capacitance_rsolv}, varying $R_\solv$ in this way has almost no effect on the predictions of the model since the position of the dielectric cavity is not changing. In contrast, it will be seen in the next section that the solvent radius has a significant impact on a surface with an adsorbed water bilayer where solvent `leaks' into the small spaces between the explicit water molecules if $R_\solv$ is too small.

Varying the ionic radius has a larger effect on the differential capacitance curve, as seen in Figure \ref{fig:capacitance_rion}. As $R_\ion$ increases, the capacitance decreases since the ionic cavity is being pushed further away from the surface. This is most pronounced at high polarization when the dielectric response begins to saturate in the solvent gap between the surface and the ionic cavity. Additionally, a larger ionic radius extends the region of ionic saturation in the ionic cavity at high polarization. It it not possible to have a \SI{1}{\molar} electrolyte with an ionic radius greater than \SI{5.25}{\angstrom}, so higher values are absent from the plot in Figure \ref{fig:capacitance_rion}. A radius of \SI{4}{\angstrom} corresponds to the \ce{K+} cation, so this value is used as a default value.
%%SR reference needed to give for this value.
%%TODO@SR
%%CP What was the reference you used for this value? I got it from your paper on cation size effects. You had a plot of solvated radius vs. ionic radius or something like that

\begin{figure}
    \includegraphics[width=.45\textwidth]{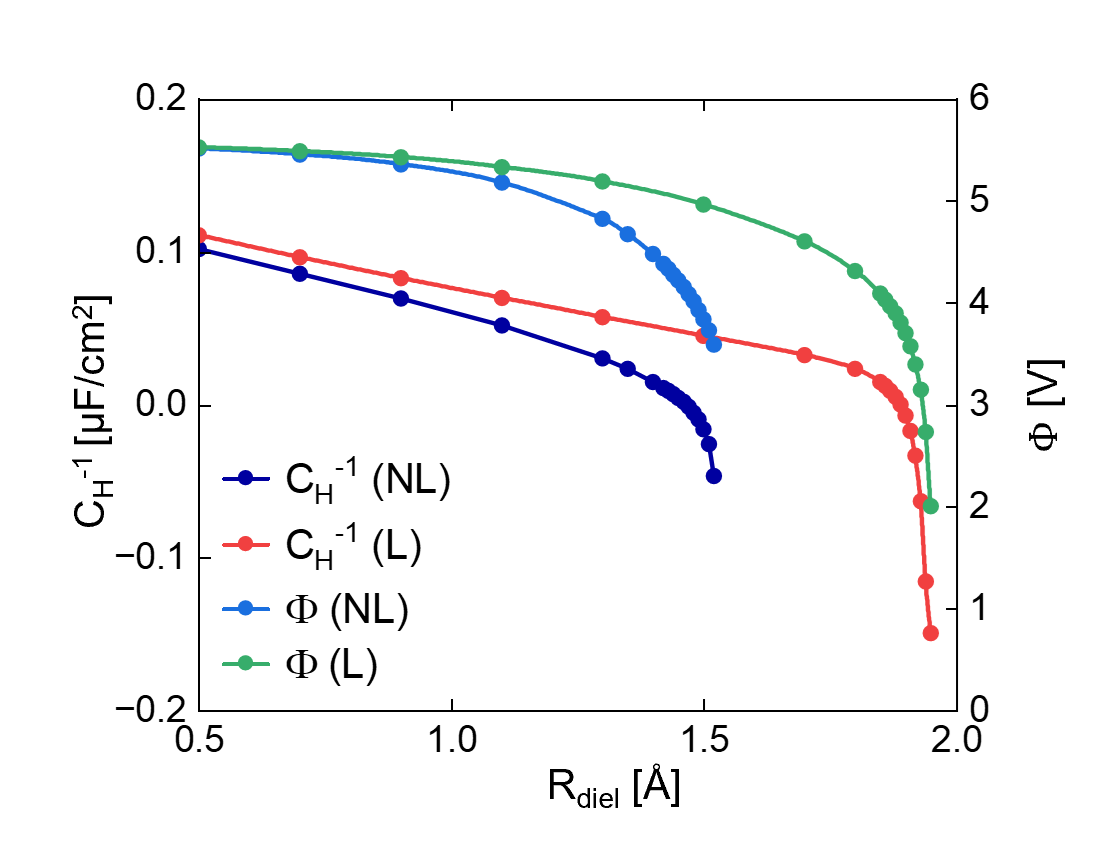}
    \caption{Inverse Helmholtz capacitance and work function computed at the PZC for the Au(111) surface in an aqueous \SI{1}{\molar} 1:1 electrolyte, using both nonlinear and linear electrolyte models.}
    \label{fig:CH_and_WF}
\end{figure}

As can be seen in Figure \ref{fig:capacitance_rdiel}, the variation of the dielectric radius has the largest effect of all three model radii ($R_\solv$, $R_\ion$, $R_\diel$) on the differential capacitance. We can also calculate the Helmholtz capacitance at the PZC by subtracting off the Gouy-Chapman capacitance ($C_\rm{GC} = \SI{228}{\micro\farad\per\cm^2}$) from the double layer capacitance $C_\rm{d}$ computed for a \SI{1}{\molar} aqueous 1:1 electrolyte,
\begin{equation}
    C_\rm{H}^{-1} = C_\rm{d}^{-1} - C_\rm{GC}^{-1}
\end{equation}
Figure \ref{fig:CH_and_WF} shows that the inverse Helmholtz capacitance at the PZC nearly halves upon increasing $R_\diel$ from \qtyrange[range-phrase={ to }]{0.6}{1.4}{\angstrom}. The reason is that the largest contribution to the potential drop occurs in the vacuum gap where there is no screening. As $R_\diel$ increases, the dielectric cavity moves closer to the surface and the vacuum gap decreases. This leads to a linear decrease in $C_H^{-1}$ up until a dielectric radius of about \SI{1.8}{\angstrom}. For $R_\diel$ larger than this, $C_H^{-1}$ begins to rapidly decrease and actually becomes negative for $R_\diel$ equal to \SI{1.9}{\angstrom} or larger. This is caused by the spurious penetration of bound charge into the electron density of the surface that was discussed in the previous section. As $R_\diel$ increases, the center of the bound charge moves closer towards the center of the surface charge until the centers overlap causing the capacitance to diverge.

%%SR from the figure it seems that that it is the NL model which diverges at a smaller radius of 1.45
%%CP no, they were just mislabeled in the figure. Need to fix
The linear model exhibits similar behavior, but the capacitance diverges at a lower dielectric radius of \SI{1.45}{\angstrom}. This occurs because the bound charge is located significantly closer to the surface in the linear model than in the nonlinear model for the same value of $R_\diel$, as seen in Figure \ref{SI-fig:SI_bound_charge} in the \SM. In the absence of dielectric saturation, significant polarization of the electrolyte occurs in the tail of the dielectric cavity. This causes the bound charge to peak at a distance where $S_\diel$ has a value only around \num{0.005}. Thus, it can be concluded that the linear model allows dielectric screening to unphysically extend much further than the dielectric radius. In contrast, the bound charge peaks in the nonlinear model at a distance where $S_\diel$ has a much larger value of around \num{0.2}. Polarization diminishes rapidly at distances further from the edge of the dielectric cavity, since the maximum polarization is limited to a value of $P_\rm{max} = S_\diel n_\rm{mol} p_\rm{mol}$ by dielectric saturation. This also explains why the Helmholtz capacitance at the PZC is significantly higher in the linear model than in the nonlinear model.

It should be noted that reasonable values of the dielectric radius ($R_\diel < R_\solv = \SI{1.4}{\angstrom}$) lead to values of the Helmholtz capacitance at the PZC that are \numrange{4}{5} times lower than the experimentally measured values in the range \qtyrange{70}{100}{\micro\farad\per\cm^2} \cite{Koper}. This high of a value would require a vacuum gap on the order of \SI{0.1}{\angstrom}, which is unlikely to be possible given the size of a water molecule (\SI{\sim 1.4}{\angstrom}) and the Thomas-Fermi screening length in typical metals (\SI{\sim 0.5}{\angstrom}). A dielectric radius above \SI{1.85}{\angstrom} is required to obtain a Helmholtz capacitance in the experimental range, which was seen to result in extreme unphysical overlap of the bound charge and surface charge densities. It is more likely that adsorption of water on the surface is responsible for this high value of the capacitance, as has been suggested for Pt(111) \cite{Pt111}. This is further evidenced by the anomalously low value of the Parsons-Zobel slope measured on Au(111) and Pt(111) \cite{Koper}.

In addition to the capacitance, the dielectric radius also has a large impact on the work function. Figure \ref{fig:CH_and_WF} shows that as the dielectric cavity moves closer to the surface, the work function decreases in both the linear and nonlinear models. The work function also appears to diverge around the same value of $R_\diel$ that $C_H^{-1}$ diverges in each model. This is caused by dielectric polarization of the solvent penetrating into the electron density of the surface. The (likely unphysical) polarization results in a dipole layer with negative charge directed towards the surface that lowers the energy required to remove an electron across the interface. When the dielectric cavity penetrates deeper into the surface, the dielectric polarization increases and so does the work function. The experimental work function of Au(111) in a dilute aqueous electrolyte is close to \SI{5.00}{\eV}, which best corresponds to the largest dielectric radius of \SI{1.4}{\angstrom}. Nonetheless, one should keep in mind the physically dubious origin of the decrease in work function in the implicit solvation model.

\subsection{Prevention of electrolyte `leakage' by the nonlocal cavity definition} \label{sec:leakage}

The main purpose of introducing a nonlocal cavity definition into our model is to prevent the `leakage' of solvent into small spaces that would normally not be able to accommodate a single water molecule. For example, it has been found that implicit water enters into the spaces between water molecules in an adjoining explicit water phase in solvation models using a local cavity definition \cite{solvent-aware}. In such a model, the cavity functions that determine the dielectric and ionic responses at a given location are based only on the local electron density of the solute at that same location. Thus, the cavity function has no `knowledge' of the solute electron density in the surrounding region. With the nonlocal cavity definition used here, the cavity function at a given location is determined based on the solute electron density everywhere in the local vicinity through use of convolutions. As such, it is able to exclude solvent from regions of space that are too small to accommodate a single water molecule.

\begin{figure}
    \includegraphics[width=.45\textwidth]{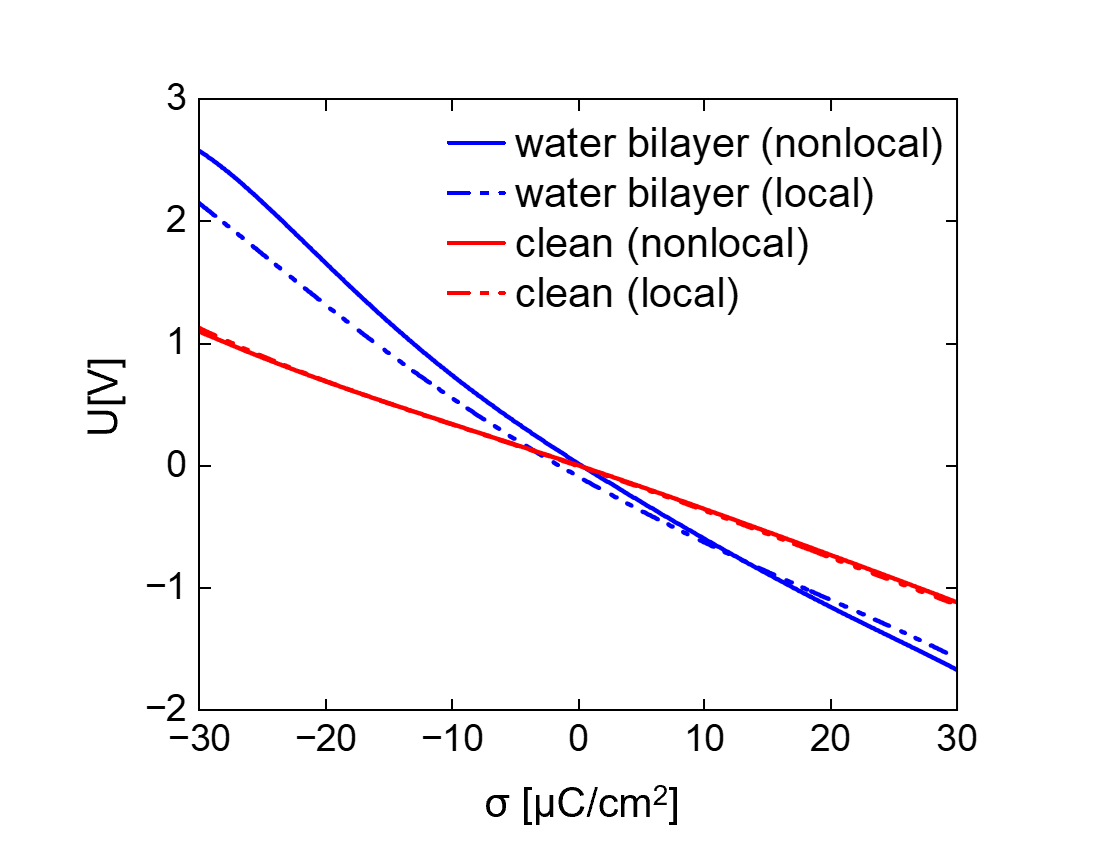}
    \caption{Charging curves computed for the Pt(111) surface with an explicit water bilayer in an aqueous \SI{1}{\molar} 1:1 electrolyte, using both nonlocal and local cavity definitions. The charging curve for the clean Pt(111) surface is also shown.}
    \label{fig:Pt_water_charge}
\end{figure}

To test the ability of the nonlocal cavity definition to exclude solvent from such small regions of space, we apply it to the case of a water bilayer on a Pt(111) surface. Such structures are proposed to form at anodic potentials where water binds strongly to the surface, and their formation may be responsible for the rapid increase in capacitance immediately anodic of the PZC \cite{Pt111}. Figure \ref{fig:Pt_water_charge} compares the charging curves calculated for this surface using both the nonlocal cavity definition and a local definition. The nonlocal dielectric cavity was calculated using a solvent radius of \SI{1.4}{\angstrom} and a dielectric radius of \SI{1.0}{\angstrom}, while the local cavity is constructed using an electron density cutoff of $n_\rm{c} = \SI{0.00336}{\angstrom^{-3}}$. This cutoff was chosen so that both local and nonlocal models give nearly the same charging response for a clean Pt(111) surface (also shown in Figure \ref{fig:Pt_water_charge}).

\begin{figure}
    \includegraphics[width=.45\textwidth]{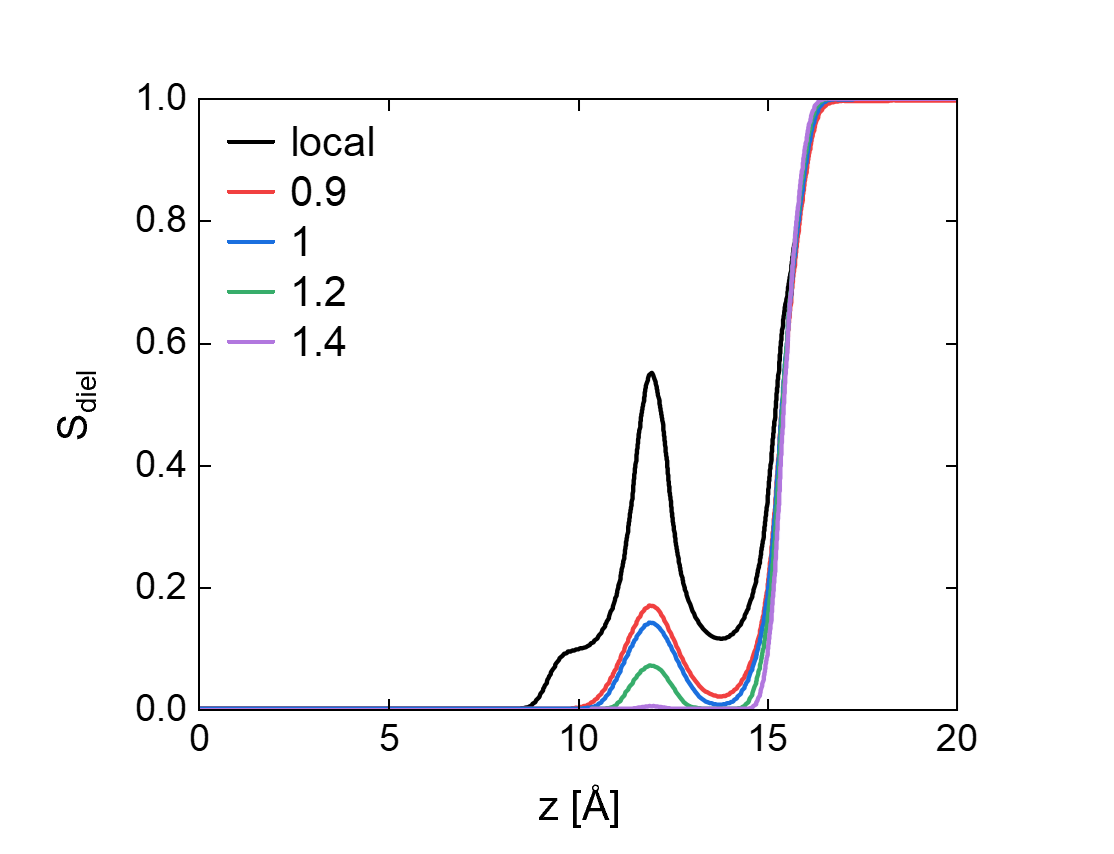}
    \caption{Dielectric cavity $S_\diel$ computed for the Pt(111) surface with an explicit water bilayer in an aqueous \SI{1}{\molar} 1:1 electrolyte, using both nonlocal and local cavity definitions. For the nonlocal cavity definition, the indicated values were used for the solvent radius $R_\solv$, with $R_\solv - R_\diel$ fixed to \SI{0.4}{\angstrom}.}
    \label{fig:Pt_water_cavity}
\end{figure}

One can see that the local cavity results in a significantly lower potential than the nonlocal cavity for a given surface charge at cathodic polarization. This is due to leakage of the solvent into the void spaces in the water bilayer, as can be seen in Figure \ref{fig:Pt_water_cavity}. The figure also shows that the nonlocal model requires a solvent radius of \SI{1.4}{\angstrom} (with $R_\solv - R_\diel$ fixed to \SI{0.4}{\angstrom}) in order to prevent solvent leakage. As $R_\solv$ decreases, the amount of implicit solvent penetrating into the water bilayer is seen to increase.

\subsection{Prediction of molecular solvation free energies} \label{sec:molecules}

The final application we will discuss is for computing solvation free energies of molecules. A large set of simple organic molecules was examined that includes alkanes, alcohols, ethers, aldehydes, and ketones. Additionally, we examine the self-solvation of water in itself.

\begin{figure}
    \includegraphics[width=.45\textwidth]{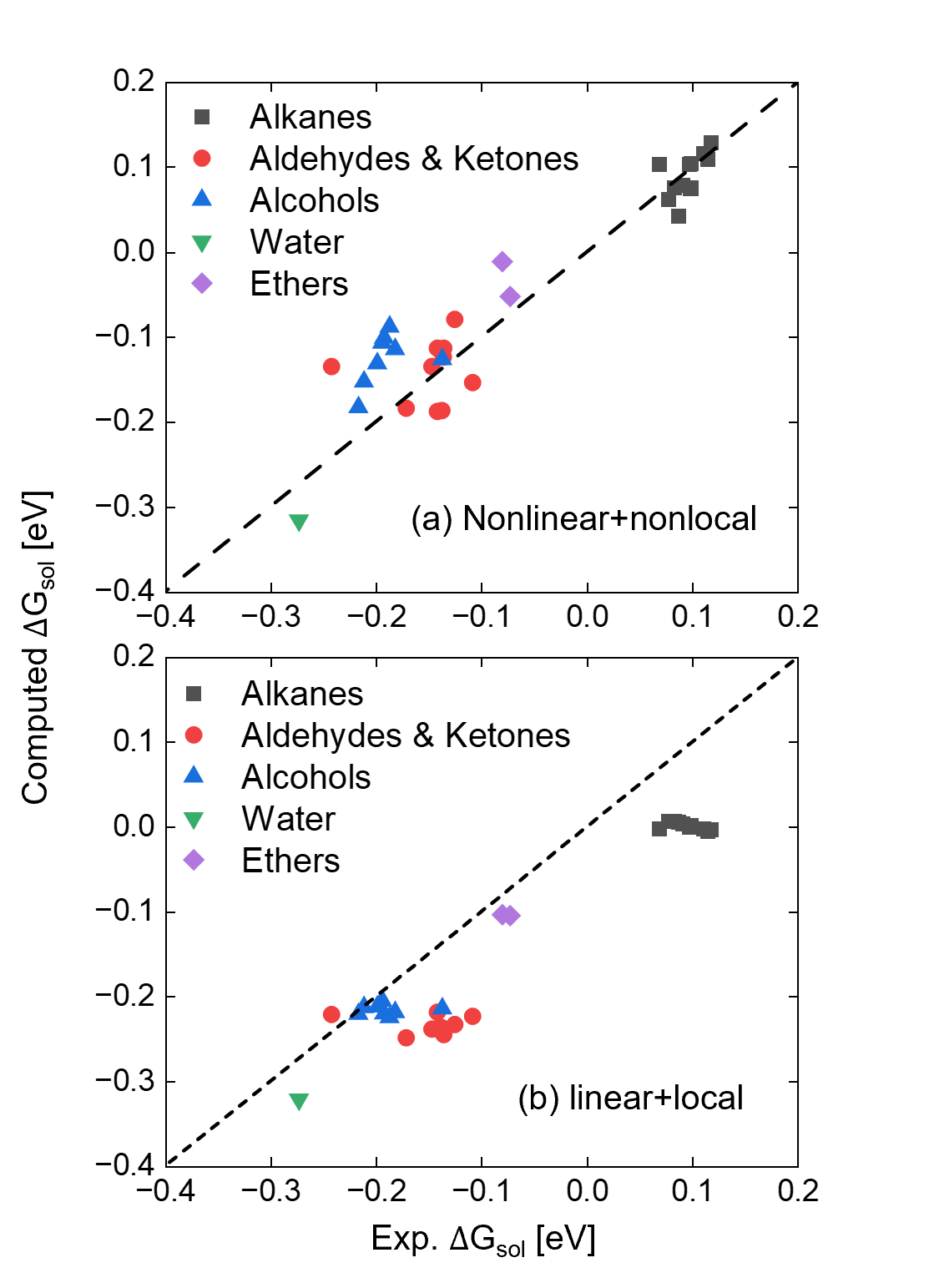}
    \caption{Parity plot comparing calculated and experimental aqueous solvation free energies for different organic molecules computed with (top) the nonlinear+nonlocal model and (bottom) the original linear+local LPCM model.}
    \label{fig:solvation_parity}
\end{figure}

A parity plot of the calculated solvation free energies for the set of organic molecules is shown in Figure \ref{fig:solvation_parity} with respect to the experimental values obtained from the UNIQUAC activity model in AspenPlus. For comparison, the same plot obtained using the original linear+local model in VASPsol is also shown. It can be seen that both models give comparable results, with the mean signed and absolute errors (MSE and MAE) of each model given in Table \ref{tab:mean_errors}.

\begin{table}
\caption{Comparison of mean signed errors (MSE) and mean absolute errors (MAE) of the solvation free energies (in \unit{\eV}) for organic molecules calculated with the linear+local and nonlinear+nonlocal models.}
\label{tab:mean_errors}
\begin{ruledtabular}
\begin{tabular}{lrr}
    & MSE & MAE \\ \hline
    lin.+loc. & \num{-0.07} & \num{0.07} \\
    nonlin.+nonloc.\footnote{Using $R_\diel = \SI{1.00}{\angstrom}$} & \num{0.02} & \num{0.04} \\
\end{tabular}
\end{ruledtabular}
\end{table}

\begin{figure}
    \includegraphics[width=.45\textwidth]{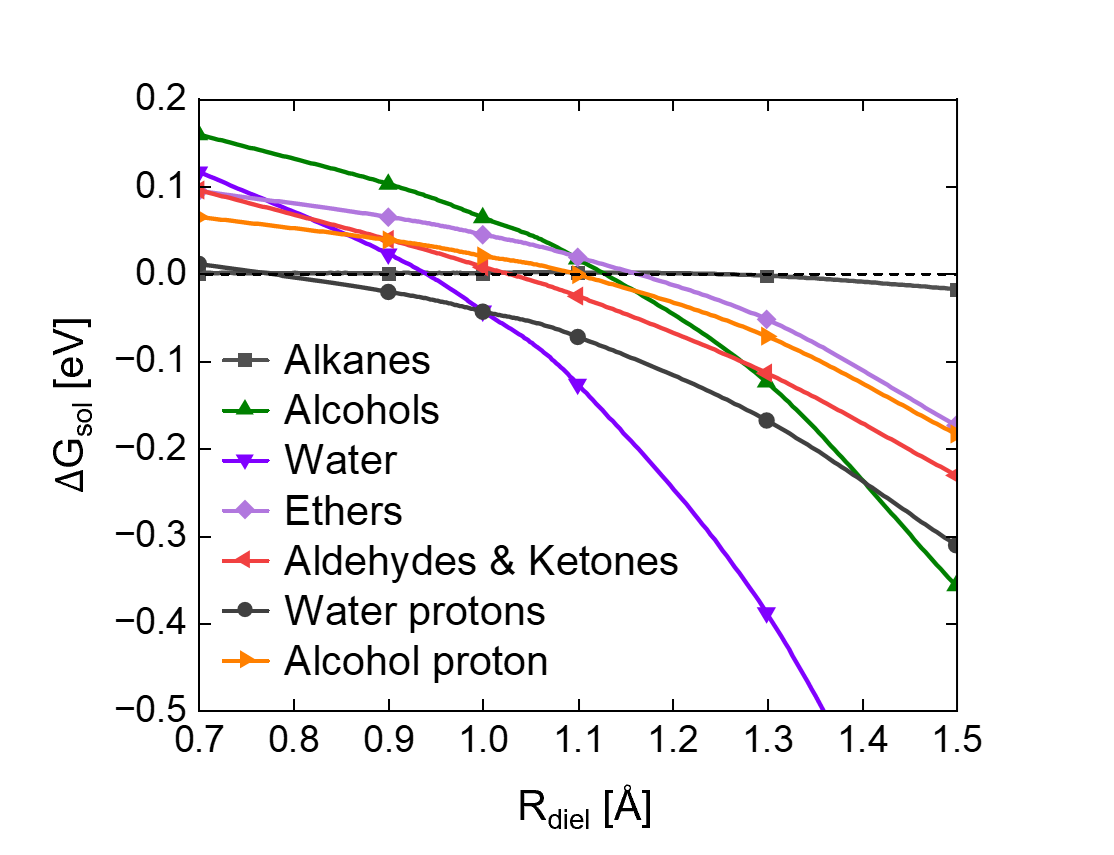}
    \caption{Errors in the computed solvation free energies of different classes of organic molecules for different values of the dielectric radius.}
    \label{fig:solvation_vs_rdiel}
\end{figure}

One key observation from Figure \ref{fig:solvation_parity} is that the new model undersolvates alcohols and ethers while oversolvating water. To explore this behavior in more depth, the MSE of each class of molecule is plotted in Figure \ref{fig:solvation_vs_rdiel} with respect to the dielectric radius. This shows that water requires the lowest dielectric radius (\SI{0.94}{\angstrom}) to reproduce the experimental self-solvation free energy. Aldehydes and ketones require roughly the same value (\SI{0.99}{\angstrom}), but alcohols and ethers require a significantly higher dielectric radius of \qtyrange{1.14}{1.16}{\angstrom}.

To understand these differences, we attempt to assign the error in solvation free energy of each molecule to individual atoms. First, we note that the the calculated solvation free energies of alkanes have almost no error since the effective surface tension $\tau$ was fit to a similar set of alkanes. This means that we can assign all of the error to the specific functional groups in the molecule. Since alcohols possess an \ce{-OH} group and ethers possess an \ce{-O-} group, we assume that the difference in the solvation error between the two groups is due to the additional proton in the alcohol. We therefore estimate the solvation error due to this proton as the difference in the MSE between alcohols and ethers. We can similarly estimate the solvation error due to the protons in water as half the difference in the MSE between water and ethers, since water has two protons. Note that with this decomposition, the solvation error assigned to the oxygen atom in alcohols and water is assumed to be equal to the MSE of ethers.

The solvation error assigned to the protons in alcohols and water is also plotted with respect to the dielectric radius in Figure \ref{fig:solvation_vs_rdiel}. One can see that at $R_\diel = \SI{1.00}{\angstrom}$, the protons in water are oversolvated and the proton in alcohols is undersolvated. The oxygen present in ethers, alcohols, and water is also undersolvated, while the oxygen in aldehydes and ketones is appropriately solvated. It is unclear why such a large difference exists between water and alcohols in the solvation error associated with the protons. We can speculate that the undersolvation of ethers is due to general undersolvation of \ce{sp^3} lone pairs in oxygen.

\subsection{Self-ionization of water and the absolute potential of the standard hydrogen electrode} \label{sec:water}

A robust solvation model should not only be capable of predicting the solvation free energies of neutral molecules but also those of cations and anions. As a test of this, we examine the performance of our model for calculating the free energy associated with self-ionization of water into hydronium and hydroxide ions. Additionally, we use these species to compute the absolute potential of the standard hydrogen electrode with respect to vacuum in order to provide a consistent reference for the electron chemical potential.

The free energy $\Delta G_\rm{w}$ for the self-ionization of water,
\begin{equation}
    \ce{2H2O <=> H3O+ + OH-}
\end{equation}
is computed as,
\begin{equation}
    \Delta G_\rm{w} = G(\ce{H3O+}) + G(\ce{OH-}) - 2 G(\ce{H2O})
\end{equation}
where $G(\ce{H2O})$, $G(\ce{H3O+})$, and $G(\ce{OH-})$ are the free energies of water, hydronium, and hydroxide computed in implicit water containing \SI{1}{\molar} of a 1:1 electrolyte. The electrolyte is included in the calculation in order to balance the solute charge on hydronium and hydroxide and should only have a small effect on the computed free energies. Further details of these calculations are reported in the \SM.

\begin{figure}
    \includegraphics[width=.45\textwidth]{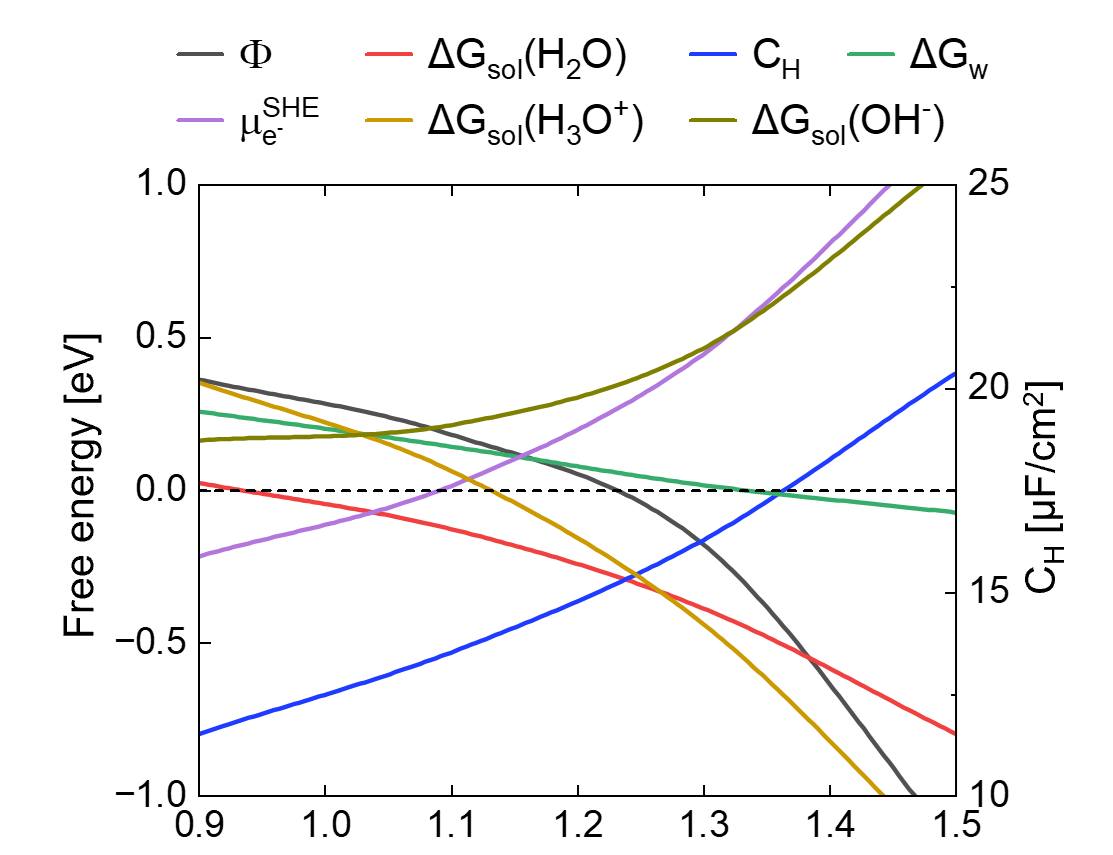}
    \caption{Errors in the self-solvation and self-ionization free energies of water ($\Delta G_\sol(\ce{H2O})$ and $\Delta G_\rm{w}$), the Helmholtz capacitance ($C_\rm{H}$) and work function ($\Phi$) at the PZC of the Au(111) surface in a \SI{1}{\molar} 1:1 aqueous electrolyte, and the absolute electron chemical potential of the standard hydrogen electrode ($\mu_{\ce{e-}}^\rm{SHE}$) computed with different values of the dielectric radius. The errors in the solvation free energies of hydronium ($\Delta G_\sol(\ce{H3O+})$) and hydroxide ($\Delta G_\sol(\ce{OH-})$) are also shown.}
    \label{fig:values_vs_rdiel}
\end{figure}

The deviation of the self-ionization free energy from the experimental value of \SI{0.83}{\eV} is plotted in Figure \ref{fig:values_vs_rdiel} for different values of the dielectric radius $R_\diel$, where it can be seen to decrease at higher values of this parameter. When using only implicit solvation, the calculated $\Delta G_\rm{w}$ were found to be significantly higher than the experimental value for all reasonable values of $R_\diel$. This is due to the fact that hydronium and hydroxide ions participate in strong hydrogen bonds with the surrounding water molecules, which is not captured in an implicit solvation model. Therefore, three explicit water molecules were included for hydrogen bonding to hydronium while four were included for hydrogen bonding to hydroxide. This significantly improves the calculated $\Delta G_\rm{w}$ so that it matches the experimental value at a dielectric radius close to \SI{1.3}{\angstrom}. The values of $\Delta G_\rm{w}$ plotted in Figure \ref{fig:values_vs_rdiel} are computed with these explicit water molecules included.

%%TODO@SMRI change y-axis scale to [-1,1]
%%SMRI Done

In calculating $\Delta G_\rm{w}$, four explicit water molecules were also included when computing the free energy of water. Additionally, an empirical hydrogen bond correction having a value of \SI{0.115}{\eV} at $R_\diel = \SI{1.00}{\angstrom}$ was applied for each explicit water included in the calculations of water, hydronium, and hydroxide. This correction is necessary to account for the entropy loss associated with formation of a hydrogen bond and is fit to reproduce the experimental self-solvation energy of water when coordinated to four other hydrogen bonding water molecules. The value of the correction ranges from \SI{0.129}{\eV} at $R_\diel = \SI{0.90}{\angstrom}$ to \SI{-0.079}{\eV} at $R_\diel = \SI{1.50}{\angstrom}$. Further details are given in the \SM.

In order to rationalize the effect of the dielectric radius on $\Delta G_\rm{w}$, we plot separately the deviation from experimental values of the solvation free energies for hydronium and hydroxide ($\Delta G_\sol(\ce{H3O+})$ and $\Delta G_\sol(\ce{OH-})$) with respect to $R_\diel$ in Figure \ref{fig:values_vs_rdiel}. By construction, the solvation free energy of water is always equal to the experimental value since this was the condition used to determine the empirical hydrogen bond correction. As one might expect, the solvation free energy of hydronium becomes more negative as $R_\diel$ increases due to the dielectric cavity moving closer to the charged solute. Surprisingly though, the solvation free energy of hydroxide becomes less negative as $R_\diel$ increases. We postulate the reason for this is that the dielectric screening is significantly stronger around H than around O, as suggested in Section \ref{sec:molecules} when comparing the solvation free energies of water, alcohols, and ethers. The difference is just more extreme in hydronium and hydroxide because they interact much more strongly with the implicit solvent. Hydronium interacts with the implicit solvent primarily through protons, while hydroxide interacts primarily through oxygen; water interacts equally through protons and oxygen. Since dielectric screening of protons appears to be stronger than for oxygen, the solvation free energy of hydronium varies more strongly with $R_\diel$ than water, while the solvation free energy of hydroxide varies more weakly than water. The dependence for hydronium is also  stronger than for hydroxide, which explains why $\Delta G_\rm{w}$ decreases for larger values of the dielectric radius.

The absolute electron chemical potential of the standard hydrogen electrode (SHE) is defined as the chemical potential of an electron (with respect to vacuum) that is in equilibrium with \SI{1}{\bar} of $\ce{H2}$ in the gas phase and protons in an electrolyte at a pH of zero according to,
\begin{equation}
    \ce{\frac12 H2 <=> e- + H+}
\end{equation}
This value can be computed by,
\begin{equation}
    \mu_{\ce{e-}}^\rm{SHE} = \frac12 G(\ce{H2}) - \mu_{\ce{H+}}
\end{equation}
where the proton chemical potential is taken as an average from the two processes involved in water self-ionization,
\begin{equation}
    \ce{H3O+ <=> H2O + H+}
\end{equation}
\begin{equation}
    \ce{H2O <=> OH- + H+}
\end{equation}
giving,
\begin{equation}
    \mu_{\ce{H+}} = \frac12 \qty[G(\ce{H3O+}) - G(\ce{OH-}) + \Delta G_\rm{w,expt}]
\end{equation}
We find this to be a more balanced definition of $\mu_{\ce{H+}}$ since it treats both cations ($\ce{H3O+}$) and anions ($\ce{OH-}$) on equal footing. Effectively, we are computing the proton chemical potential at neutral pH and then shifting it to a pH of zero with the additional term $\Delta G_\rm{w,expt}$, the experimental value of $\Delta G_\rm{w}$.

The resulting deviation of $\mu_{\ce{e-}}^\rm{SHE}$ from the experimental value of \SI{-4.44}{\eV} is plotted in Figure \ref{fig:values_vs_rdiel} with respect to the dielectric radius, showing a strong increase as $R_\diel$ becomes larger. This is related to the fact that the model predicts solvation of hydronium to become stronger as $R_\diel$ increases while solvation of hydroxide becomes weaker. While these effects partially canceled out in the dependence of $\Delta G_\rm{w}$, they add together in the dependence of $\mu_{\ce{e-}}^\rm{SHE}$ which leads to the much stronger dependence of the latter on $R_\diel$.

\subsection{Choosing a value for the dielectric radius} \label{sec:rdiel}

As mentioned in Section \ref{sec:parameterization}, the dielectric radius is the model parameter that is least straightforward to determine. It was seen in the previous sections that this parameter has a large effect on several predicted values such as the capacitance and work function of a metal surface, the solvation free energies of molecules, the self-ionization free energy of water, and the absolute electron chemical potential of the standard hydrogen electrode (SHE). We now discuss the trade offs in choosing an appropriate value of the dielectric radius and relate these to fundamental limitations of the model.

Returning to Figure \ref{fig:values_vs_rdiel}, we now compare the effect of the dielectric radius on five relevant quantities: the capacitance and work function of the Au(111) surface, the self-solvation and self-ionization free energies of water ($\Delta G_\sol(\ce{H2O})$ and $\Delta G_\rm{w}$), and the absolute chemical potential of the SHE ($\mu_{\ce{e-}}^\rm{SHE}$). It can be immediately seen that no single value of $R_\diel$ is able to reproduce the experimental values for all of these properties. 

The self-solvation free energy of water requires the smallest dielectric radius (\SI{0.93}{\angstrom}) to reproduce the experimental value of \SI{-0.27}{\eV}. This also sets an upper limit on the dielectric radius since an implicit solvation model is expected to underestimate the hydrogen bonding interactions in water. Using a dielectric radius greater than \SI{1.00}{\angstrom} would lead to an unacceptable degree of oversolvation for water and other polar molecules. For this reason, we restrict $R_\diel$ to a maximum value of \SI{1.00}{\angstrom} where the self-solvation free energy is \SI{-0.32}{\eV}.

The work function of Au(111) required the largest dielectric radius (\SI{1.47}{\angstrom}) to reproduce the experimental value of \SI{5.00}{\eV}, which is likely due to the unphysical mechanism by which the implicit solvation modifies the work function as discussed in Section \ref{sec:Au111_radii}. Instead of compressing the electron density tail by Pauli repulsion and thus reducing the inwardly directed surface dipole layer, the implicit solvent penetrates into the electron density and polarizes to create its own outwardly directed dipole layer. The implicit solvation model also neglects chemical interactions between water and the surface that lead to chemisorption or preferential orientation of water molecules at the interface.

The self-ionization free energy of water also requires a larger value of $R_\diel$ (\SI{1.34}{\angstrom}) 
to reproduce the experimental value of \SI{0.83}{\eV}, 
while the absolute electron chemical potential of the SHE requires a smaller dielectric radius (\SI{1.04}{\angstrom}) 
to reproduce the experimental value of \SI{-4.44}{\eV}. 
To rationalize these values, it is more transparent to examine the computed solvation free energies of the hydronium and hydroxide ions. Figure \ref{fig:values_vs_rdiel} indicates that both species are undersolvated at values of the dielectric radius appropriate for describing the solvation of water ($R_\diel < \SI{1.00}{\angstrom}$). When computing $\mu_{\ce{e-}}^\rm{SHE}$, the solvation free energies of hydronium and hydroxide are subtracted from one another so that most of the error cancels out and the experimental value is reproduced by a reasonably low dielectric radius. 
When computing $\Delta G_\rm{w}$, however, the solvation free energies are added together so that the experimental result is only reproduced at a high value of the dielectric radius where hydronium actually becomes oversolvated to such a degree that it cancels out the undersolvation of hydroxide. 
Correcting this issue in the model would require addressing the general undersolvation of anions compared to cations that is well known to occur in most implicit solvation models that use only the electron density of the solute to construct the cavities \cite{CANDLE}.

%work function: 1.47
%H2O solvation: 0.93
%H3O+ solvation: 1.13 (oversolvated wrt H2O)
%OH- solvation: <0 (undersolvated wrt H2O)
%mu_e: 1.04
%G_w: 1.34

Unlike the other quantities, the Helmholtz capacitance of Au(111) at the PZC only matches the experimental result for unreasonably large values of the dielectric radius. As discussed in Section \ref{sec:Au111_radii}, a large part of this is due to neglect of specific adsorption of water on the surface that becomes more favorable at anodic polarization. A more inert surface like Hg is not expected to partake in any specific adsorption and is measured to have a significantly lower Helmholtz capacitance of \SI{29}{\micro\farad\per\cm^2} \cite{Koper}. Even this is more than twice the capacitance predicted by the model using a dielectric radius of \SI{1.0}{\angstrom}. One possible reason for this underprediction may be the aforementioned neglect of compression of the electron density into the metal surface by Pauli repulsion with the solvent. This compression would allow the dielectric cavity to approach closer to the plane of the surface and increase the capacitance of the vacuum gap.

Considering that all quantities plotted in Figure \ref{fig:values_vs_rdiel} except for the self-solvation free energy of water require a dielectric radius larger than \SI{1.00}{\angstrom} to reproduce the experimental values, we choose $R_\diel = \SI{1.00}{\angstrom}$ as the `recommended' value. As mentioned earlier in this section, using a higher value would result in an unacceptable level of oversolvation of water in itself. The calculated quantities using this value are reported in Table \ref{tab:values} alongside the results obtained using the original linear+local model in VASPsol. The nonlinear+nonlocal model outperforms the linear+local model for calculating the work function of Au(111) and the absolute electron chemical potential of the SHE, while the opposite is true for the self-ionization free energy of water and the Helmholtz capacitance at the PZC. Both models perform equally for calculating the self-solvation free energy of water.

\begin{table}
\caption{Computed and experimental values for the self-solvation and self-ionization free energies of water ($\Delta G_\sol(\ce{H2O})$ and $\Delta G_\rm{w}$), the absolute electron chemical potential of the SHE ($\mu_{\ce{e-}}^\rm{SHE}$), and the work function and Helmholtz capacitance at the PZC ($\Phi$ and $C_\rm{H}$) of the Au(111) surface in an aqueous 1:1 electrolyte.}
\label{tab:values}
\begin{ruledtabular}
\begin{tabular}{lrrr}
    & nonlin.+nonloc. & lin.+loc. & expt. \\ \hline
    $\Delta G_\sol(\ce{H2O})$\footnotemark[1]  & \num{-0.32} & \num{-0.33} & \num{-0.27} \\
    $\Delta G_\rm{w}$\footnotemark[1]  & \num{1.03} & \num{0.87} & \num{0.83} \\
    $\mu_{\ce{e-}}^\rm{SHE}$\footnotemark[1]  & \num{-4.47} & \num{-4.52} & \num{-4.44} \\
    $\Phi$\footnotemark[1]  & \num{5.27} & \num{5.35} & \num{5.00} \\
    $C_\rm{H}$\footnotemark[2]  & \num{13} &  & \numrange{70}{100} \\
\end{tabular}
\end{ruledtabular}
\footnotetext[1]{\unit{\eV}}
\footnotetext[2]{\unit{\micro\farad\per\cm^2}}
\end{table}

\section{Conclusions} \label{sec:conclusions}

In summary, we have developed an implicit electrolyte model that captures the nonlinear dielectric and ionic response characteristic of electrodes at realistic operating conditions for important electrocatalytic processes such as the oxygen evolution reaction and electrochemical \ce{CO2} reduction. In addition, the regions of space in which the dielectric and ionic responses occur are determined by cavities having a nonlocal dependence on the surface or solute electron density. This prevents the `leakage' of electrolyte into regions that are too small to accommodate a single water molecule or solvated ion. The model is implemented into the Vienna Ab initio Simulation Package (VASP) by significantly extending the functionality of the VASPsol code developed by Mathew \etal \cite{VASPsol1,VASPsol2}. An additional modification allows for performing DFT calculations at constant potential rather than constant charge. We call the new implementation VASPsol++.

The nonlinear+nonlocal electrolyte model is based on a free energy functional for the solute/surface plus electrolyte. The ground state is obtained by finding the stationary point of this functional with respect to eight functional degrees of freedom: the solute atomic coordinates and electron density, the electrostatic potential, the rotational distribution function and intramolecular polarization of the solvent, and the `site occupancies' of the electrolyte species based on a translationally invariant lattice gas approximation. Analytical expressions are obtained for the ground state electrolyte degrees of freedom, while the electrostatic potential is determined by solving a nonlinear Poisson-Boltzmann (NLPB) equation and the solute electron density is determined by solving a modified Kohn-Sham equation. We implement a numerically efficient and robust algorithm for solving the NLPB equation that is based on Newton's method with a line search, the latter being found necessary for numerical stability. We also carefully control the smoothness of all quantities to allow their representation on the same finite Fourier transform (FFT) grids used in VASP to represent the potential and electron density, without requiring an increase in the density of these grids. The resulting algorithm is found to be exceptionally robust and stable while requiring only marginally more computational effort than the original VASPsol implementation.

The model was applied to several systems including an electrified aqueous Au(111) interface, an explicit water bilayer on Pt(111) in the presence of an implicit electrolyte, a large set of solvated organic molecules, and solvated hydronium and hydroxide ions. The nonlinear dielectric response was found to be necessary for reproducing the `double hump' shape experimentally observed for differential capacitance curves of metal electrodes. As with other implicit solvation models, the capacitance is underpredicted due to the neglect of specific chemisorption of water on the metal surface. Even so, the capacitance is still underpredicted compared to nonadsorbing electrodes such as Hg, likely due to an overly diffuse electron density tail extending from the surface. In reality, Pauli repulsion with the solvent electron density would compresses this tail, shrinking the vacuum gap at the surface and increasing the capacitance. The work function is overpredicted for the same reason. Thus, an improved model would explicitly include Pauli repulsion between the solvent and the electrons in the surface. When applied to at Pt(111) surface with an explicit water bilayer, the nonlocal cavity definition is found to prevent the unphysical `leakage' of the solvent into the bilayer.

The nonlinear+nonlocal model also performs well for computing solvation free energies of organic molecules in water, with the calculated values being closer to experimental values than the original LPCM model in VASPsol. Additionally, the model computes reasonable values of the self-solvation and self-ionization free energies of water and the absolute electron chemical potential of the standard hydrogen electrode. The accuracy of these computed values is comparable to the original LPCM model.

In conclusion the accuracy, computational efficiency, and numerical robustness of the nonlinear+nonlocal electrolyte model we have developed and implemented should allow for the hassle-free calculation of free energies and activation barriers of electrocatalytic processes at constant potential. Our hope is that this will open the door for routine simulation of these processes in a proper electrochemical environment, leading to new insights into important processes such as the oxygen evolution reaction and electrochemical \ce{CO2} reduction.

\section{Code availability}
The software is freely available as a patch to the original VASP code at \url{https://github.com/VASPsol}.

\section{Supporting Material} \label{SM}
See {\SM} for the derivation of the prefactor of the density kernel used for cavity construction, the derivation of the cavity potential corrections, details of the conjugate gradient algorithm used for solving the linearized Poisson-Boltzmann equation, details of the line search algorithm used in the Newton's method solver for the nonlinear Poisson-Boltzmann equation, additional results, and details of the VASP calculations.

\begin{acknowledgments}
We thank Eric Fonseca and Prof.\ Richard Hennig for insightful discussions about the original VASPsol implementation. F.K.\ acknowledges funding by the Louisiana Board of Regents (LEQSF(2019-22)-RD-A-14). Portions of this research were conducted with high performance computing resources provided by Louisiana State University (\url{http://www.hpc.lsu.edu}).
\end{acknowledgments}

\section{REFERENCES}

\bibliography{main.bib}

\end{document}

%\preprint{}

\title{A nonlocal and nonlinear implicit electrolyte model for plane wave density functional theory} %Title of paper



\author{S M Rezwanul Islam}
\affiliation{Department of Chemical Engineering, Louisiana State University, Baton Rouge, Louisiana 70803, USA}
\author{Foroogh Khezeli}
\affiliation{Department of Chemical Engineering, Louisiana State University, Baton Rouge, Louisiana 70803, USA}
\author{Stefan Ringe}
\affiliation{Department of Chemistry, Korea University, Seoul 02841, Republic of Korea}
\author{Craig Plaisance}
%\thanks{}
\affiliation{Department of Chemical Engineering, Louisiana State University, Baton Rouge, Louisiana 70803, USA}


\date{\today}

\pacs{}% insert suggested PACS numbers in braces on next line

\maketitle %\maketitle must follow title, authors, abstract and \pacs


\section{Derivation of the prefactor of the density kernel used for cavity construction}

The density kernel in eq \eqref{main-eq:w} of the main text is,
\begin{equation}
    w_i(\vbr) = \qty(\frac{2b}{2b+R_i}) \qty(\frac{1}{4\pi b^3}) \exp(-\frac{r-R_i}{b})
    \quad ,
\end{equation}
where $b = a/\sigma$. Now, consider a cavity function $S(\vbr)$ defined by a plane passing through $z=0$ so that $S(z<0)=1$ and $S(z>0)=0$,
\begin{equation}
    S(z) = 
    \begin{dcases}
        1 & z<0 \\
        0 & z>0
    \end{dcases}
    \quad .
\end{equation}
We will show that the effective density given by,
\begin{equation} \label{eq:n_i}
    n_i(\vbr) = \nc \qty(w_i \ast S)(\vbr)
    \quad ,
\end{equation}
has the value of $\nc$ for a point $\vbr$ lying in the plane $z = R_i$.
%%SR I think without the factor of n_c before the convolution, the convolution is just 1, I think this should be corrected then here?
%%CP yes, you are right. I fixed it to include the factor of n_c

To evaluate the convolution in eq \eqref{eq:n_i}, we first write the density kernel in cylindrical coordinates $(\rho,\theta,z)$ using $r=\sqrt{z^2+\rho^2}$ and integrate in the radial and angular directions to define a new function $w_i(z)$,
\begin{equation}
    w_i(z) = \int_0^{2\pi} \dd{\theta} \int_0^\infty \dd{\rho} \rho \, w_i\qty(\sqrt{z^2+\rho^2})
    =  \frac{1}{2b}\qty(\frac{2b+2\abs{z}}{2b+R_i}) \exp(-\frac{\abs{z}-R_i}{b})
    \quad .
\end{equation}
We can then integrate in the $z$ direction to obtain the effective density $n_i$ given by eq \eqref{eq:n_i},
\begin{equation}
    n_i(z) = \nc \int_z^\infty \dd{z'} w_i(z') S(z-z')
    \quad .
\end{equation}
For $z>0$ we obtain,
\begin{equation}
    n_i(z>0) = \nc \qty(\frac{2b+\abs{z}}{2b+R_i}) \exp(-\frac{\abs{z}-R_i}{b})
    \quad ,
\end{equation}
while for $z<0$ we get,
\begin{equation}
    n_i(z<0) = 2\nc \qty(\frac{2b}{2b+R_i}) \exp(\frac{R_i}{b}) - \nc \qty(\frac{2b+\abs{z}}{2b+R_i}) \exp(-\frac{\abs{z}-R_i}{b})
    \quad .
\end{equation}
It can be seen that this expression evaluates to $n_i = \nc$ for $z=R_i$.

It can also be seen that deep inside the cavity ($z<<0$) the effective electron density approaches the limiting value given by,
\begin{equation} \label{eq:ni_limit}
    n_i(z\to -\infty) = 2\nc \qty(\frac{2b}{2b+R_i}) \exp(\frac{R_i}{b})
    \quad .
\end{equation}
If a new cavity function is being constructed from $n_i$ -- for example, when constructing $S_\diel$ from $S_\solv$ according to eqs \eqref{main-eq:n_diel} and \eqref{main-eq:S_diel} in the main text -- then the limiting value given by \eqref{eq:ni_limit} must be several orders of magnitude greater than $\nc$ in order for the new cavity function to approach the limiting values of 0 and 1 far from the interface. This becomes an issue when using values of $R_i$ that are too small compared to $b$. For example, we have found that using a dielectric radius $R_\diel < \SI{0.5}{\angstrom}$ leads to a dielectric cavity that approaches a value less than \num{0.99} in the bulk electrolyte. For this reason, we do not recommend using values for any of the $R_i$ parameters less then $4a$ (assuming $\sigma = \num{0.6}$).

\section{Derivation of cavity potential corrections}

To derive the corrections to the Kohn-Sham potential arising from dependence of the van der Waals cavity on the solute electron density, we start with the expression for the total free energy given in eq \eqref{main-eq:Atot_opt} in the main text,
\begin{equation} \label{eq:A_tot_SI}
    A_\rm{tot} = A_\rm{TXC} + \intr \phi(\vbr) \rho_\sol(\vbr) + \frac{\eps}{8\pi} \intr \phi(\vbr) \laplacian \phi(\vbr) + A_\cav + A_\diel + A_\ion
    \quad ,
\end{equation}
where,
\begin{subequations}
\begin{align}
    A_\cav &= \tau \intr \abs{\grad S_\cav(\vbr)} \\
    A_\diel &= n_\rm{mol} \intr S_\diel(\vbr) \lambda_\diel(\vbr) \\
    A_\ion &= n_\rm{max} \intr S_\ion(\vbr) \lambda_\ion(\vbr)
    \quad .
\end{align}
\end{subequations}
The last three terms in eq \eqref{eq:A_tot_SI} indirectly contain the dependence of $S_\vdW$ on $\nel$, leading to the three Kohn-Sham potential corrections $v_\cav$, $v_\diel$, and $v_\ion$ defined in Section \ref{main-sec:nel_min} of the main text. These potential corrections are defined as functional derivatives of the three terms with respect to the solute electron density,
\begin{subequations}
\begin{align}
    v_\cav(\vbr) &= \fdv{A_\cav}{\nel(\vbr)} \\
    v_\diel(\vbr) &= \fdv{A_\diel}{\nel(\vbr)} \\
    v_\ion(\vbr) &= \fdv{A_\ion}{\nel(\vbr)}
\end{align}
\end{subequations}

Before proceeding, we note that the variation of $A_\cav$ can be rewritten using the relation,
\begin{equation}
    \var\abs{\grad S_\cav} = \var\sqrt{\grad S_\cav \vdot \grad S_\cav} = \frac{\grad S_\cav}{\abs{\grad S_\cav}} \vdot \grad \var{S_\cav}
    \quad ,
\end{equation}
to give,
\begin{equation}
    \var{A_\cav} = \tau \intr \frac{\grad S_\cav}{\abs{\grad S_\cav}} \vdot \grad \var{S_\cav} = -\tau \intr \var{S_\cav} \div{\frac{\grad S_\cav}{\abs{\grad S_\cav}}}
    \quad ,
\end{equation}
where integration by parts is used to obtain the form on the right hand side. By defining the quantity $\lambda_\cav(\vbr)$,
\begin{equation}
    \lambda_\cav(\vbr) = -\frac{\tau}{n_\rm{mol}}\div{\frac{\grad S_\cav}{\abs{\grad S_\cav}}}(\vbr)
    \quad ,
\end{equation}
the variations of all three terms can be written in analogous forms,
\begin{subequations}
\label{eq:delta_A}
\begin{align}
    \var{A_\cav} &= \intr \var{S_\cav(\vbr)} \lambda_\cav(\vbr) \\
    \var{A_\diel} &= \intr \var{S_\diel(\vbr)} \lambda_\diel(\vbr) \\
    \var{A_\ion} &= \intr \var{S_\ion(\vbr)} \lambda_\ion(\vbr)
    \quad .
\end{align}
\end{subequations}

Through a series of transformations, each of the terms in eq \eqref{eq:delta_A} can be cast in terms of the variation in $\nel(\vbr)$. For $A_\cav$ we obtain,
\begin{equation}
\begin{aligned}
    \var{A_\cav}
    &&= &\intr  \var{S_\cav(\vbr)} \qty(n_\rm{mol} \lambda_\cav(\vbr)) &= &\intr \var{n_\cav(\vbr)} v''_\cav(\vbr) \\
    &&= &\intr \var{S_\solv(\vbr)} \qty(\nc w_\cav \ast v''_\cav)(\vbr) &= &\intr \var{n_\solv(\vbr)} v'_\cav(\vbr) \\
    &&= &\intr \var{S_\vdW(\vbr)} \qty(-\nc w_\solv \ast v'_\cav)(\vbr) &= &\intr \var{\nel(\vbr)} v_\cav(\vbr)
    \quad ,
\end{aligned}
\end{equation}
for $A_\diel$ we obtain,
\begin{equation}
\begin{aligned}
    \var{A_\diel}
    &&= &\intr  \var{S_\diel(\vbr)} \qty(n_\rm{mol} \lambda_\diel(\vbr)) &= &\intr \var{n_\diel(\vbr)} v''_\diel(\vbr) \\
    &&= &\intr \var{S_\solv(\vbr)} \qty(\nc w_\diel \ast v''_\diel)(\vbr) &= &\intr \var{n_\solv(\vbr)} v'_\diel(\vbr) \\
    &&= &\intr \var{S_\vdW(\vbr)} \qty(-\nc w_\solv \ast v'_\diel)(\vbr) &= &\intr \var{\nel(\vbr)} v_\diel(\vbr)
    \quad ,
\end{aligned}
\end{equation}
and for $A_\ion$ we obtain,
\begin{equation}
\begin{aligned}
    \var{A_\ion}
    &&= &\intr  \var{S_\ion(\vbr)} \qty(n_\rm{max} \lambda_\ion(\vbr)) &= &\intr \var{n_\ion(\vbr)} v'_\ion(\vbr) \\
    &&= &\intr \var{S_\vdW(\vbr)} \qty(-\nc w_\ion \ast v'_\ion)(\vbr) &= &\intr \var{\nel(\vbr)} v_\ion(\vbr)
    \quad .
\end{aligned}
\end{equation}
In deriving the above expressions, we have used the definitions in Section \ref{main-sec:cavity} of the main text to write the variations of the cavity functions,
\begin{subequations}
\begin{align}
    \var{S_\cav(\vbr)} &= -\frac{S'_\cav(\vbr)}{n_\cav(\vbr)} \var{n_\cav(\vbr)} \\
    \var{S_\diel(\vbr)} &= -\frac{S'_\diel(\vbr)}{n_\diel(\vbr)} \var{n_\diel(\vbr)} \\
    \var{S_\solv(\vbr)} &= \frac{S'_\solv(\vbr)}{n_\solv(\vbr)} \var{n_\solv(\vbr)} \\
    \var{S_\ion(\vbr)} &= \frac{S'_\ion(\vbr)}{n_\ion(\vbr)} \var{n_\ion(\vbr)} \\
    \var{S_\vdW(\vbr)} &= \frac{S'_\vdW(\vbr)}{\nel(\vbr)} \var{\nel(\vbr)}
    \quad ,
\end{align}
\end{subequations}
and effective electron densities,
\begin{subequations}
\begin{align}
    \var{n_\cav(\vbr)} &= \nc \qty(w_\cav \ast \var{S_\solv}) \\
    \var{n_\diel(\vbr)} &= \nc \qty(w_\diel \ast \var{S_\solv}) \\
    \var{n_\solv(\vbr)} &= -\nc \qty(w_\solv \ast \var{S_\vdW}) \\
    \var{n_\ion(\vbr)} &= -\nc \qty(w_\ion \ast \var{S_\vdW})
    \quad ,
\end{align}
\end{subequations}
These expressions are written in terms of the effective potential corrections arising from the cavity formation free energy,
\begin{subequations}
\begin{align}
    v''_\cav(\vbr) &= v_\rm{S}\oparg{-n_\rm{mol} \lambda_\cav, n_\cav}(\vbr) \\
    v'_\cav(\vbr) &= v_\rm{S}\oparg{\nc \qty(w_\cav \ast v''_\cav), n_\solv}(\vbr) \\
    v_\cav(\vbr) &= v_\rm{S}\oparg{-\nc \qty(w_\solv \ast v'_\solv), \nel}(\vbr)
    \quad ,
\end{align}
\end{subequations}
the dielectric free energy,
\begin{subequations}
\begin{align}
    v''_\diel(\vbr) &= v_\rm{S}\oparg{-n_\rm{mol} \lambda_\diel, n_\diel}(\vbr) \\
    v'_\diel(\vbr) &= v_\rm{S}\oparg{\nc \qty(w_\diel \ast v''_\diel), n_\solv}(\vbr) \\
    v_\diel(\vbr) &= v_\rm{S}\oparg{-\nc \qty(w_\solv \ast v'_\solv), \nel}(\vbr)
    \quad ,
\end{align}
\end{subequations}
and the ionic free energy,
\begin{subequations}
\begin{align}
    v'_\ion(\vbr) &= v_\rm{S}\oparg{n_\rm{max} \lambda_\ion, n_\diel}(\vbr) \\
    v_\ion(\vbr) &= v_\rm{S}\oparg{-\nc \qty(w_\ion \ast v'_\ion), \nel}(\vbr)
    \quad ,
\end{align}
\end{subequations}
with the general cavity potential given by eq \eqref{main-eq:vS} in the main text,
\begin{equation}
    v_S\oparg{\mathcal{A},n} \equiv \fdv{n} \int \dd[3]{\vbr'} S\oparg{n}(\vbr') \, \mathcal{A}(\vbr')
    \equiv \frac{1}{n} S'\mathcal{A}
    \quad .
\end{equation}

\section{Conjugate gradient algorithm used for solving the linearized Poisson-Boltzmann equation}

The linearized Poisson-Boltzmann (LPB) equation is solved using a modified conjugate gradient method to determine the Newton step direction $\Delta\phi_i$. The modification involves how the $G=0$ component of $\Delta\phi_i$ (in the reciprocal space representation) is determined and leads to faster convergence. The first step is to define the residual $r$ of the LPB equation,
\begin{equation}
     r\oparg{\Delta\phi_i} \equiv R_i - \hat{L}_\rm{LPB}\oparg{\phi_i} \Delta\phi_i = 0
     \quad .
\end{equation}
A step is then taken in the $G=0$ direction to eliminate the $G=0$ component of the initial LPB residual $r_0 = R_i$,
\begin{equation}
    \Delta\phi_i = \frac{\expval{r_0}}{\lambda_0}
    \quad .
\end{equation}
The $G=0$ component $\lambda_0$ of the LPB operator is equivalent to the $G=0$ component of the ionic response function $\epsilon_\bulk \kappa^2$,
\begin{equation}
    \lambda_0 \equiv \expval{\hat{L}_\rm{LPB}} = \expval{\epsilon_\bulk \kappa^2}
    \quad .
\end{equation}
(the notation $\expval{\cdots}$ indicates the $G=0$ component of a field or operator) 

Each subsequent step $j$ of the conjugate gradient method is taken along a search direction $p_j$ having the property that $\expval{\hat{L}_\rm{LPB}\,p_j} = 0$. This property ensures that the $G=0$ component of the LBP residual $r$ remains zero after it is eliminated in the initial step. Such a step direction is given by,
\begin{equation}
    p_j = z_j - \frac{\lambda_j}{\lambda_0} + \beta_j p_{j-1}
    \quad ,
\end{equation}
where $z_j = \hat{K}r_j$ is the preconditioned LPB residual and $\beta_j$ is the usual $\beta$ parameter of the conjugate gradient method,
\begin{equation}
    \beta_j = \frac{\expval{r_j z_j}}{\expval{r_{j-1} z_{j-1}}}
    \quad .
\end{equation}
The parameter $\lambda_j$ is the $G=0$ component of the ionic response in the direction $\Delta\phi=z_j$,
\begin{equation}
    \lambda_j \equiv \expval{\hat{L}_\rm{LPB}\,z_j} = \expval{\epsilon_\bulk \kappa^2 z_j}
    \quad ,
\end{equation}
and the preconditioner $\hat{K}$ is the inverse of the Poisson operator expressed in a diagonal reciprocal space representation as,
\begin{equation}
    \hat{K}_{\vb{G}} = 
    \begin{dcases}
        \hfil 1 & G=0 \\
        \hfil\frac{1}{G^2} & G>0
    \end{dcases}
    \quad .
\end{equation}
Once the step direction is determined, $\Delta\phi_i$ is updated according to the usual formula,
\begin{equation}
    \Delta\phi_i \gets \Delta\phi_i + \alpha_j p_j
    \quad ,
\end{equation}
with the step size given by,
\begin{equation}
    \alpha_j = \frac{\expval{r_j z_j}}{\expval{p_j\,\hat{L}_\rm{LPB}\,p_j}}
\end{equation}
Convergence of the LPB solver is obtained when the RMS of the LPB residual $r$ is a factor of 10 smaller than the RMS of the NLPB residual $R_i$. This ensures that the Newton step direction $\Delta\phi_i$ is sufficiently accurate relative to the NLPB residual while minimizing the number of iterations spent in the LPB solver.

\section{Line search algorithm used in the Newton's method solver for the nonlinear Poisson-Boltzmann equation}

A a backtracking line search is used to determine the step size $\alpha$ taken during each iteration of the Newton's method solver for the NLPB equation,
\begin{equation}
    \phi_{i+1} = \phi_i + \alpha \Delta\phi_i
    \quad .
\end{equation}
The following condition is used for determining whether the current step size is acceptable,
\begin{equation} \label{eq:ls_condition}
A_\rm{tot}[\phi_{i+1}] \geq A_\rm{tot}[\phi_i] + \alpha cm   
\quad ,
\end{equation}
where $c = \num{0.1}$ and $m$ is the first order change in the free energy $A_\rm{tot}$ with respect to the step size,
\begin{equation}
m = \dv{A_\rm{tot}}{\alpha} = \intr \Delta\phi_i \fdv{A_\rm{tot}}{\phi}\oparg{\phi_i} = \intr \Delta\phi_i(\vbr) R_i(\vbr)
\quad .
\end{equation}
The line search begins with $\alpha = 1$, and reduces it sequentially by factors of two ($\alpha \gets \frac12\alpha$) until the condition in eq \eqref{eq:ls_condition} is satisfied.

\section{Details of the VASP calculations}

All electronic structure calculations were performed using spin-unpolarized density functional theory (DFT) as implemented in the Vienna Ab-initio Simulation Package (VASP) \cite{VASP}. The exchange-correlation energy was calculated at the generalized gradient approximation (GGA) level using the Bayesian error estimation functional with van der Waals correlation (BEEF-vdW) functional \cite{BEEF}. The wave functions were constructed in a plane wave basis up to an energy cutoff of \SI{400}{\eV}, with the projector augmented wave (PAW) method \cite{PAW1,PAW2} used to describe the higher energy features. For the determination of the Kohn-Sham orbital populations, the second order Methfessel-Paxton method \cite{MP} with a width of \SI{0.2}{\eV} was used for metal surfaces while an error function distribution with a width of \SI{0.02}{\eV} was used for molecular species. The Brillioun zone was sampled using a \numproduct{5 x 5} $\Gamma$-centered k-point mesh for the metal surfaces and only the $\Gamma$-point for molecular species. The SCF cycle used a convergence criterion of \SI{1e-4}{\eV} except for calculations used to compute differential capacitance curves. These calculations require a well-converged Fermi level, so a tighter SCF convergence criterion of \SI{1e-6}{\eV} was used. Geometry optimization was performed using the conjugate gradient algorithm with a convergence criterion of \SI{0.05}{\eV\per\angstrom} for the force on each atom, except for differential capacitance calculations where the atoms were frozen at the positions optimized at the PZC. Again, this was necessary to ensure a well-converged Fermi level and was found to have negligible impact on the computed capacitance.

Atom and molecular species were places in a \qtyproduct{16 x 16 x 16}{\angstrom} box. The Au(111) surface was modeled as a 6-layer slab using a \numproduct{2 x 2} surface unit cell with the middle two layers frozen to the experimental bulk positions. The Pt(111) surface was modeled as a 4-layer slab using a \numproduct{2 x 2} surface unit cell with the bottom two layers frozen to the experimental bulk positions. The unit cells for Au(111) and Pt(111) were \SI{64}{\angstrom} and \SI{36.81}{\angstrom}, respectively, in the direction parallel to the surface. A larger electrolyte width was required for the Au(111) surface since calculations were performed at dilute electrolyte concentrations where the Debye length approaches \SI{30}{\angstrom}.

The hybrid solvation detailed in ref \citenum{Foroogh} was used to compute the free energies of water, hydronium, and hydroxide. This required the use of a hydrogen bonding correction $G_\rm{W,corr}$ that was added to the free energy of each explicit hydrogen bonding water. Using this approach, the free energy of a solute \ce{A} was computed as,
\begin{equation}
    G(\ce{A}) = G(\ce{A.nW}) + n G_\rm{W,corr} - n G(\ce{W})
    \quad ,
\end{equation}
where $G(\ce{A.nW})$ is the free energy computed for A hydrogen bonded to $n$ explicit water and $G(\ce{W})$ is the free energy of a single implicitly solvated water molecule. The hydrogen bonding correction $G_\rm{W,corr}$ was determined by requiring the self-solvation free energy of water, hydrogen bonded to four explicit water molecules, be equal to the experimental value of \SI{-0.27}{\eV}. The resulting correction is reported in Table \ref{tab:Gwcorr} for different values of the dielectric radius.

\begin{table}
\caption{Values of the hydrogen bonding correction computed for different values of $R_\diel$ in the nonlinear+nonlocal model. The value computed in the linear+local model is also given.}
\label{tab:Gwcorr}
\begin{ruledtabular}
\begin{tabular}{lrrrrrr}
    $R_\diel$\footnote{\unit{\angstrom}} & \num{0.9} & \num{1.0} & \num{1.1} & \num{1.3} & \num{1.5} & linear+local \\
    $G_\rm{w,corr}$\footnote{\unit{\eV}} & \num{0.129} & \num{0.115} & \num{0.095} & \num{0.028} & \num{-0.079} & \num{0.120} \\
\end{tabular}
\end{ruledtabular}
\end{table}

\section{Additional results}

\begin{figure}[H]
    \includegraphics[width=\textwidth]{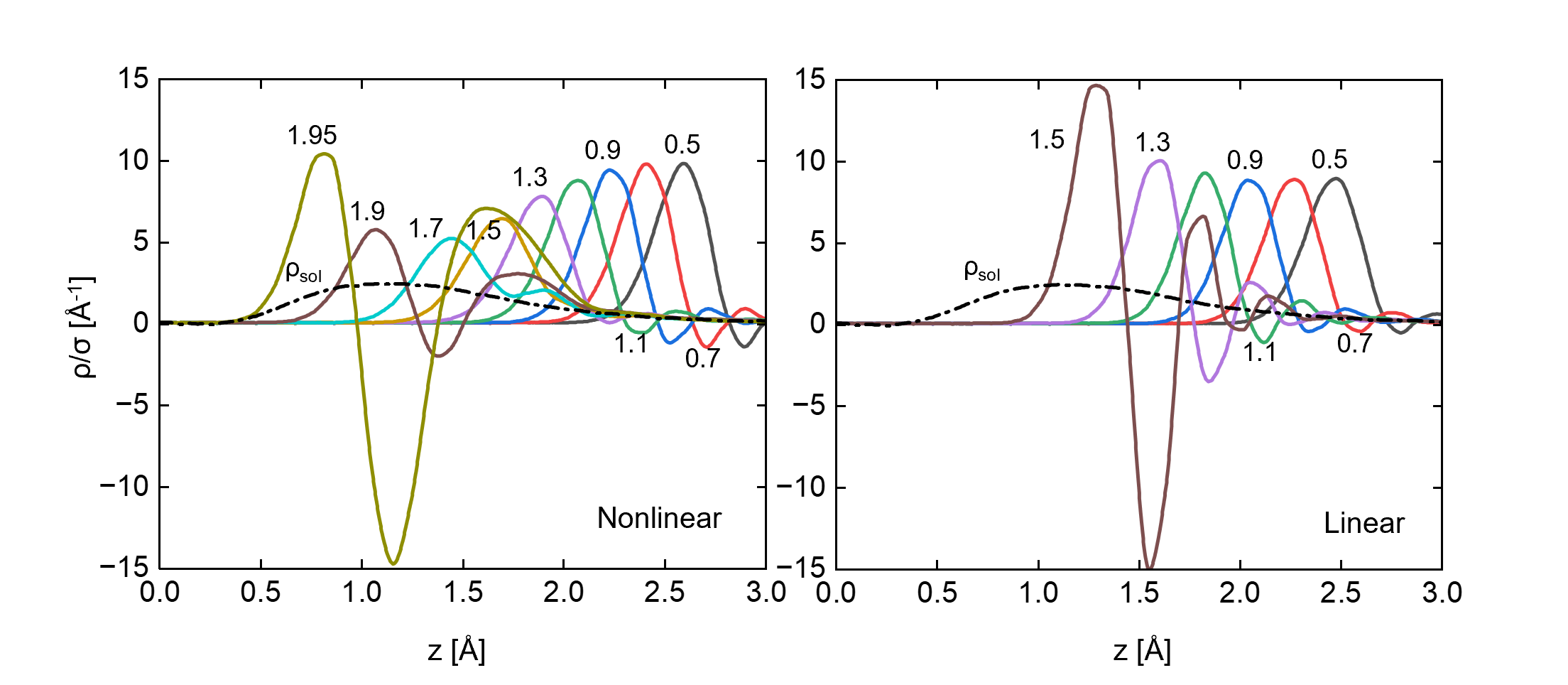}
    \caption{Bound charge distributions for the Au(111) surface in an aqueous \SI{1}{\molar} 1:1 electrolyte calculated at the PZC for the indicated values of $R_\diel$. Results using both nonlinear (left) and linear (right) dielectric screening models are shown. The horizontal axes correspond to the distance normal to the plane of the surface.}
    \label{fig:SI_bound_charge}
\end{figure}

\begin{figure}[H]
    \includegraphics[width=.45\textwidth]{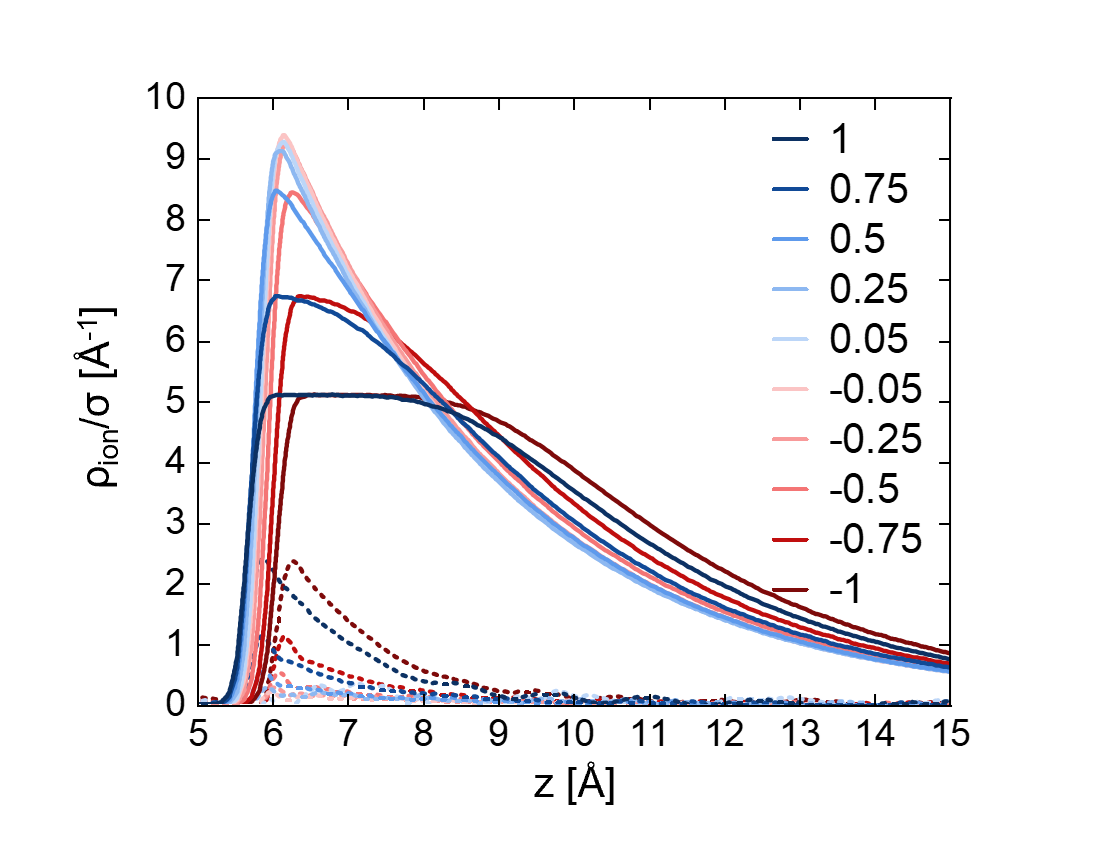}
    \caption{Total (solid lines) and unscreened (dashed lines) ionic charge distributions for the Au(111) surface in an aqueous \SI{1}{\molar} 1:1 electrolyte calculated at different surface charge densities. The unscreened ionic charge density is defined as $\rho_\ion+\rho_\rm{b}$. All ionic charge densities have been normalized by the surface charge density, with each curve labeled by the charge (in units of \unit{\e}) in a \numproduct{2 x 2} unit cell of a six-layer slab. The horizontal axis corresponds to the distance normal to the plane of the surface.}
    \label{fig:SI_ionic_charge}
\end{figure}

\begin{figure}[H]
    \includegraphics[width=\textwidth]{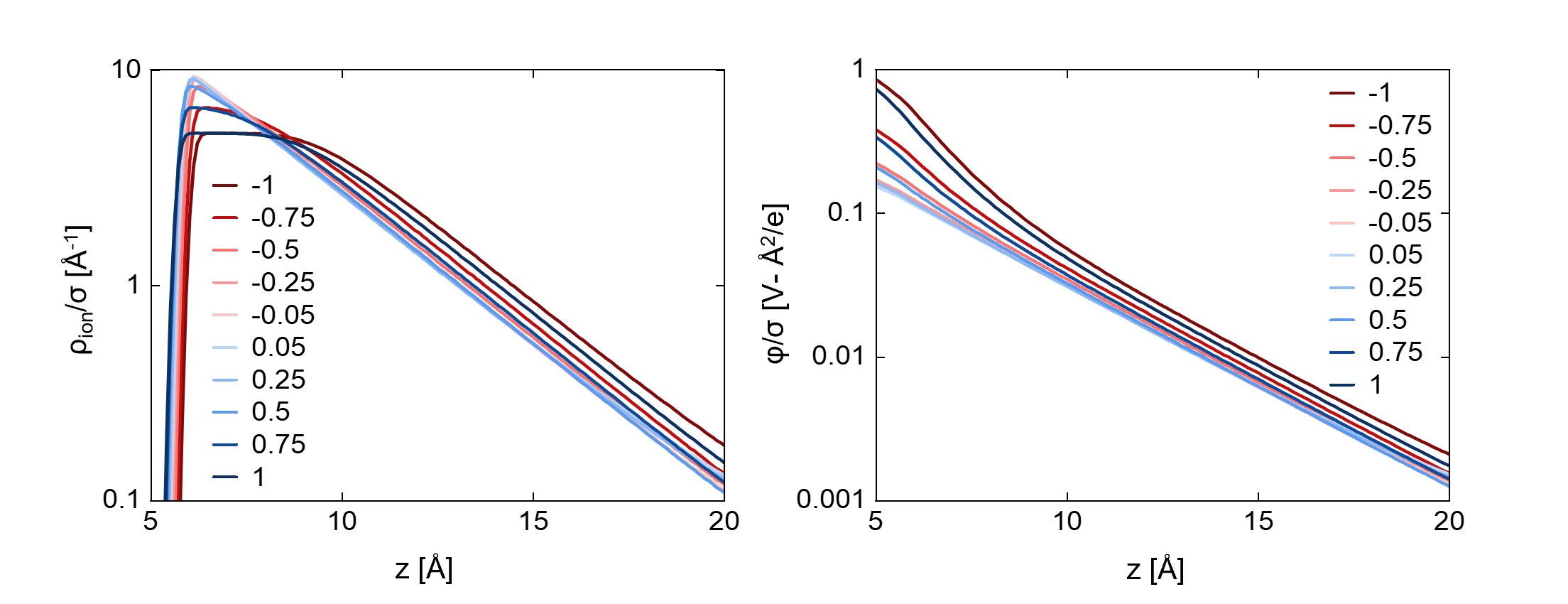}
    \caption{Ionic charge distribution (left) and electrostatic potential (right) on a logarithmic scale, showing the linear region corresponding to the diffuse layer and the nonlinear region where dielectric and/or ionic saturation occur.
    The ionic charge density and electrostatic potential have both been normalized by the surface charge density, with each curve labeled by the charge (in units of \unit{\e}) in a \numproduct{2 x 2} unit cell of a six-layer slab. The horizontal axes correspond to the distance normal to the plane of the surface.}
    \label{fig:log_ionic_cavity}
\end{figure}

\bibliography{main.bib}